\documentclass[fleqn,usenatbib,linenumbers]{mnras}

\usepackage[T1]{fontenc}
\usepackage{ae,aecompl}
\usepackage{graphicx}	
\usepackage{amsmath}	
\usepackage{amssymb}	
\usepackage{subfigure}
\usepackage{threeparttable}

\title[Measuring cosmological parameters with GRBs]{Measuring cosmological parameters with a luminosity-time correlation of gamma-ray bursts}
\author[Hu, Wang \& Dai]{
J. P. Hu$^{1}$,
F. Y. Wang$^{1,2}$\thanks{E-mail: fayinwang@nju.edu.cn}, Z. G. Dai$^{3,1}$
\\
$^{1}$School of Astronomy and Space Science, Nanjing University, Nanjing 210093, China\\
$^{2}$Key Laboratory of Modern Astronomy and Astrophysics (Nanjing University), Ministry of Education, Nanjing 210093, China\\
$^{3}$Department of Astronomy, School of Physical Sciences, University of Science and Technology of China, Hefei 230026, Anhui, China\\
}

\date{Accepted XXX. Received YYY; in original form ZZZ}

\pubyear{2021}
\def\ltb{$L_{0}$ -- $t_{\rm b}$}

\begin{document}
\label{firstpage}
\pagerange{\pageref{firstpage}--\pageref{lastpage}}
\maketitle

\begin{abstract}
Gamma-ray bursts (GRBs), as a possible probe to extend the Hubble diagram to high redshifts, have attracted much attention recently. In this paper, we select two samples of GRBs that have a plateau phase in X-ray afterglow. One is short GRBs with plateau phases dominated by magnetic dipole (MD) radiations. The other is long GRBs with gravitational-wave (GW) dominated plateau phases. These GRBs can be well standardized using the correlation between the plateau luminosity $L_0$ and the end time of plateau $t_b$. The so-called circularity problem is mitigated by using the observational Hubble parameter data and Gaussian process method. The calibrated \ltb ~correlations are also used to constrain $\Lambda$CDM and $w(z)$ = $w_{0}$ models. Combining the MD-LGRBs sample from Wang et al. (2021) and the MD-SGRBs sample, we find $\Omega_{m} = 0.33_{-0.09}^{+0.06}$ and $\Omega_{\Lambda}$ = $1.06_{-0.34}^{+0.15}$ excluding systematic uncertainties in the nonflat $\Lambda$CDM model. Adding type Ia supernovae from Pantheon sample, the best-fitting results are $w_{0}$ = $-1.11_{-0.15}^{+0.11}$ and $\Omega_{m}$ = $0.34_{-0.04}^{+0.05}$ in the $w=w_0$ model. These results are in agreement with the $\Lambda$CDM model. Our result supports that selection of GRBs from the same physical mechanism is crucial for cosmological purposes.
	
\end{abstract}



\section{Introduction}\label{sec:intro}

Gamma-ray bursts (GRBs) are among the most energetic phenomena in our Universe \citep{Meszaros2006,2015PhR...561....1K}. Therefore, GRBs can be used as promising tools to probe the properties of high-redshift universe, including cosmic expansion and dark energy \citep{2004ApJ...612L.101D,2004ApJ...613L..13G,2006MNRAS.369L..37L,2007ApJ...660...16S,2008MNRAS.391L...1K,2013IJMPD..2230028A}, reionization epoch \citep{2004ApJ...601...64B,2006PASJ...58..485T}, star formation rate \citep{1997ApJ...486L..71T,2002ApJ...564...23B,2008ApJ...683L...5Y,2013A&A...556A..90W,2017ApJ...850..161T,Zhang2018,2020MNRAS.498.5041L}, and metal enrichment history of the Universe \citep{Wang2012,2015MNRAS.447.2575V,2015A&A...580A.139H}. More details information about the cosmological implications of GRBs can be found in reviews \citep{2007AIPC..937..532B,2015NewAR..67....1W}.

According to the duration time $T_{90}$, GRBs are usually divided into two categories: long GRBs (LGRBs; $T_{90}>2$\,s) and short GRBs (SGRBs; $T_{90}<2$\,s) \citep{1993ApJ...413L.101K}. LGRBs are thought to arise when a massive star undergoes core collapse. Meanwhile, the progenitors of SGRBs are associated with the mergers of double neutron stars (NS-NS) or a neutron star and a black hole binary (NS-BH) \citep{2006ARA&A..44..507W}. GRBs are observed up to redshift $z = 9.4$ \citep{2011ApJ...736....7C}, much more distant than type Ia supernovae (SNe Ia). Therefore they can fill the gap between SNe Ia and cosmic microwave background (CMB).

Similar to SNe Ia, it has been proposed to use GRB correlations to standardize their energies and/or luminosities. Until now, a lot of GRB correlations have been proposed for cosmological purpose. They can be divided into three categories, such as prompt correlations, afterglow correlations and prompt-afterglow correlations \citep[for reviews, see][]{2015NewAR..67....1W,2017NewAR..77...23D}. The prompt correlations include the Amati correlation \citep{2002A&A...390...81A}, the Ghirlanda correlation \citep{2004ApJ...616..331G}, the $L_{\rm iso}$ - $t_{\rm lag}$ correlation \citep{2000ApJ...534..248N}, and the Yonetoku correlation \citep{2004ApJ...609..935Y}. Afterglow correlations mainly include the \ltb~ correlation \citep{2008MNRAS.391L..79D}, the $L_{\rm O, 200s} - \alpha_{\rm O, >200s}$ correlation \citep{2012MNRAS.426L..86O} and the $L_{\rm X}(T_{\rm a})-T_{\rm X,a}^{*}$ and $L_{\rm O}(T_{\rm a})-T_{\rm O,a}^{*}$ correlations \citep{2009MNRAS.393..253G}. Like the Liang-Zhang correlation \citep{2005ApJ...633..611L} $E_{\rm \gamma,afterglow}-E_{\rm X,prompt}$ correlation \citep{2007ApJ...670..565L}, the $L_{\rm X,afterglow}-E_{\rm \gamma,prompt}$ correlation \citep{2007ApJ...670.1254B}, the $L_{\rm X}(T_{\rm a})-L_{\rm \gamma,iso}$ correlation \citep{2011MNRAS.418.2202D}, and the Combo correlation \citep{2015A&A...582A.115I}, these relations belong to the prompt-afterglow correlation. The \ltb~ correlation is similar to the Combo correlation \citep{2015A&A...582A.115I}. The main difference between these two correlations is that the Combo correlation includes an additional parameter, spectral peak energy $E_{p}$. These correlations might also serve as discriminating factors among different GRB classes, as several of them hold different forms for SGRBs and LGRBs. Some of these correlations could be used to explore the high-redshift universe \citep{2008MNRAS.391..577A,2008MNRAS.391L..79D,2008ApJ...680...92L,2008PhRvD..78l3532W,2017A&A...598A.112D,2017heas.confE...2D,2019MNRAS.486L..46A,2020MNRAS.499..391K,2021ApJ...908..181M,2021MNRAS.501.3515M,Khadka21}.

If the central engine of GRBs is a magnetar, energy injection from the magnetar can cause a shallow decay phase in afterglows \citep{1998A&A...333L..87D,2001ApJ...552L..35Z}. This prediction has been confirmed by $Swift$ observations. A subset of GRBs shows a plateau in early X-ray light curves from \emph{Swift} observations \citep{2006ApJ...642..354Z,2006ApJ...642..389N,2006ApJ...647.1213O,2007ApJ...670..565L}. Similar to supernovae, only SNe Ia formed from the same mechanism, i.e., white dwarfs accrete mass from a binary companion, can be regarded as standard candles. Therefore, for GRBs with plateau phases, only the plateaus caused by the same physical mechanism should be considered. Different ways of the losing rotational energy of magnetar can explain the decaying indices of the X-ray light curves. For instance,  decay indices -2 and -1 following plateau phases have been explained by magnetic dipole (MD) radiation and gravitational wave (GW) emission \citep{2001ApJ...552L..35Z,2016MNRAS.458.1660L}, respectively. In order to standardize GRBs, a more detailed classification of GRBs could be performed using different decay indices. Recently, \cite{2021prepWang} selected LGRBs showing plateau phases dominated by MD radiations from newly born magnetars. These LGRBs are standardized using the $L_0-t_b$ correlation. Besides the MD radiation \citep{2007ApJ...665..599T,2010MNRAS.409..531R}, the rotational energy can also be lost by GW emission \citep{2013PhRvD..88f7304F,2015ApJ...798...25D}. However, plateau phases of SGRBs dominated by MD radiation and LGRBs dominated by GW radiation have not been studied.

In this paper, we investigate SGRBs whose spin-down is dominated by MD radiations (MD-SGRBs sample) and LGRBs whose spin-down is dominated by GW radiations (GW-LGRBs sample), and explore the $L_0-t_b$ correlation. Then we measure cosmological parameters in $\Lambda$CDM model and $w(z) = w_{0}$ models. The organization of this paper is as follows. In next section, we briefly introduce the process of our sample selection. Section 3 is devoted to studying the fits and calibration of the \ltb{} ~correlation. In Section 4, we show the cosmological implications of the \ltb{} ~correlation. Finally, conclusions and discussions are given in the last section.

\section{Sample selection}\label{sec:selec}

A newly born magnetar could spin-down through a combination of the electromagnetic dipole and gravitational wave quadrupole emissions \citep{2016MNRAS.458.1660L}. If GW radiation comes from non-axisymmetry in the neutron star, the spin-down law is \citep{1983bhwd.book.....S}
\begin{eqnarray}
	-I\Omega \dot{\Omega} = \frac{B_{\rm P}^{2}R^{6}\Omega^{4}}{6c^{3}} + \frac{32GI^{2}\epsilon^{2}\Omega^{6}}{5c^{5}},
	\label{q_gw}
\end{eqnarray}
where $I$ is the moment of inertia, $\Omega$ and $\dot{\Omega}$ are the angular frequency and its time derivative, $B_{\rm P}$ is the dipole component of the magnetic field at the pole, $R$ is the neutron star radius, $\epsilon$ is the ellipticity, and $c$ and $G$ are the speed of light and Newton's gravitational constant, respectively. Considering the case when MD radiation or GW emission dominates, the spin-down luminosity evolves with time as \citep{2016MNRAS.458.1660L,2020ApJ...898L...6L}

\begin{eqnarray}
	L(t) = \left\{
	\begin{array}{ll}
		L_{\rm em,0} \times (1 + \frac{t}{\tau_{\rm em}})^{-2},  &  \textnormal{MD-dominated}, \\
		L_{\rm em,0} \times (1 + \frac{t}{\tau_{\rm gw}})^{-1},  &  \textnormal{GW-dominated},
	\end{array}
	\right.
	\label{q_ev}
\end{eqnarray}
where
\begin{eqnarray}
   \tau_{\rm em} = \frac{3c^{3}I}{B_{\rm P}^{2}R^{6}\Omega^{2}_{0}},\ \tau_{\rm gw} = \frac{5c^{5}}{128GI\epsilon^{2}\Omega^{4}_{0}}, \	L_{\rm em,0} = \frac{\eta I\Omega_{0}^{2}}{2\tau_{\rm em}},
	\label{q_l0}
\end{eqnarray}
where $L_{\rm em,0}$ and $\Omega_{0}$ are the luminosity and angular frequency at $t = 0$, respectively. $\eta$ accounts for the
efficiency in converting spin-down energy into electromagnetic
radiation. Equation (\ref{q_ev}) can be rewritten as

\begin{eqnarray}
	L(t) = \left\{
	\begin{array}{ll}
		L_{0} \times \frac{1} {(1 + t/t_{\rm b})^{-2}}, \ \  &  \textnormal{MD-dominated}, \\
		L_{0} \times \frac{1} {(1 + t/t_{\rm b})^{-1}}, \ \  &  \textnormal{GW-dominated},
	\end{array}
	\right.
	\label{q_rev}
\end{eqnarray}
where parameters $L_{0}$ and $t_{\rm b}$ represent the plateau luminosity and the end time of the plateau. The values of $L_{0}$ and $t_{\rm b}$ can be obtained by fitting the plateau phase in the X-ray afterglow light curve using equation (\ref{q_rev}).

In this paper, we select two samples: the SGRBs whose spin-down is dominated by MD radiation (MD-SGRBs sample) and the LGRBs whose spin-down is dominated by GW radiation (GW-LGRBs sample). These two samples are selected from the total \emph{Swift} GRBs up to October 2020. The XRT data is downloaded from the UK Swift Science Data Center\footnote{https://www.swift.ac.uk/}. In the MD-SGRBs sample, all the well sampled X-ray afterglows have a plateau phase with a constant luminosity followed by a decay index of $-2$ in X-ray light curves. There is a similar behavior in the GW-LGRBs sample but with a decay index of $-1$ in X-ray light curves. These behaviors are well predicted by the newly born magnetars as central engine. All GRBs in our samples are selected using the following five criteria.

\begin{itemize}	
	\item Enough data points at plateau and decay phases are required and their distribution is relatively homogeneous.
	\item There are no weak flares, especially during the plateau. Weak flares maybe yield wrong values of the plateau luminosity $L_{0}$ and the end time of plateau $t_{\rm b}$.
	\item Combining the XRT and BAT data, there exists an expected plateau.
	\item The decay phase should span a long time. The duration of the decay phase ($T_{\rm decay}$) can be described by a simple function $(t_{\rm last} - t_{\rm b})/t_{\rm b}$, where $t_{\rm last}$ is the time of the last point.
	\item The plateau phase is followed by light curves with $t^{-2}$ and $t^{-1}$ in the MD-SGRBs sample and the GW-LGRBs sample, respectively.
\end{itemize}
The first two criteria are to ensure the fitting results are reliable. The other three criteria are utilized to pick out the SGRBs and LGRBs which we need. In the process of selecting the GW-LGRBs sample, we also consider a special case that the early time of the spin-down is dominated by the GW emission ($t^{-1}$) then dominated by the DM radiation ($t^{-2}$) or the external shock emission ($t^{-1.2}$) \citep{Sari1998}.

\begin{figure}
	\centering
	\includegraphics[width=0.31\hsize]{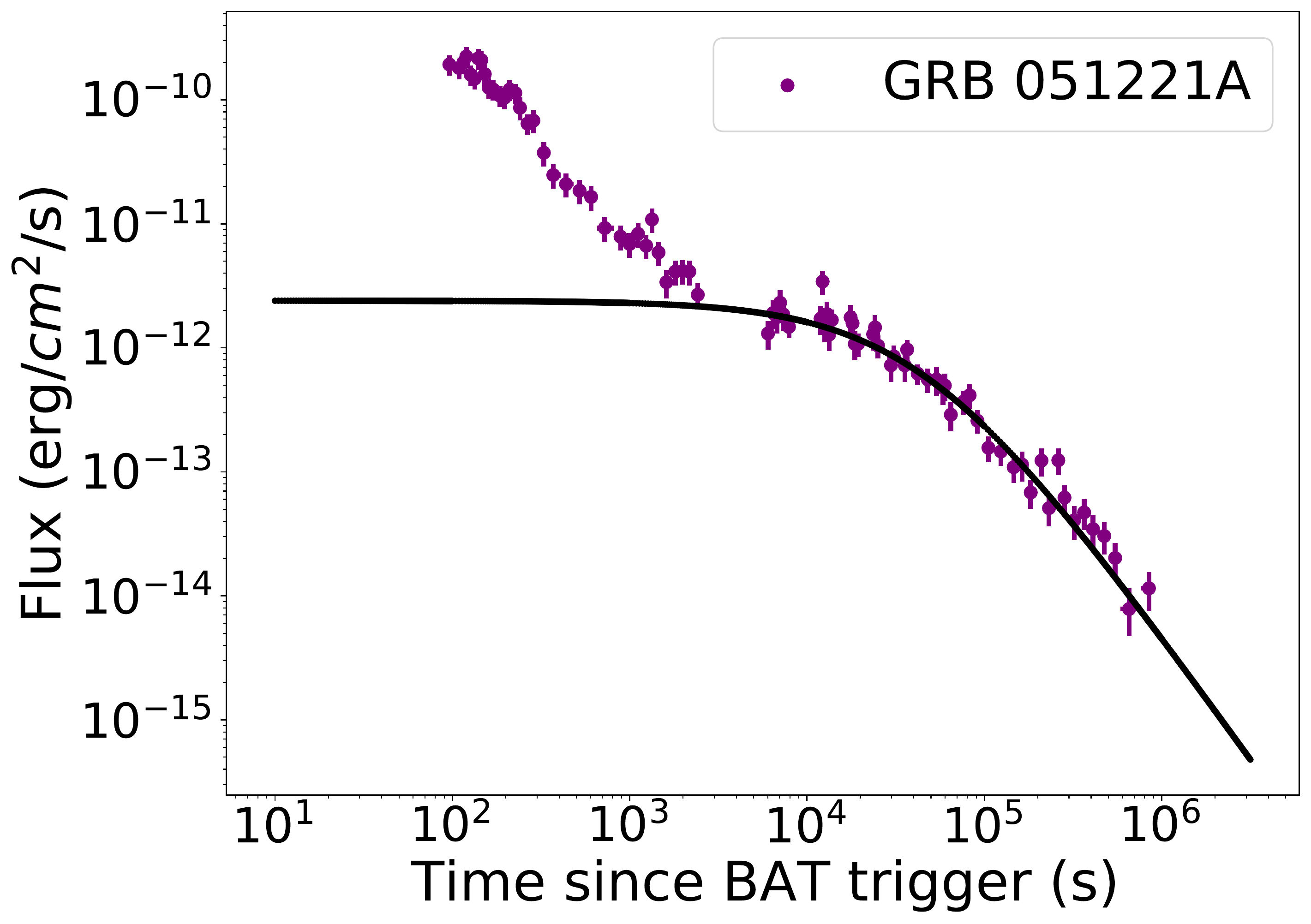}
	\includegraphics[width=0.31\hsize]{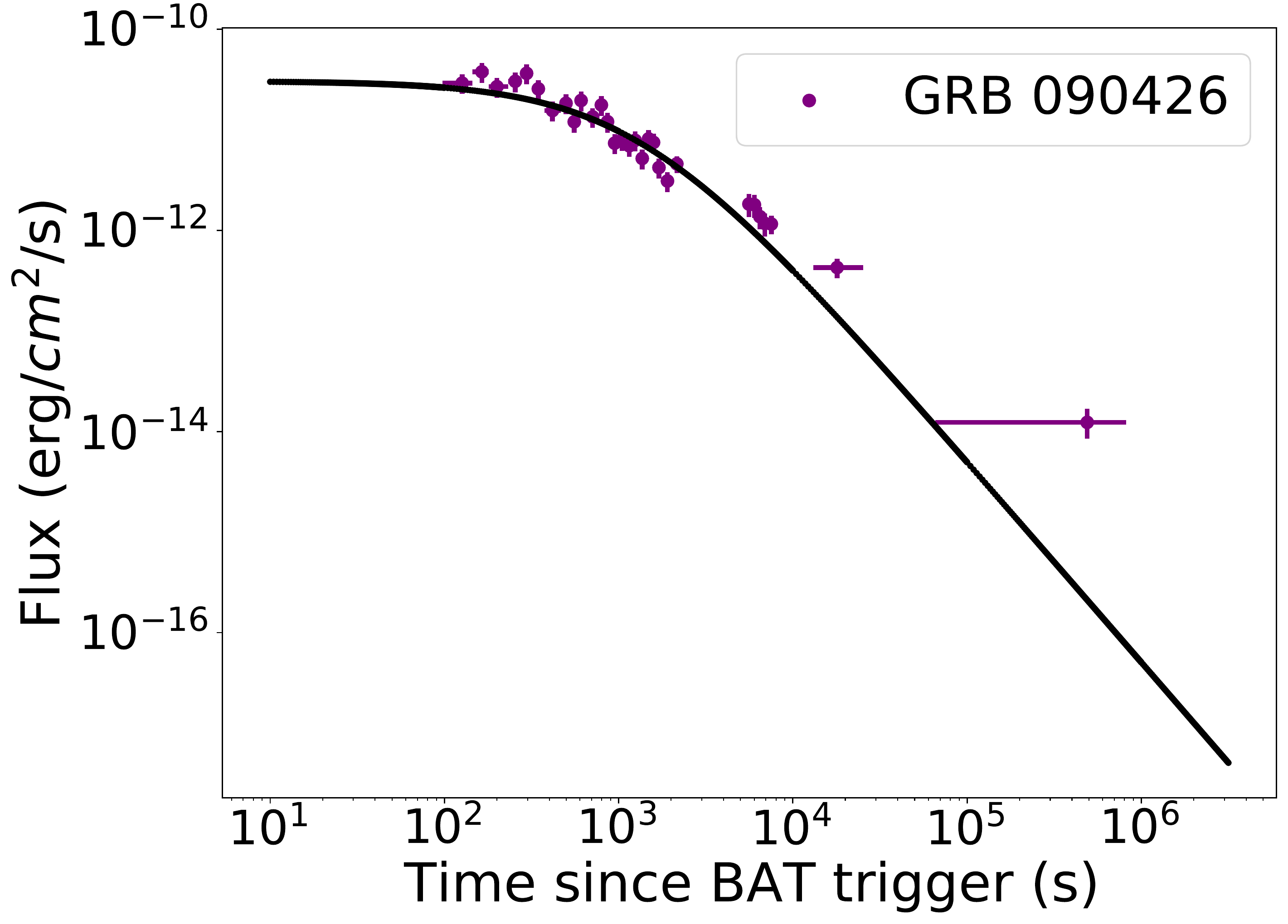}
	\includegraphics[width=0.31\hsize]{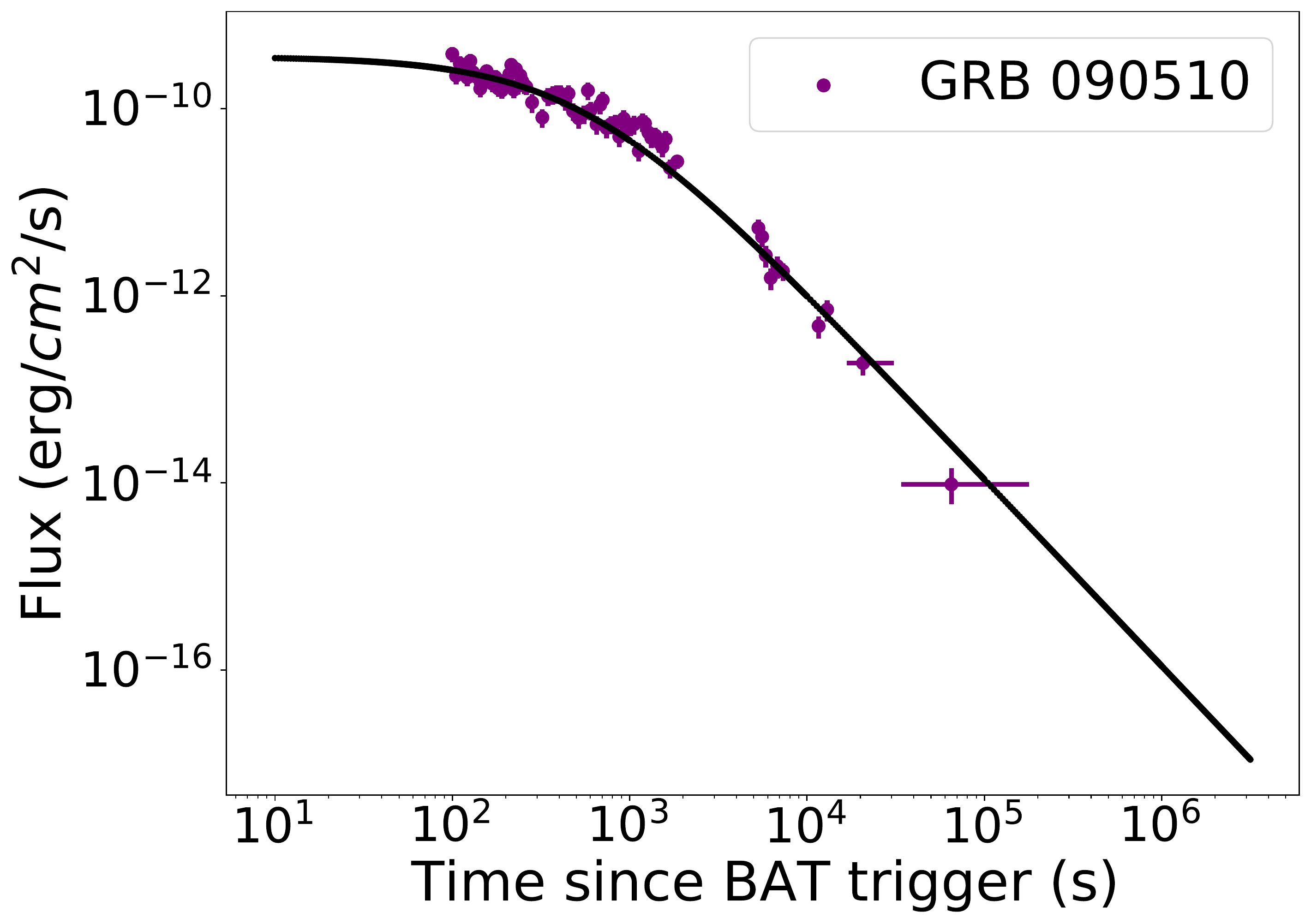}
	\includegraphics[width=0.31\hsize]{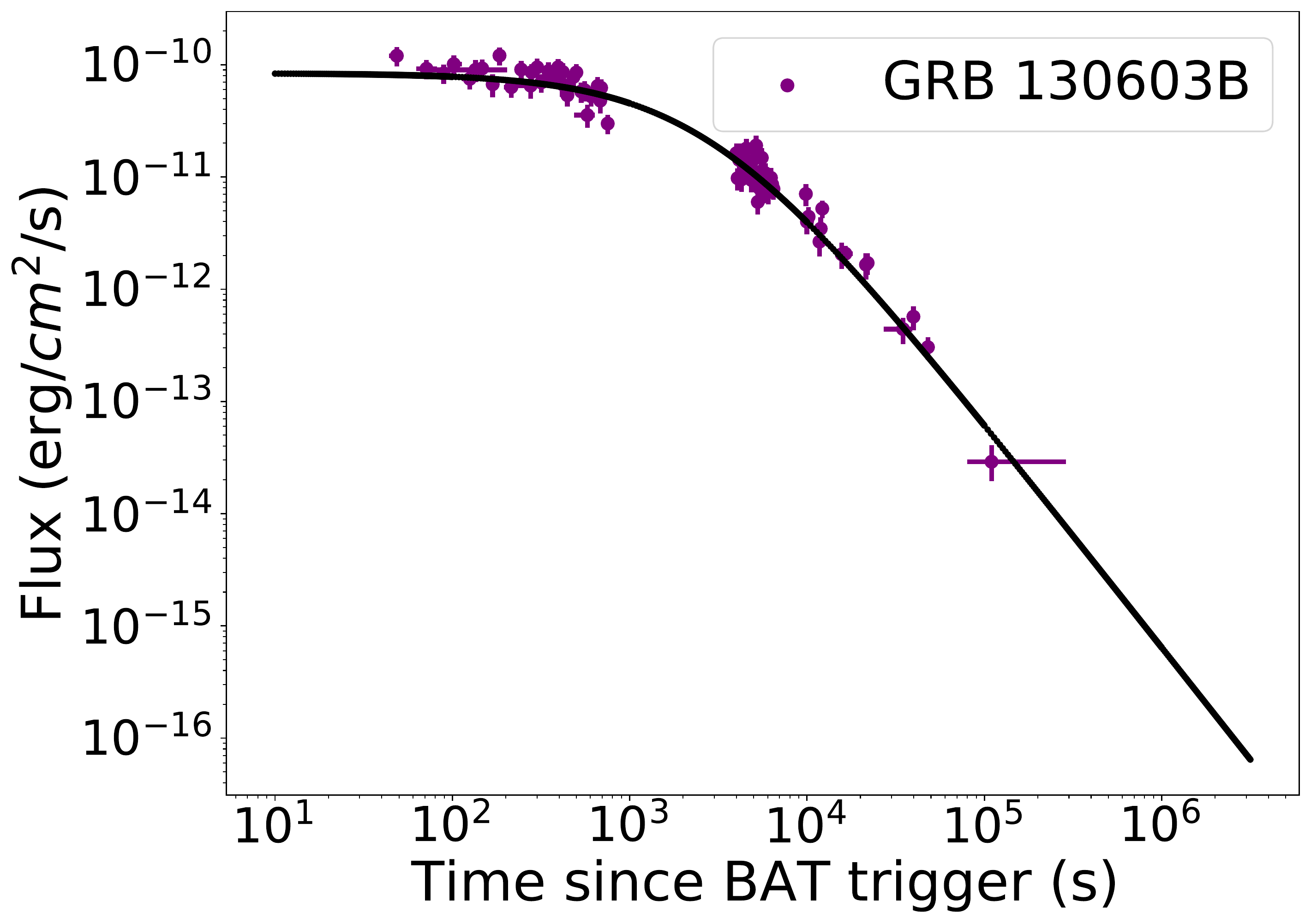}
	\includegraphics[width=0.31\hsize]{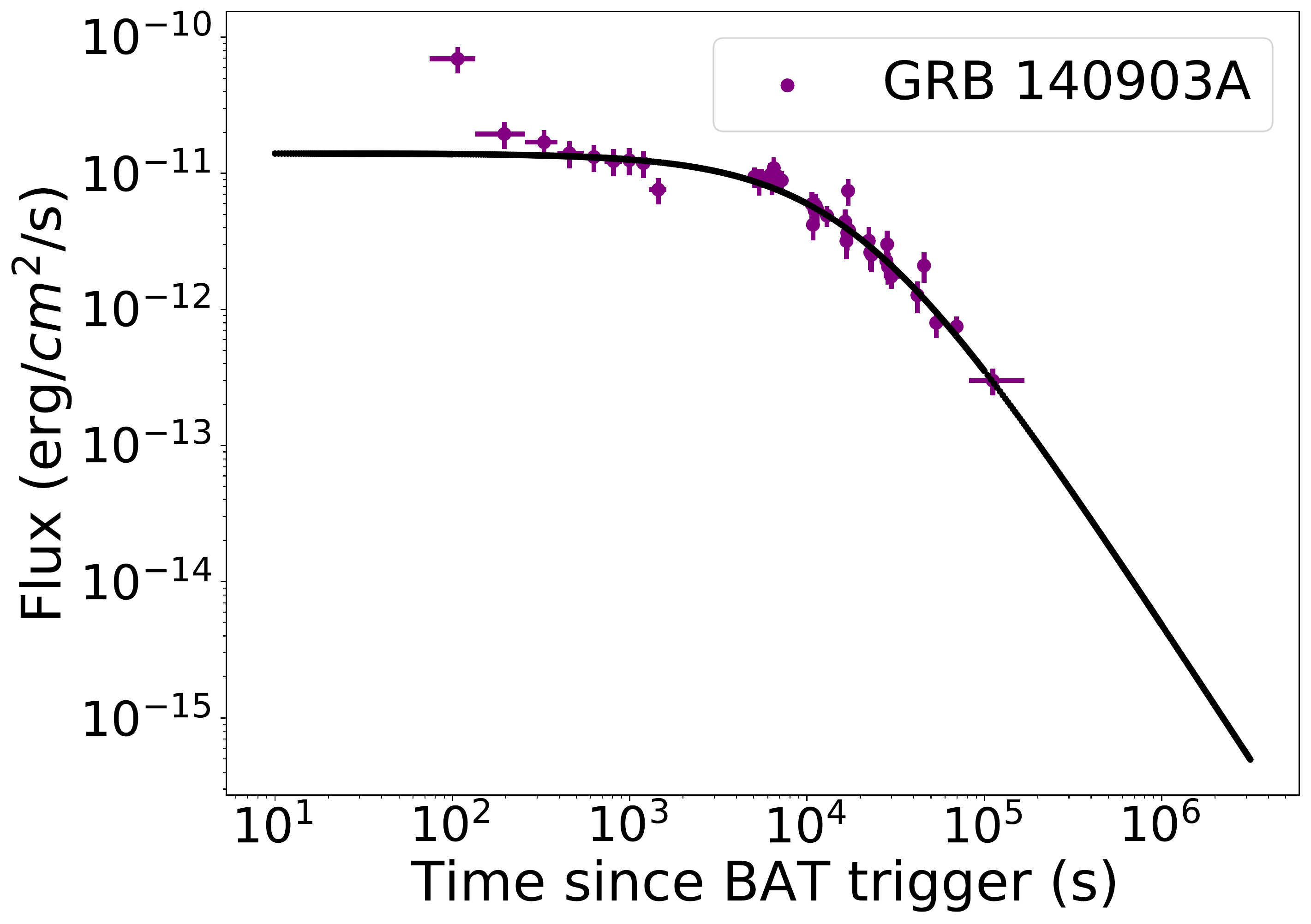}
	\caption{XRT light curves (0.3-10 keV) of short GRBs in the MD-SGRBs sample. The black solid curves are the best fits with a
		smooth power-law model to the data (purple points).}
	\label{F:SGRB}       
\end{figure}

There are five SGRBs in the MD-SGRBs sample, with the redshift range (0.3, 2.6). The GW-LGRBs sample consists 24 LGRBs. The maximum redshift is 4.81. Adopting equation (\ref{q_rev}), we get the best-fitting values of $L_{0}$ and $t_{b}$ of all GRBs in our two samples. The best-fitting results are summarized in Table \ref{T1}. The luminosity $L_{0}$ is obtained from the measured flux assuming a flat $\Lambda$CDM model with $\Omega_{m}$ = 0.3 and $H_{0}$ = 70 km/s/Mpc. The X-ray light curves of our two samples along with the broken power-law fittings using the equation (\ref{q_rev}) (black curves) are shown in Figures \ref{F:SGRB} and \ref{F:LGRB1}, respectively.
The corresponding $\chi^{2}_{\rm r}$ values obtained from the MD-SGRBs sample and the GW-LGRBs sample are within the range of (0.84, 1.86) and (0.60, 2.83), respectively. From the values of $\chi^{2}_{\rm r}$ and the analyses of Figures \ref{F:SGRB} and \ref{F:LGRB1}, we can find the quality of the MD-SGRBs sample is better than that of GW-LGRBs sample. The reason is perhaps that the MD-SGRBs sample is clean and dependent of the complex physics of external shock emission ($t^{-1.2}$), while the GW-LGRBs sample is not. The decay index of GW emission and external shock emission is analogous. The MD-SGRBs sample is small, as expected. The total number of short GRBs observed by \emph{Swift} is not large and even fewer with measured redshifts. For example, there are only five short GRBs detected by \emph{Swift} in 2020, of which only two have redshifts. In addition, the light curve of these short GRBs also need fulfill our five selection criteria.

\section{The \ltb{} correlation}
\subsection{Fitting the \ltb{} correlation}
With the plateau flux $F_{0}$ being derived from above, the luminosity of the plateau phase, $L_{0}$, can be derived as
\begin{eqnarray}
	\label{eq:L}
	L_0  =  \frac{4\pi d_{\rm L}^{2}F_0}{(1+z)^{1-\beta}},
\end{eqnarray}
where $z$ is the redshift, $\beta$ is the spectral index of the plateau phase and $d_{\rm L}$ is the luminosity distance. The term $(1+z)^{1-\beta}$ is used to perform the K-correction \citep{2001AJ....121.2879B}, which converts the luminosity to 0.3$-$10 keV range in the rest frame of GRBs. The values of $z$ and $\beta$ for all GRBs in our two samples are listed in Table \ref{T1}. In a flat universe, luminosity distance $d_{\rm L}$ can be written as
\begin{eqnarray}
	\label{eq:dL}
	d_{\rm L} = \frac{c(1+z)}{H_{0}} \int_{0}^{z} \frac{dz'}{\sqrt{\Omega_{m} (1+z')^{3} + \Omega_{\Lambda}}},
\end{eqnarray}
here, $\Omega_{m}$, $\Omega_{\Lambda}$, $H_{0}$ and $c$ represent the cosmic matter density, dark energy density, Hubble constant and the speed of light, respectively. For a set of GRBs, if their distance can be obtained from observations directly, the deriving \ltb{} correlation is model-independent. Unfortunately, we lack low-redshift GRBs ($z$ $<$ 0.1). Therefore, in order to obtain the value of $d_{\rm L}$ using equation (\ref{eq:dL}), the flat $\Lambda$CDM model with $\Omega_{m}$ = 0.3 and $H_{0}$ = 70 km/s/Mpc is employed.

We write the \ltb{} correlation in the following form
\begin{eqnarray}
	\label{eq:logL}
	\log \left (\frac{L_0}{10^{47}~\rm erg/s} \right) = k\times \log \frac{t_{\rm b}}{10^3(1+z)~ \rm s}  + b.
\end{eqnarray}
The term $(1+z)$ is the relativistic time dilation factor to transfer the time into the source's rest frame. According to equation (\ref{eq:logL}), the \ltb{} correlation can be expressed as $y = kx + b$. The best-fitting values of $k$, $b$ and the intrinsic scatter $\sigma_{\rm int}$ are achieved by adopting the techniques presented in the literature \citep{2005physics..11182D,2011MNRAS.415.3423W,2021prepWang}. The corresponding likelihood function is
\begin{eqnarray}
	\label{eq:likelihood}
	L(k,b,\sigma _{{\mathop{\rm int}} } ) &\propto&
	\prod\limits_i {\frac{1}{{\sqrt {\sigma ^2 _{{\mathop{\rm int}} }  +
					\sigma ^2 _{y_i }  + k^2 \sigma ^2 _{x_i } } }}} \nonumber \\
				&\times&
	\exp [ - \frac{{(y_i - kx_i - b )^2 }}{{2(\sigma ^2 _{{\mathop{\rm
						int}} }  + \sigma ^2 _{y_i }  + k^2 \sigma ^2 _{x_i } )}}].
\end{eqnarray}

The minimization is performed employing a Bayesian Monte Carlo Markov Chain (MCMC) \citep{2013PASP..125..306F} with the emcee package\footnote{https://emcee.readthedocs.io/en/stable/}. All the fittings in this paper are obtained adopting this python package.

\begin{figure}
	\centering
	\includegraphics[width=0.4\hsize]{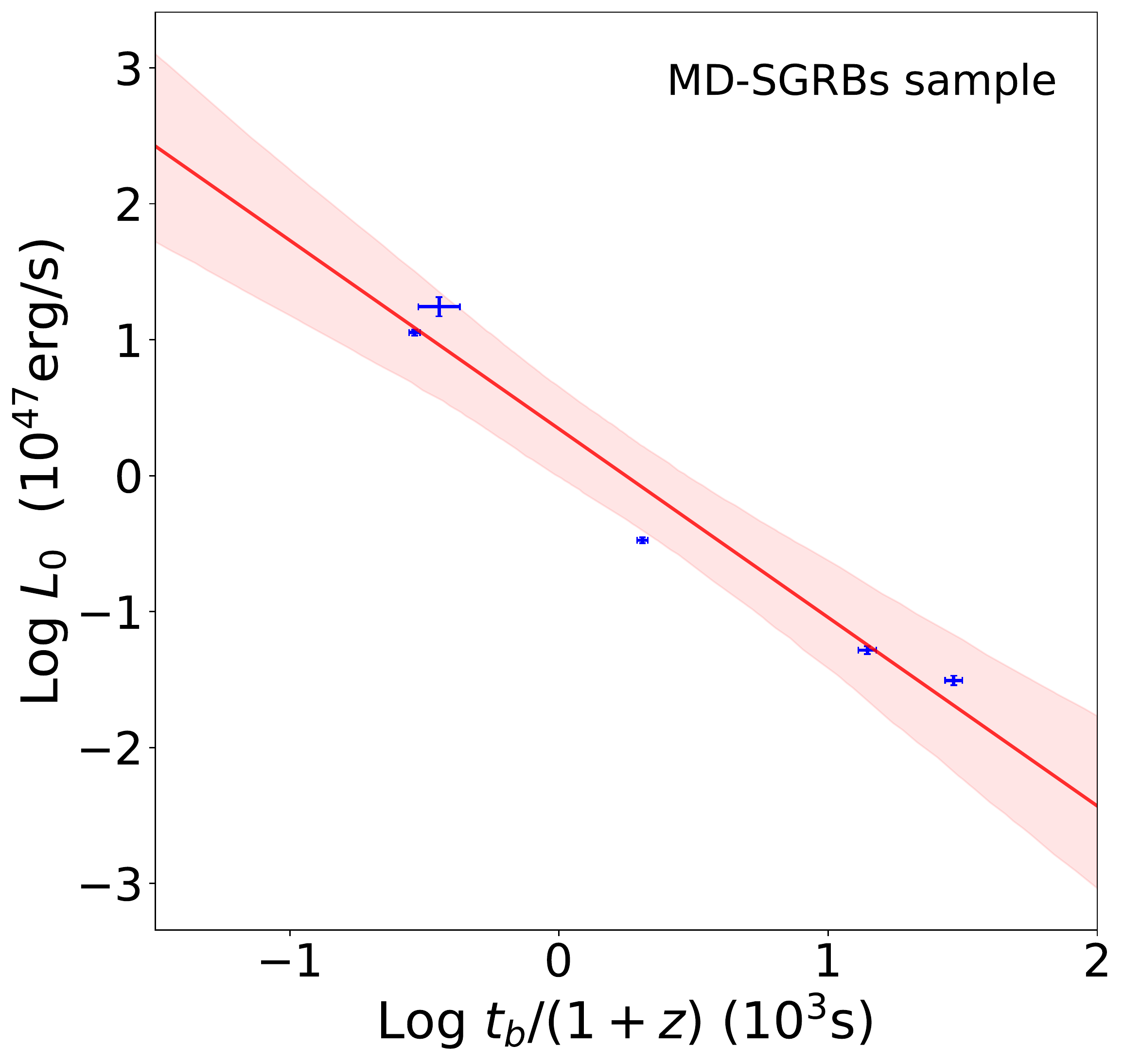}
	\includegraphics[width=0.4\hsize]{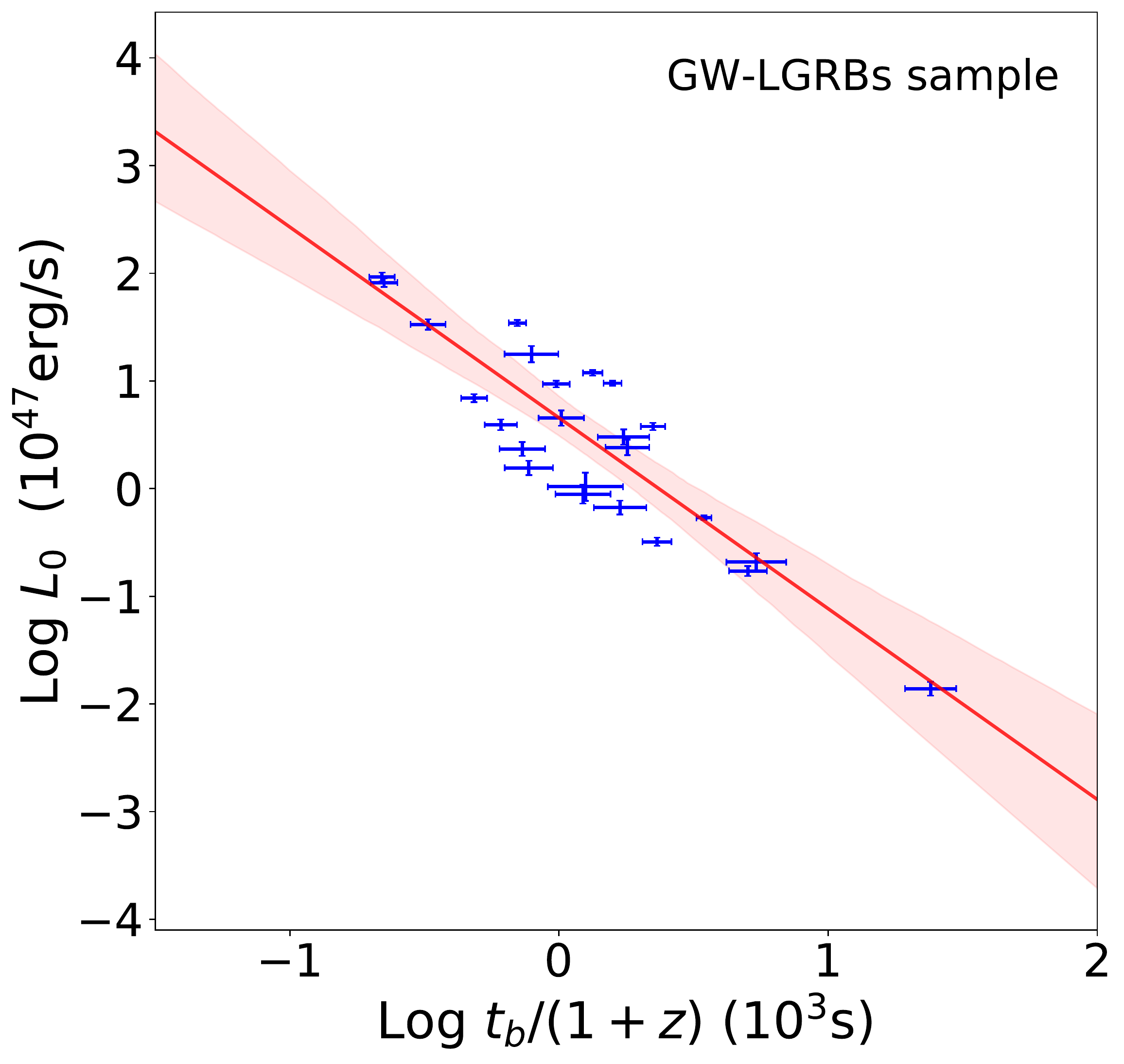}
	\caption{The $L_0$-$t_b$ correlation for the MD-SGRBs sample (left panel) and the GW-LGRBs sample (right panel). The luminosities are obtained from the measured fluxes assuming a standard $\Lambda$CDM model with $\Omega_{m}$ = 0.3 and $H_{0}$ = 70 km/s/Mpc. Blue points are the observational GRBs data. Red solid line and dark-red region represent the best fit and 2$\sigma$ confidence level. Left panel shows the best-fitting result $k = -1.38_{-0.19}^{+0.17}$, $b = 0.33_{-0.16}^{+0.17}$ and $\sigma_{\rm int}$ = $0.35_{-0.12}^{+0.20}$ for the MD-SGRBs sample. Right panel shows the best-fitting results $k = -1.77_{-0.20}^{+0.20}$, $b = 0.66_{-0.01}^{+0.01}$ and $\sigma_{\rm int}$ = $0.42_{-0.06}^{+0.08}$ for the GW-LGRBs sample. }
	\label{L_tb}       
\end{figure}

For the MD-SGRBs sample, the best-fitting results are $k = -1.38_{-0.19}^{+0.17}$, $b = 0.33_{-0.16}^{+0.17}$ and $\sigma_{\rm int}$ = $0.35_{-0.12}^{+0.20}$. The \ltb  correlation obtained from the GW-LGRBs sample is $k = -1.77_{-0.20}^{+0.20}$ and $b = 0.66_{-0.01}^{+0.01}$. The \ltb{} ~diagrams of the MD-SGRBs sample and the GW-LGRBs sample are presented in Figure \ref{L_tb}. Interestingly, the slope $k$ of the \ltb ~correlation from our two samples is a little different with that from the previous works \citep{2008MNRAS.391L..79D,2010ApJ...722L.215D,2013MNRAS.436...82D,2019ApJS..245....1T,2019ApJ...883...97Z}. Their results are about -1.00. The main reason is caused by different types of GRBs, which used to obtain the \ltb ~correlation.
Our samples are pick out from the GRB with a plateau phase based on the different decay indices. Furthermore, comparing with the results of \cite{2021prepWang}, we find that the slopes $k$ of the \ltb{} correlation from the MD-LGRBs sample and the MD-SGRBs sample are different. The slope obtained from the MD-LGRBs sample is close to $-1$. This means that the values of $L_{0} \times t_{b}$ is a constant, i.e. the initial energy of newly born magnetar is a constant. However, the slope $k$ obtained from the MD-SGRBs sample is $-1.38^{+0.17}_{-0.19}$. This means that the properties of magnetars powering LGRBs and SGRBs are different.

In addition, some selection effects (i.e., the redshift dependence of $t_{\rm b}$ and $L_0$, the threshold of detector) also affect the GRBs correlation. A lot of work has been devoted to study selection effects on this correlation \citep{2013ApJ...774..157D,2018PASP..130e1001D}. Fortunately, the \ltb{} ~correlation has been tested against selection bias robustly by \cite{2013ApJ...774..157D}. We confirm that the \ltb{} correlation really exists in GRBs with different origins. The slopes $k$ of the \ltb{} correlation are different, which indicates that the properties of newborn magnetars produced by different processes are different.
\subsection{Standardizing GRBs using $L_0-t_b$ correlation}

\begin{figure}
	\centering
	\includegraphics[width=0.67\hsize]{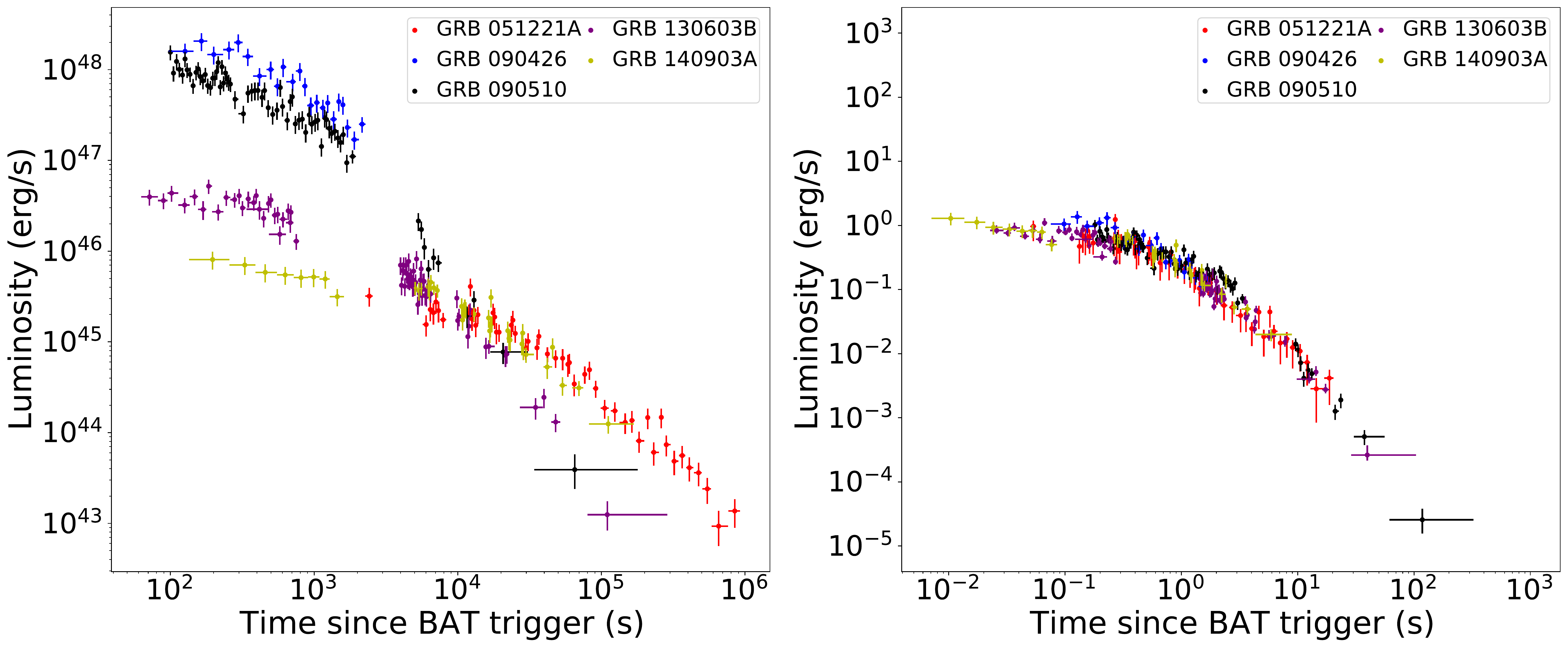}
	\caption{Original and scaled plateau light curves of short GRBs in the MD-SGRBs sample. Left panel shows the original X-ray (0.3-10 keV) light curves. Right panel shows the scaled light curves. The luminosity dispersion of the scaled light curves is about 0.4 dex. }
	\label{F:S_SGRB}       
\end{figure}

\begin{figure}
	\centering
	\includegraphics[width=0.67\hsize]{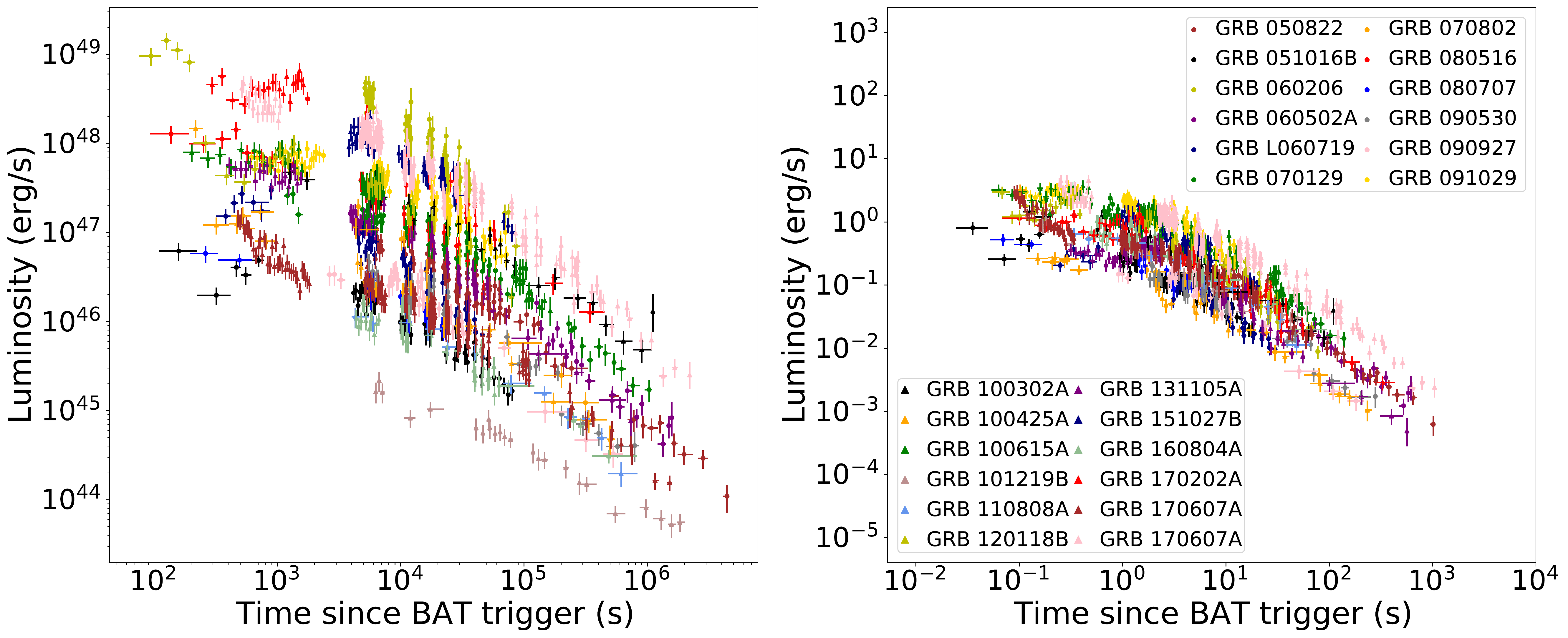}
	\caption{Original and scaled plateau light curves of long GRBs in GW-LGRBs sample. Left panel shows the original X-ray (0.3-10 keV) light curves. Right panel shows the scaled light curves. The luminosity dispersion of the scaled light curves is about 1.0 dex.}
	\label{F:S_LGRB}       
\end{figure}

Below, we use the \ltb{} correlation obtained from the MD-SGRBs and GW-LGRBs sample to standardize the corresponding afterglow light curves, respectively. First, all GRBs are scaled to the same time using $t/t_{\rm b}$, where $t$ is the observed time. Then the scale luminosity $L/L_{0}$ can be derived, where $L_{0}$ is the plateau luminosity fitted from XRT light curves. This method proposed by \cite{2021prepWang} is similar to standardize type Ia supernovae (SNe Ia) using Phillips correlation \citep{1993ApJ...413L.105P}. Standardized results of the MD-SGRBs sample and the GW-LGRBs sample are shown in Figures \ref{F:S_SGRB} and \ref{F:S_LGRB}. The left and right panels show the original and standardized light curves, respectively. In Figure \ref{F:S_SGRB}, we can find that the original light curves of afterglows are diverse, i.e., the luminosity spans more than 3 orders of magnitude. The scaled light curves show a universal behavior with the dispersion of 0.4 dex for luminosity. This hints that the plateau phase of the SGRBs whose spin-down luminosity dominated by MD radiations can be used as a standard candle, similar as the LGRBs in \citep{2021prepWang}. The \ltb{} correlation from the GW-LGRBs sample can also well standardized. Comparing the results of the MD-SGRBs and GW-LGRBs samples, the ranges of the original luminosity are similar, but the scaled result of the GW-LGRBs sample is a little bit worse. The reason is caused by a larger $\sigma_{\rm int}$, i.e. $\sigma_{\rm int,GW}$ $>$ $\sigma_{\rm int,MD}$. The spin-down of some long GRBs in GW sample may be not dominated by the GW radiation ($t^{-1}$) but the external shock emission ($t^{-1.2}$). We have taken a brief explanation at the end of Section \ref{sec:intro}. Only based on the analysis of the X-ray light curve behaviors, it is difficult to get rid of the GRBs dominated by external shock emissions. However, with the continuous upgrading of gravitational wave detectors, if we can detect the corresponding GW signal, a clean LGRBs sample dominated by GW emissions can be achieved.

\subsection{Calibrating \ltb{} correlation}

Due to the lack of low-redshift GRBs ($z < 0.1$), some methods have been proposed to calibrate correlations of GRBs \citep{2008A&A...490...31C,2008MNRAS.391L...1K,2008ApJ...685..354L,2011A&A...536A..96W,2016A&A...585A..68W,2019MNRAS.486L..46A,2021ApJ...908..181M,2021prepWang}. In this paper, we use the Hubble parameter data $H(z)$ compiled by \cite{2018ApJ...856....3Y} to calibrate the \ltb{} correlation. The calibrating process needs the Gaussian process (GP) method \citep{2012JCAP...06..036S}. GP Regression is implemented by employing the package GaPP\footnote{https://github.com/carlosandrepaes/GaPP} in the Python environment. For a given set of data, we can reconstruct a continuous function $f(x)$ that is the best representative of a discrete set of measurements $f(x_{i}) \pm \sigma_{i}$ at $x_{i}$, where $i$ = 1,2,..., $N$ and $\sigma_{i}$ is the 1$\sigma$ error bar. The GP method assumes that the value of function at any point $x$ is a Gaussian random variable with mean $m(x)$ and standard deviation $\sigma(x)$, which are determined from the discrete data through a defined covariance function (or kernel function) $k(x,x_{i})$. They can be derived from
\begin{eqnarray}
	\label{eq:mu_x}
	m(x) = \sum_{i,j = 1}^{N} k(x, x_{i})(M^{-1})_{ij}f(x_{j}),
\end{eqnarray}
 and
\begin{eqnarray}
	\label{eq:sigma}
	\sigma(x) = k(x, x_{i}) - \sum_{i,j = 1}^{N} k(x, x_{i})(M^{-1})_{ij}k(x_{j},x),
\end{eqnarray}
where the matrix $M_{ij} = k(x_{i}, x_{j}) + c_{ij}$ and $c_{ij}$ is the covariance matrix of the observational data. For uncorrelated data we have $c_{ij} = diag(\sigma^{2}_{i}$). Equations (\ref{eq:mu_x}) and (\ref{eq:sigma}) specify the posterior distribution of the extrapolated points. The kernel function, called Mat\'{e}rn, is used. Its form is written as
\begin{eqnarray}
	\label{eq:matern}
	k(x,\tilde{x}) = \sigma^{2}_{f}(1+\frac{\sqrt{3}|x - \tilde{x}|}{l}) \exp (-\frac{\sqrt{3}|x - \tilde{x}|}{l}),
\end{eqnarray}
where $\sigma_{f}$ and $l$ are two hyperparameters which can be constrained from the observational data. Some work investigated the influence of different kernels on the reconstructed function $f(x_{i})$ \citep{2018ApJ...856....3Y,2020EPJC...80..632M}. We also test the reconstruction predictions employing different kernels, such as the Matern, Radial-basis function (RBF) and Rational Quadratic kernels. The results are consistent with each other in 1$\sigma$ confidence level.

37 $H(z)$ data is used and its maximum redshift is 2.36. So, we can estimate Hubble parameter $H(z_{i})$ at any redshift where $z_{i}$ $<$ 2.5, then obtain the corresponding luminosity distance from
\begin{eqnarray}
	\label{eq:NdL}
	d_{\rm L}(z) = c(1+z)\int_{0}^{z} \frac{dz'}{H(z')}.
\end{eqnarray}

\begin{figure}
	\centering
	\includegraphics[width=0.5\hsize]{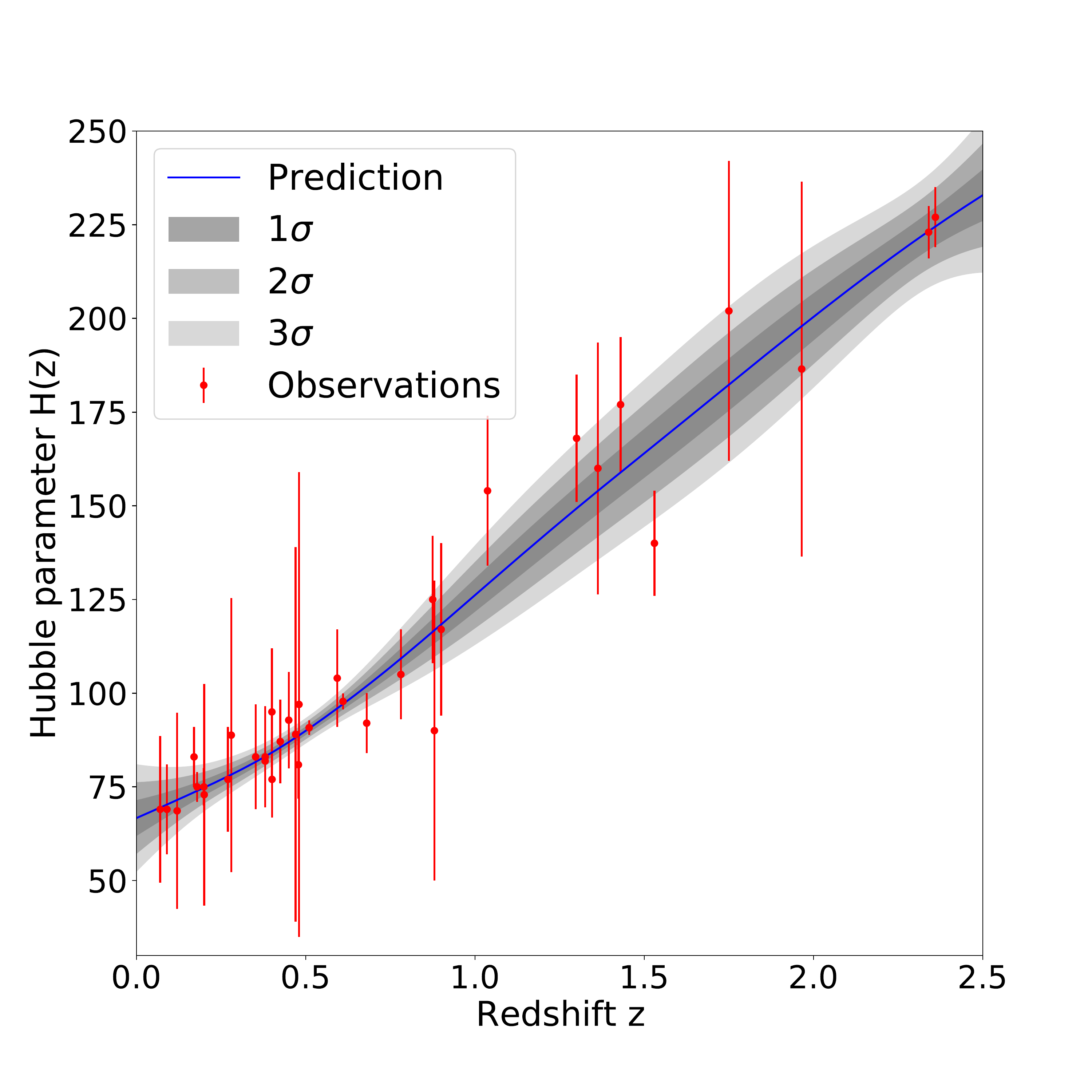}
	\caption{Smoothed $H(z)$ function (blue solid line) with 3$\sigma$ errors (gray regions) obtained from the observational Hubble data (37 red points with vertical error bars) employing GP method. The redshift of $H(z)$ data is from 0.07 to 2.36.}
	\label{F:H_z}       
\end{figure}

\begin{figure}
	\centering
	\includegraphics[width=0.5\hsize]{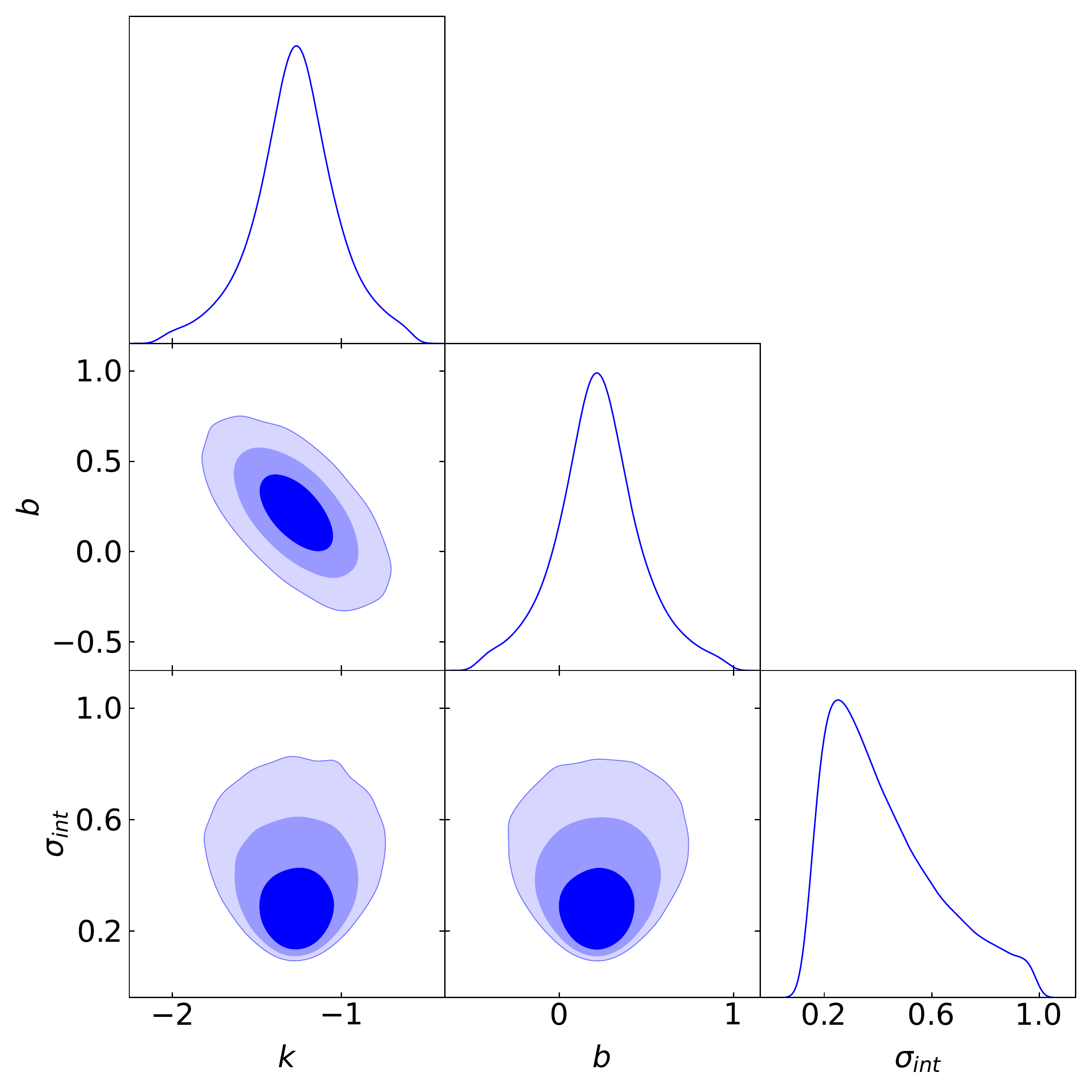}
	\includegraphics[width=0.4\hsize]{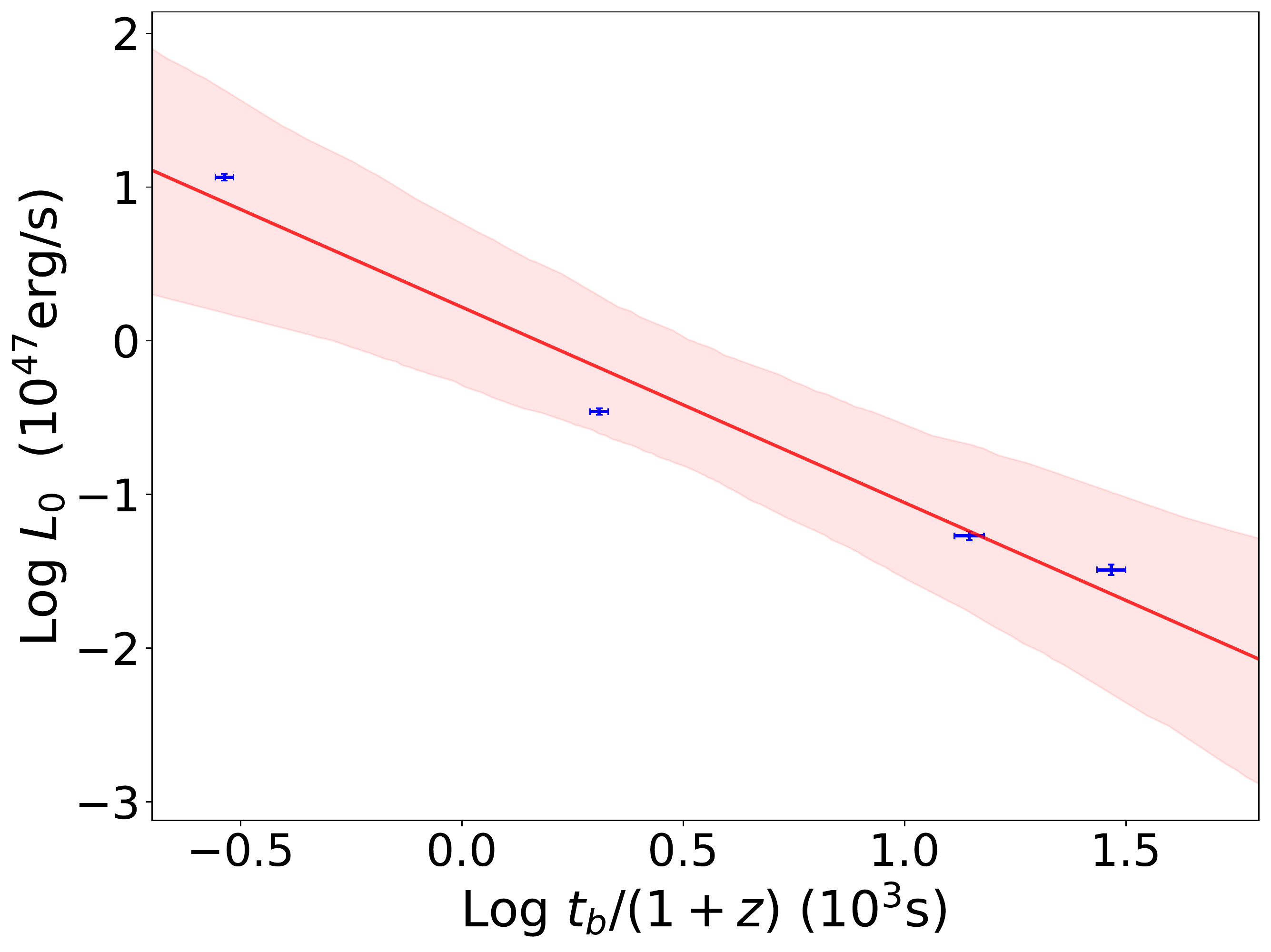}
	\caption{The calibrated $L_0-t_b$ correlation for the MD-SGRBs sample. Left panel: confidence contours ($1\sigma,2\sigma$ and $3\sigma$) and marginalized likelihood distributions for $k$, $b$, and $\sigma_{\rm int}$. Right panel: the red line shows the best-fitting result and the shaded regions corresponds to 2$\sigma$ confidence region, $k = -1.27_{-0.25}^{+0.24}$ $b = 0.22_{-0.23}^{+0.24}$ and $\sigma_{\rm int}$ = $0.27_{-0.12}^{+0.18}$. }
	\label{S_kb}       
\end{figure}

\begin{figure}
	\centering
	\includegraphics[width=0.5\hsize]{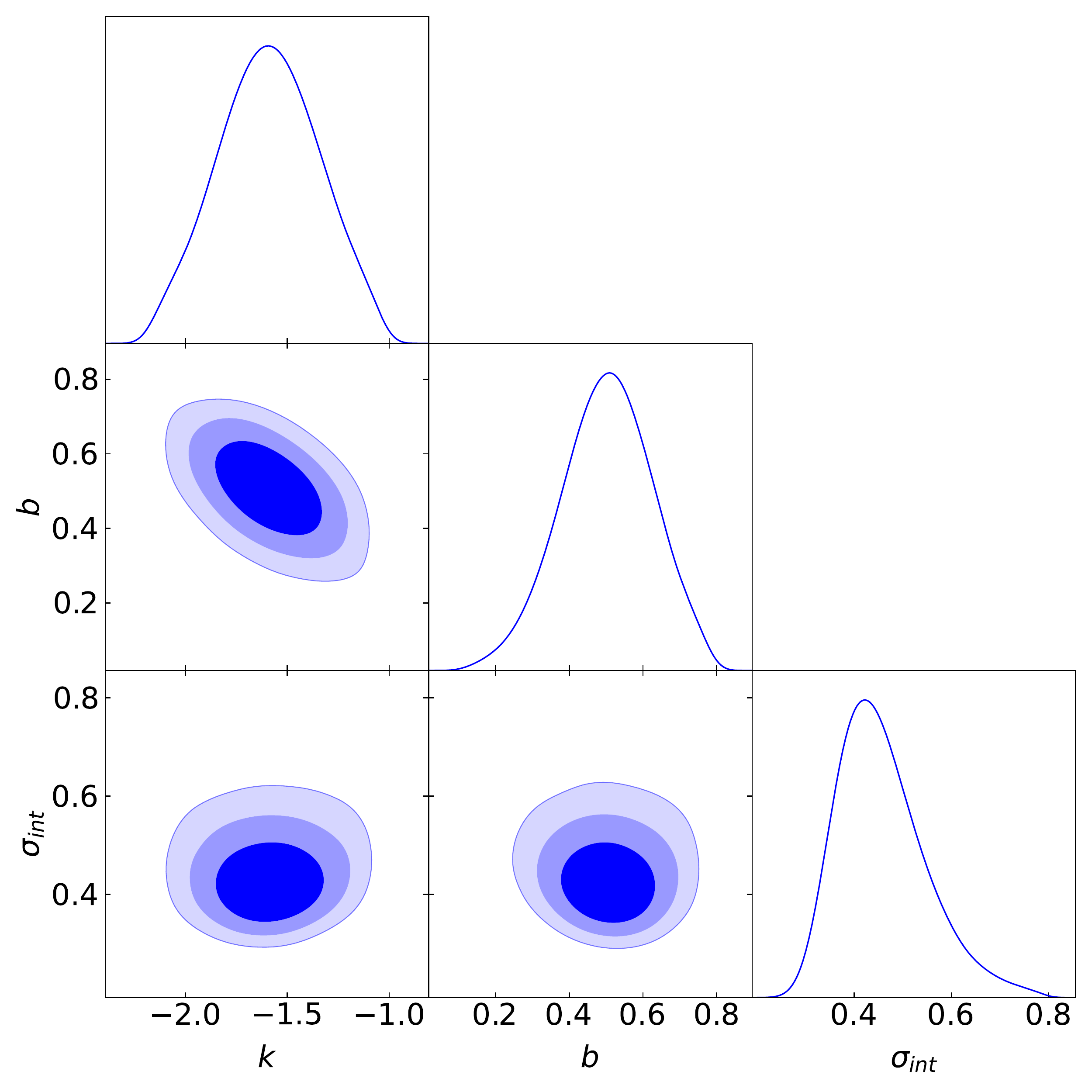}
	\includegraphics[width=0.4\hsize]{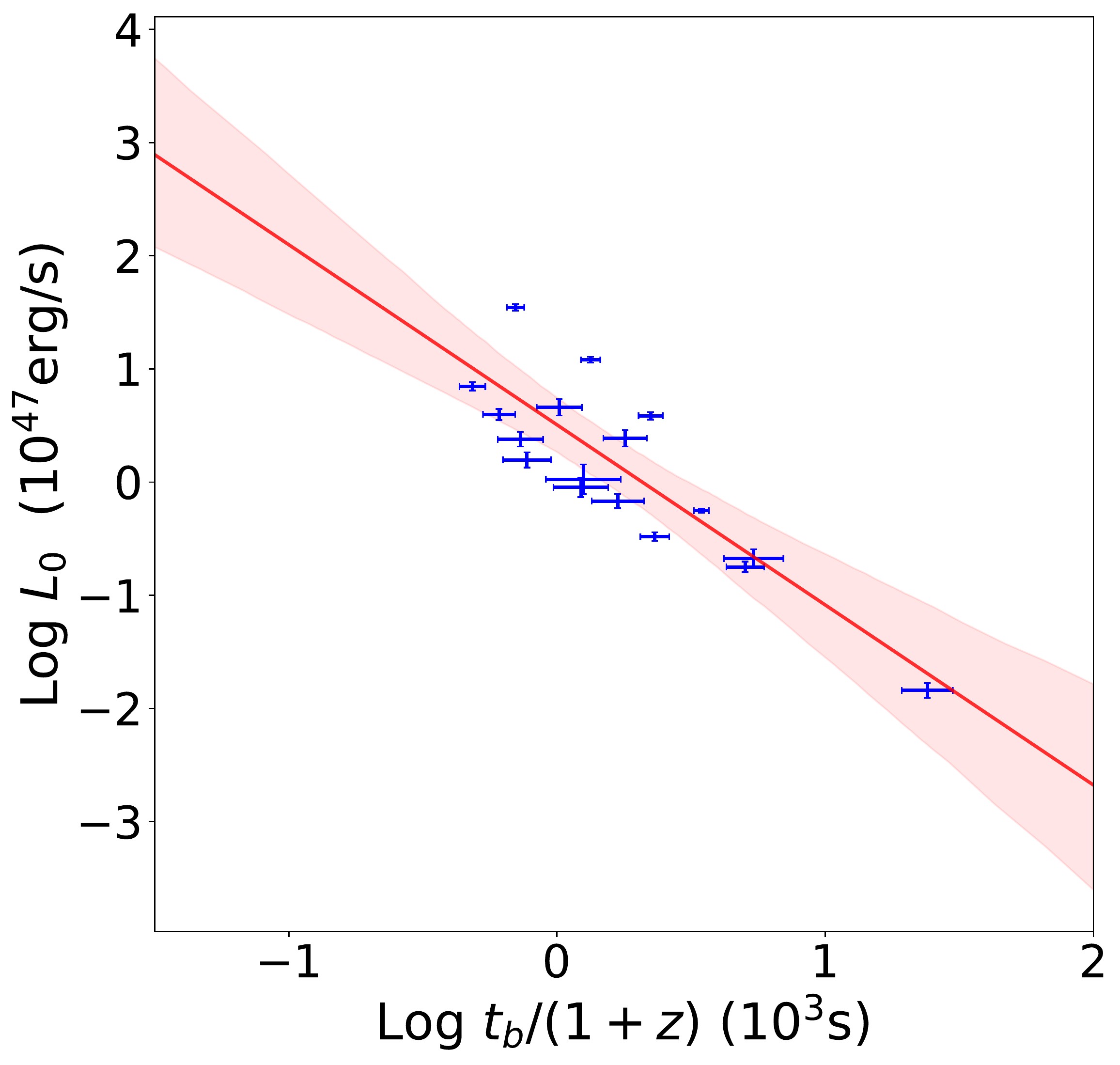}
	\caption{The calibrated $L_0-t_b$ correlation for the GW-LGRBs sample. Left panel: confidence contours ($1\sigma,2\sigma$ and $3\sigma$) and marginalized likelihood distributions for $k$, $b$, and $\sigma_{\rm int}$. Right panel: the red line shows the best-fitting result and the shaded regions corresponds to 2$\sigma$ confidence region, $k = -1.59_{-0.26}^{+0.26}$ $b = 0.50_{-0.13}^{+0.12}$ and $\sigma_{\rm int}$ = $0.42_{-0.08}^{+0.09}$. }
	\label{L_kb}       
\end{figure}

Reconstruction results of the smooth $H(z)$ curve are given in Figure \ref{F:H_z}. There are 4 SGRBs and 17 LGRBs that can be used to calibrate the \ltb{} correlation of the MD-SGRBs and GW-LGRBs samples, respectively. The corresponding MCMC results and the \ltb{} diagram are shown in Figures \ref{S_kb} and \ref{L_kb}, respectively. For the MD-SGRBs sample, the best-fitting results are $k = -1.27_{-0.25}^{+0.24}$ $b = 0.22_{-0.23}^{+0.24}$ and $\sigma_{\rm int}$ = $0.27_{-0.12}^{+0.18}$. The best-fitting results from the GW-LGRBs sample are $k = -1.59_{-0.26}^{+0.26}$ $b = 0.50_{-0.13}^{+0.12}$ and $\sigma_{\rm int}$ = $0.42_{-0.08}^{+0.09}$. The calibrated \ltb{} correlations are model-independent and can be utilized to constrain cosmological parameters.

\section{Cosmological implications of \ltb{} correlation}
\subsection{Constraints on $\Lambda$CDM model}
For the standard cosmological model, the distance modulus is
\begin{eqnarray}
	\label{eq:mu}
	\mu = 5\log\frac{d_{\rm L}}{\rm {Mpc}} + 25.
\end{eqnarray}
The flat $\Lambda$CDM model is considered. Replacing $d_{\rm L}$ by a function $\sqrt{L_{0}(1+z)^{(1-\beta)}(4\pi F_{0})^{-1}}$ derived from equation (\ref{eq:L}), and combining the calibrated \ltb{}
correlation, the observed distance modulus and its uncertainty can be derived from
\begin{eqnarray}
	\label{eq:muobs}
	\mu_{\rm obs} = \frac{5}{2} (\log{L_0} - \log \frac{4\pi F_0}{(1+z)^{1-\beta}} - 24.49) + 25,
\end{eqnarray}
and
\begin{eqnarray}
	\label{eq:muerr}
	\sigma_{\rm obs} &=& \frac{5}{2} [(\log^{2} (\frac{t_{\rm b}}{1+z}) -3) \sigma_{\rm k}^{2} + k^{2}  (\frac{\sigma_{t_{\rm b}}}{t_{\rm b} \ln{10}})^{2} \nonumber \\
		&+& \sigma_{\rm b}^{2}+ (\frac{\sigma_{F_0}}{F_0 \ln{10}})^{2} + \sigma_{\rm int}^{2}]^{1/2}.
\end{eqnarray}
Here, $\sigma_{\rm int}$ is the intrinsic scatter of the \ltb{} correlation. The likelihood function for the parameter $\Omega_{m}$ can be determined from
\begin{eqnarray}
	\label{eq:chi}
	\chi^{2}(\Omega_i) = \sum_{j=1}^{N}\frac{(\mu_{\rm obs}(z) - \mu_{\rm th}(\Omega_{i},z))^{2}}{\sigma_{\rm obs}^{2}},
\end{eqnarray}
where $\mu_{\rm th}(\Omega_{i},z)$ is the theoretical distance modulus calculated by taking equation (\ref{eq:dL}) into equation (\ref{eq:mu}), $N$ is the number of data, and $\Omega_{i}$ represents cosmological parameters. Considering a flat universe, the best-fitting result of $\Omega_{m}$ can be given by minimizing the equation (\ref{eq:chi}).

For the nonflat $\Lambda$CDM model, the luminosity distance $d_{\rm L}$ should be written as
\begin{eqnarray}
	\label{eq:dl_k}
	d_{\rm L}=
	\begin{cases}
		\frac{c(1+z)}{H_{0}}(-\Omega_{\rm k})^{-\frac{1}{2}} \sin{[(-\Omega_{\rm k})^{\frac{1}{2}} \int_{0}^{z} \frac{dz'}{E(z')}]}, &\Omega_{\rm k} < 0, \\
		\frac{c(1+z)}{H_{0}} \int_{0}^{z} \frac{dz'}{E(z')}, &\Omega_{\rm k} = 0,\\
		\frac{c(1+z)}{H_{0}}\Omega_{\rm k}^{-\frac{1}{2}} \sinh{[\Omega_{\rm k}^{\frac{1}{2}} \int_{0}^{z} \frac{dz'}{E(z')}]}, &\Omega_{\rm k} > 0,
	\end{cases}
\end{eqnarray}

\begin{eqnarray}
	\label{eq:ez}
	E(z') = \sqrt{\Omega_{m} (1+z')^{3} + (1-\Omega_{m}-\Omega_{\Lambda})(1+z')^{2} + \Omega_{\Lambda}}.
\end{eqnarray}
Then, we can get the corresponding theoretical distance modulus $\mu_{\rm th} (\Omega_{m},\Omega_{\Lambda})$ by taking equation (\ref{eq:dl_k}) into equation (\ref{eq:mu}). The constraint results of $\Omega_{m}$ and $\Omega_{\Lambda}$ in the nonflat $\Lambda$CDM model can be derived from the minimization of equation (\ref{eq:chi}).

\begin{figure}
	\centering
	\includegraphics[width=0.5\hsize]{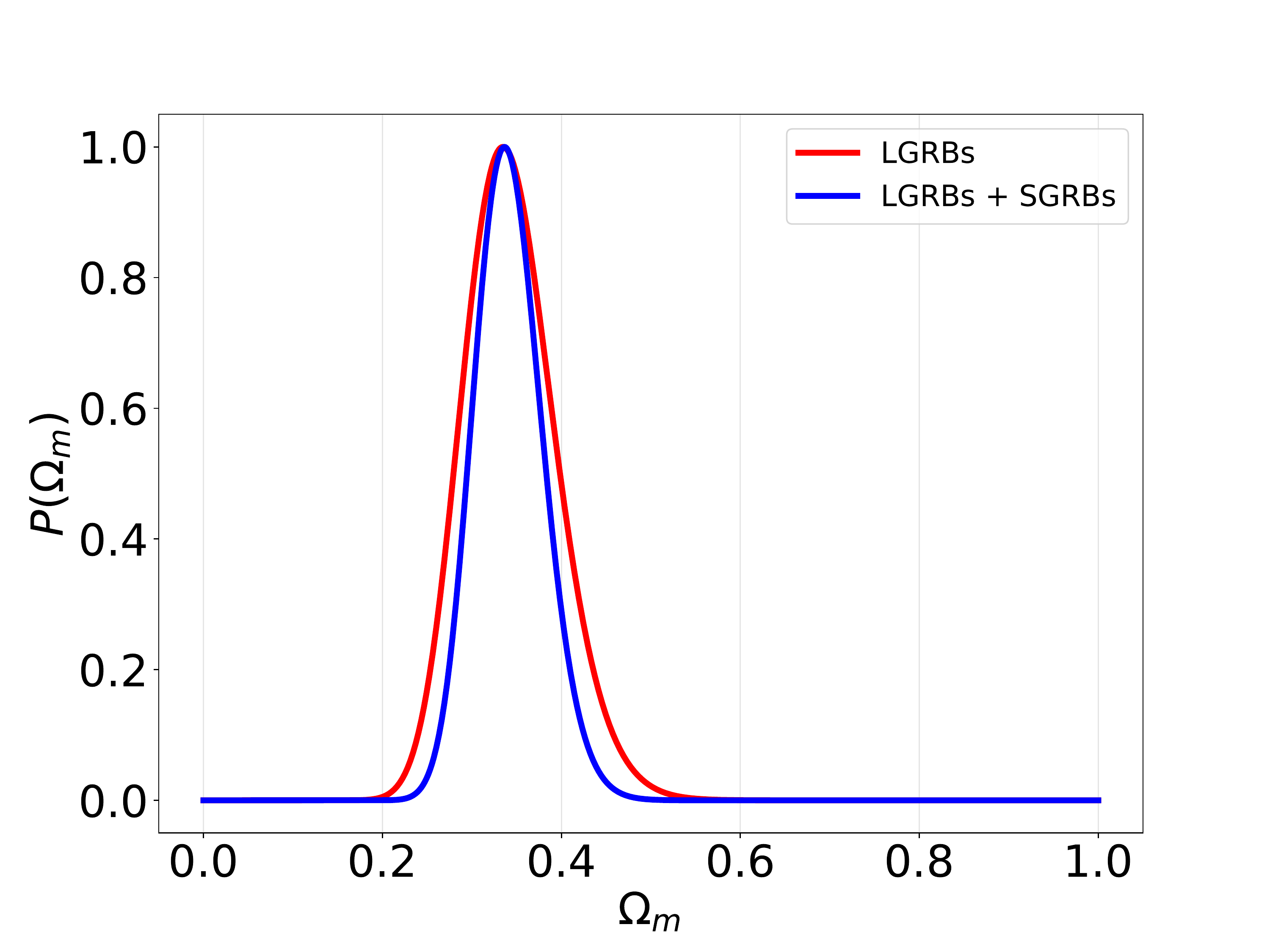}
	\includegraphics[width=0.45\hsize]{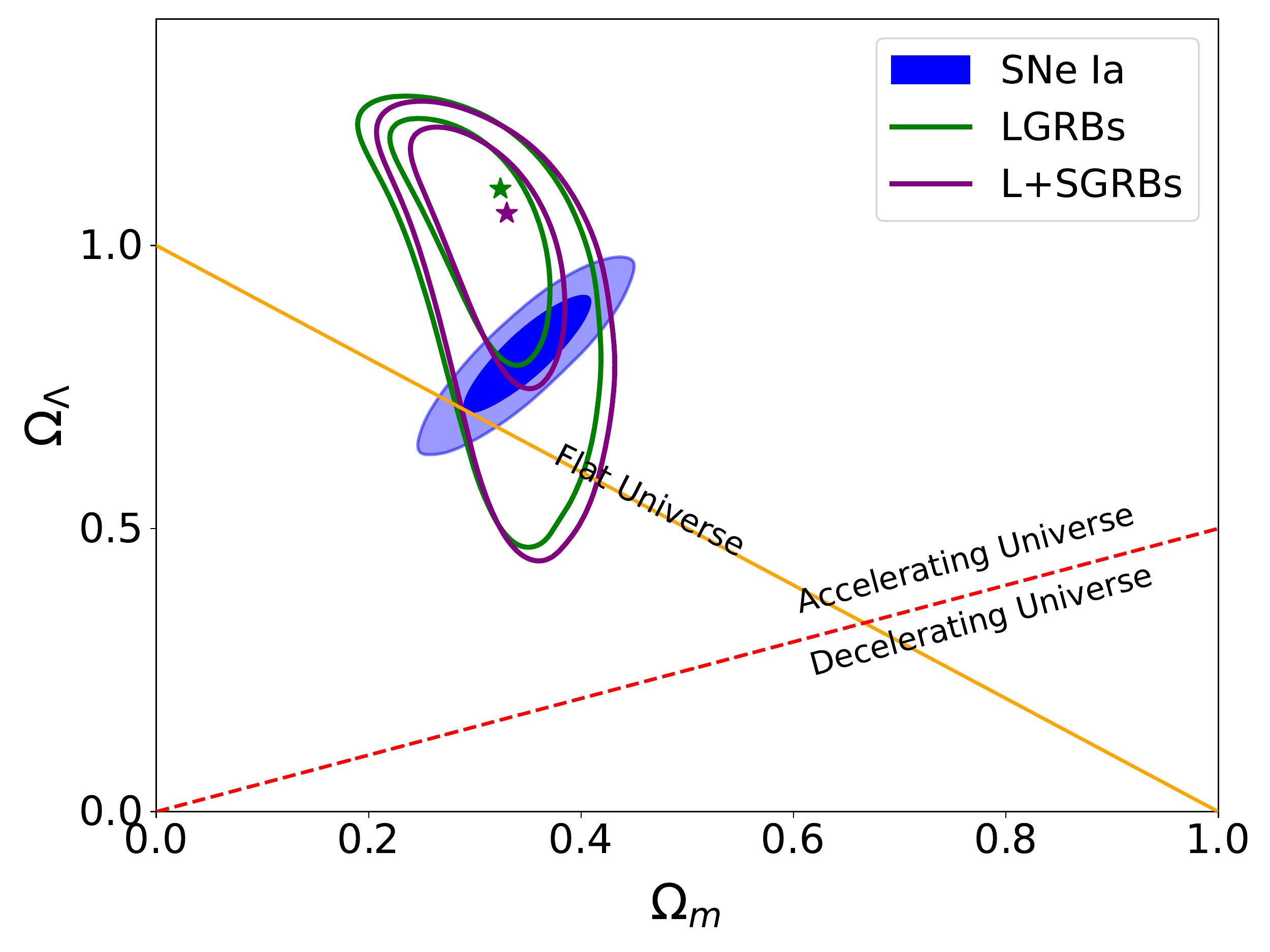}
	
	\caption{Constraints on $\Lambda$CDM model. Left panel: the probability distribution of $\Omega_{m}$ in the flat $\Lambda$CDM universe. The red and blue solid lines are the constraints on $\Omega_{m}$ of MD-LGRBs sample and combined sample of the MD-LGRBs and the MD-SGRBs (MD-total sample). The best-fitting value from the MD-total sample for $\Omega_{m}$ is $0.34_{-0.07}^{+0.08}$. Right panel: constraint on the non-flat $\Lambda$CDM universe. The orange line indicates the flat universe $\Omega_{m}+\Omega_\Lambda=1$. The red dashed line shows the separation between accelerating and decelerating universe, i.e., the deceleration parameter $q_0$ is equal to 0. The blue contour is the best-fitting result with Pantheon sample. The green and purple contours are the constraints using the MD-LGRBs sample and the MD-total sample, respectively. The best-fitting results obtained from the MD-LGRBs sample are $\Omega_{m} = 0.32_{-0.10}^{+0.05}$ and $\Omega_{\Lambda}$ = $1.10_{-0.31}^{+0.12}$ \citep{2021prepWang}. The best-fitting results obtained from the MD-total sample are $\Omega_{m} = 0.33_{-0.09}^{+0.06}$ and $\Omega_{\Lambda}$ = $1.06_{-0.34}^{+0.15}$. }
	\label{F:MD}
\end{figure}

\begin{figure}
	\centering
	\includegraphics[width=0.5\hsize]{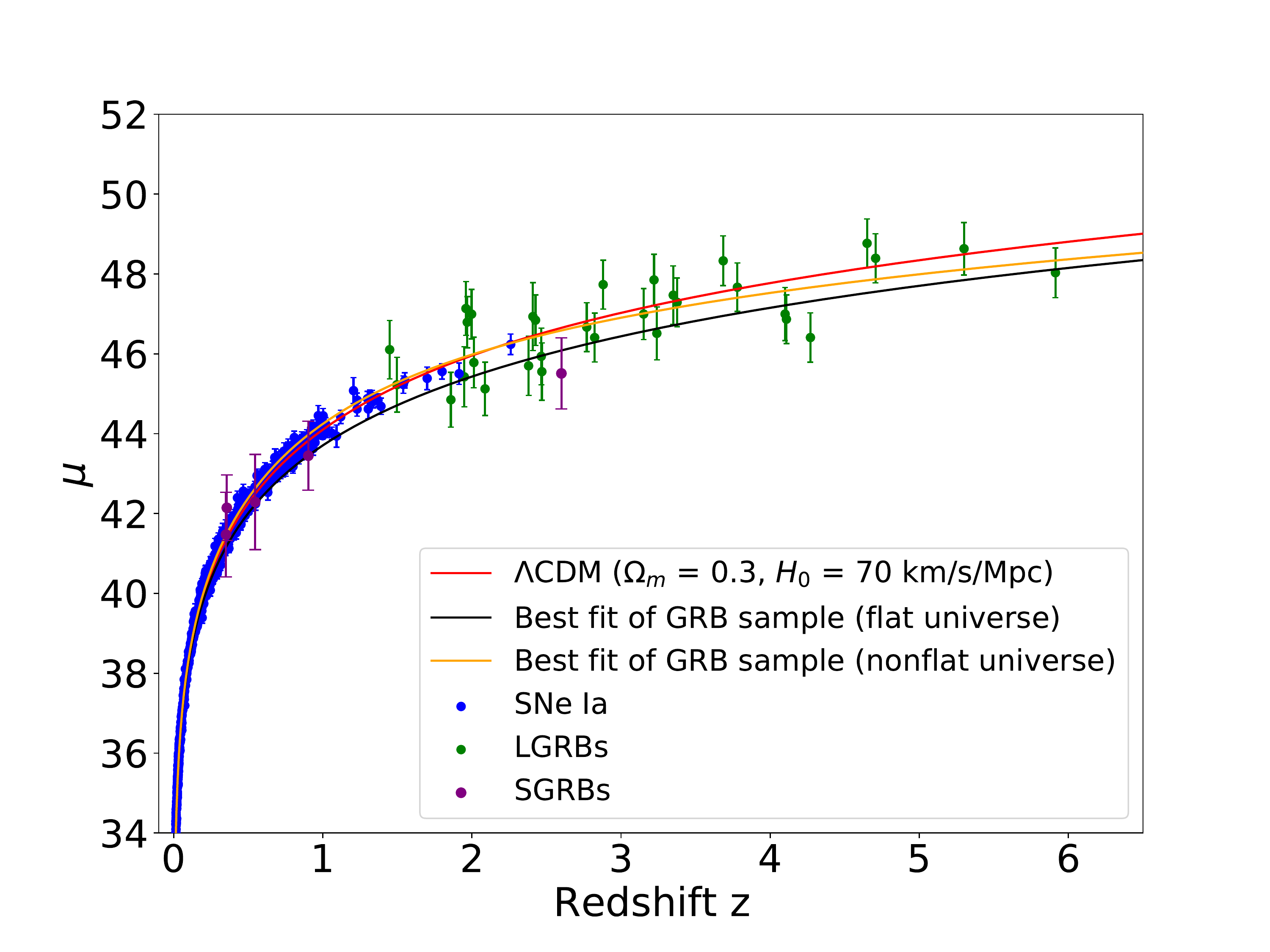}
	\caption{Hubble diagram of SNe Ia and GRBs. Blue points are SNe Ia from the
		Pantheon sample. Green points are 31 LGRBs in MD-LGRBs sample. Purple points are 5 SGRBs in MD-LGRBs sample. The red solid line is the flat $\Lambda$CDM model with $\Omega_{m}=0.3$ and Hubble constant $H_0=70 $ km s$^{-1}$ Mpc$^{-1}$. For a flat universe, the best fit from the MD-total sample is shown as a black line.
		The orange solid line shows the best fit from the MD-total sample for the nonflat $\Lambda$CDM universe. }
	\label{F:S_mu}       
\end{figure}

Combining the MD-LGRBs sample from \cite{2021prepWang} with the MD-SGRBs sample, we obtain a total sample that includes 31 LGRBs and 5 SGRBs. We refer to this dataset as MD-total sample. The redshift distribution of the MD-total sample covers 0.2 to 5.91. LGRBs are mainly distributed at high redshifts (1.45, 5.91). While SGRBs are mainly distributed at low redshifts (0.35, 2.6). Employing the MD-total sample, we give the best-fitting results, $\Omega_{m}$ = $0.34_{-0.07}^{+0.08}$ for a flat universe, $\Omega_{m} = 0.33_{-0.09}^{+0.06}$ and $\Omega_{\Lambda}$ = $1.06_{-0.34}^{+0.15}$ for a nonflat universe, as shown in the left panel and right panel of Figure \ref{F:MD}. From the left panel of Figure \ref{F:MD}, we find that adding the MD-SGRBs sample makes the constraint on $\Omega_{m}$ tighter than that of \cite{2021prepWang}. From the right panel, it is easy to find that the best-fitting results (purple line) support a flat universe at about 2$\sigma$ confidence level. Figure \ref{F:S_mu} shows the cosmological constraints, Hubble diagram of GRBs and SNe Ia.

\begin{figure}
	\centering
	\includegraphics[width=0.5\hsize]{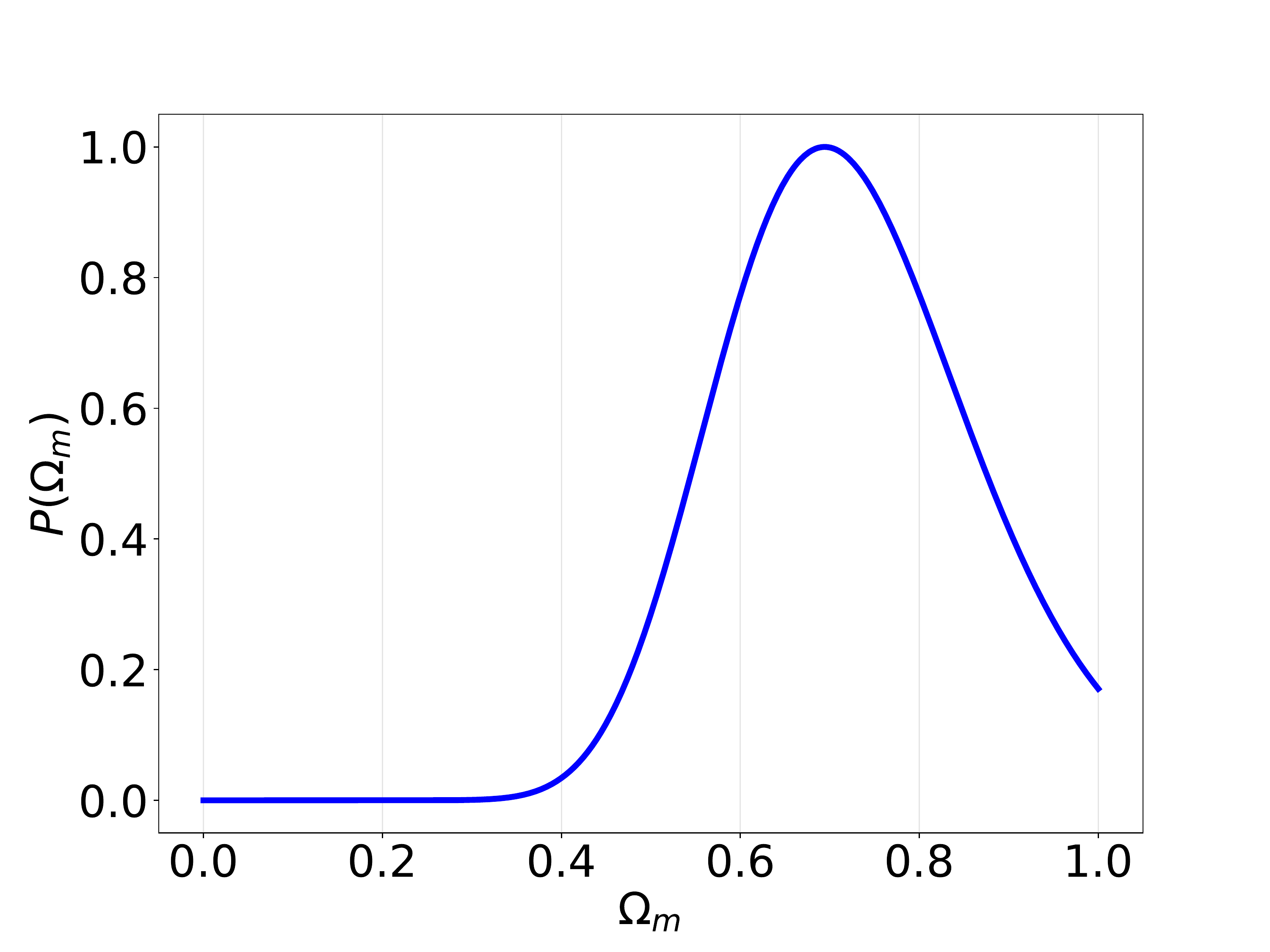}
	\includegraphics[width=0.45\hsize]{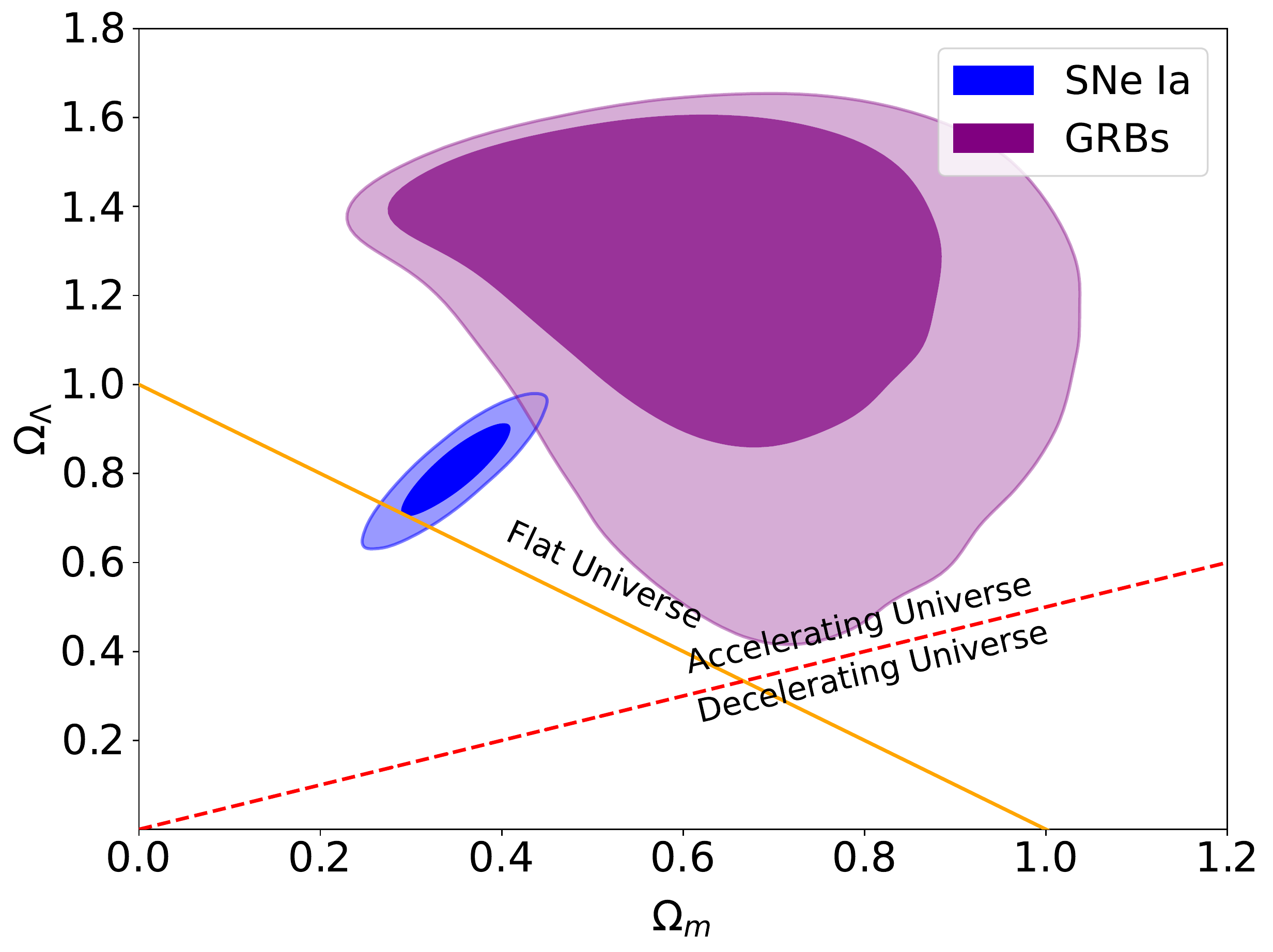}
	\caption{Constraints on cosmological parameters using the GW-LGRBs sample and the Pantheon sample. Left panel: the probability distribution of $\Omega_{m}$ in the flat $\Lambda$CDM model. The best-fitting value of $\Omega_{m}$ equals to $0.69_{-0.18}^{+0.24}$. Right panel: constraints on the nonflat $\Lambda$CDM universe. The best-fitting results are $\Omega_{m} = 0.59_{-0.27}^{+0.30}$ and $\Omega_{\Lambda}$ = $1.38_{-0.52}^{+0.23}$. The orange line indicates the flat universe $\Omega_{m}+\Omega_\Lambda=1$. The red dashed line shows the separation between accelerating and decelerating universe. The blue and purple contours are the results of $\Omega_{m}$ and $\Omega_{\Lambda}$ from the Pantheon sample and the combined sample of Pantheon sample and GW-LGRBs sample, respectively. }
	\label{F:GW}       
\end{figure}

\begin{figure}
	\centering
	\includegraphics[width=0.5\hsize]{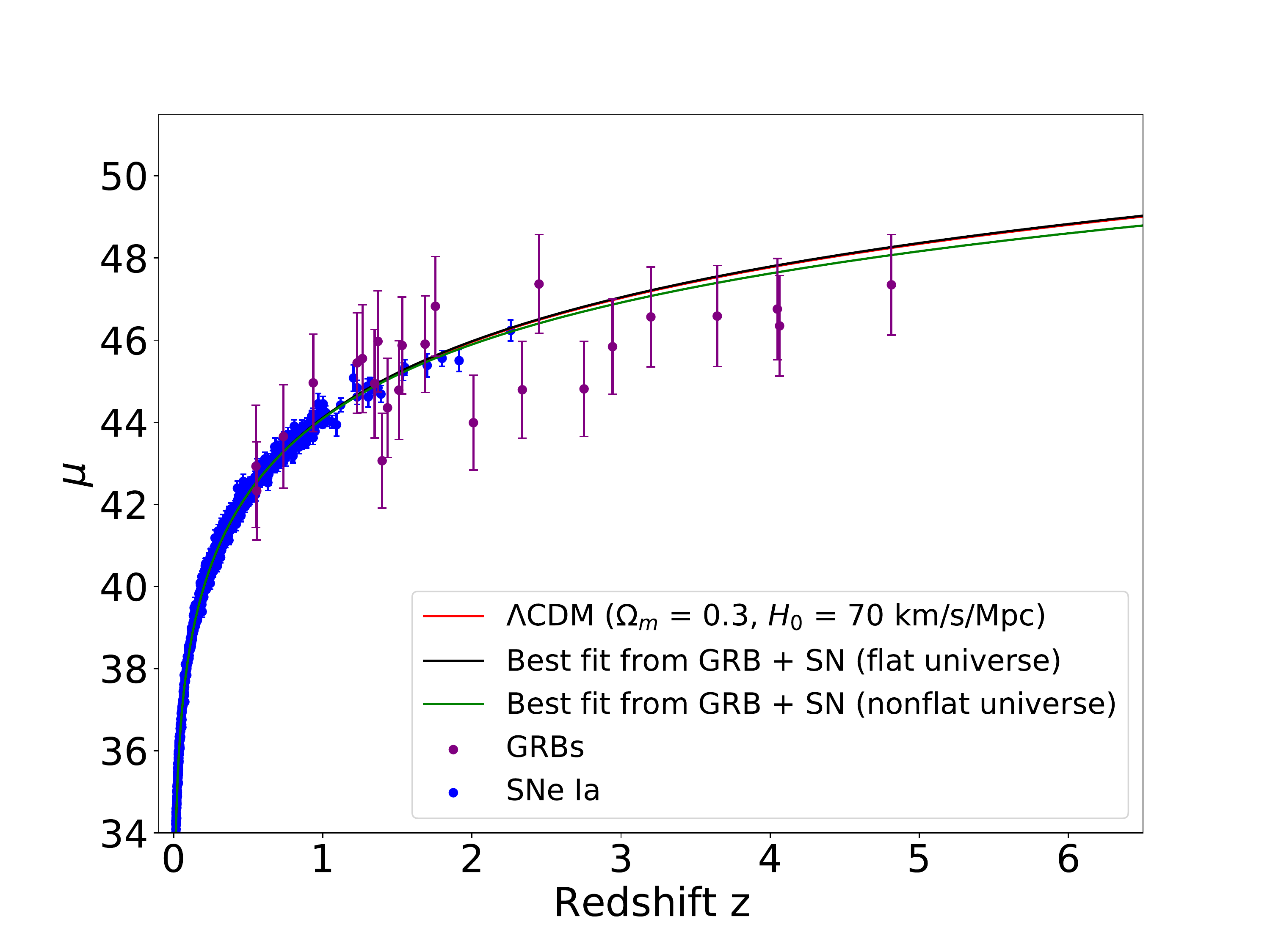}
	\caption{Hubble diagram of SNe Ia and LGRBs. Blue points are supernovae from the Pantheon sample. Purple points are 24 long GRBs in the GW-LGRBs sample. The red solid line is the $\Lambda$CDM model with $\Omega_{m}=0.3$ and $H_0=70 $ km s$^{-1}$ Mpc$^{-1}$. For a flat $\Lambda$CDM universe, the best fit from a combined sample of the GW-LGRBs sample and Pantheon sample is shown as black line with $\Omega_{m}$=$0.29$. The orange solid line shows the best fit from the combined sample for a nonflat universe with $\Omega_{m} = 0.35_{-0.05}^{+0.05}$ and $\Omega_{\Lambda} = 0.79_{-0.06}^{+0.07}$. }
	\label{F:L_mu}       
\end{figure}

We also use the GW-LGRBs sample to constrain the parameters of a flat universe and a nonflat universe, respectively. The best-fitting results are obtained by minimizing the equation (\ref{eq:chi}). For a flat universe, we give $\Omega_{m}$ = $0.69_{-0.18}^{+0.24}$, as shown in the left panel of figure \ref{F:GW}. The right panel shows the constraint on $\Omega_{m}$ and $\Omega_{\Lambda}$ in a nonflat universe from the combining sample. The results are $\Omega_{m} = 0.59_{-0.27}^{+0.30}$ and $\Omega_{\Lambda}$ = $1.38_{-0.52}^{+0.23}$. The corresponding Hubble diagram of GRBs and SNe Ia are shown in figure \ref{F:L_mu}. The results are not tight. The main reason is that the GW-LGRBs sample might be not clean. The decay of some long GRBs in the GW sample may be not dominated by the GW radiation ($t^{-1}$) but the external shock emission ($t^{-1.2}$).

\subsection{Constraints on $w(z)$ = $w_{0}$ model}

Recent works show that high-redshift Hubble diagrams of supernovae, quasars and GRBs deviate from the
flat $\Lambda$CDM at 4$\sigma$ confidence level \citep{2019NatAs...3..272R}. But a different voice was presented by \cite{2020MNRAS.492.4456K,2020MNRAS.497..263K}. They made a joint analysis of quasar, $H(z)$ and baryon acoustic oscillation data, and found that the results are still consistent with the flat $\Lambda$CDM model. In addition, \citet{2021ApJ...908..181M} found that $w$ evolves with redshift from a sample of 174 GRBs, which is similar as that of \citet{2011A&A...536A..96W}. The result of $w(z)$ at $z<$1.2 is agreement with the standard value $w = -1$ but disagreement at larger $z$. The best fix $w(z)$ seems to deviate from $w$ = -1 at 2 to 4$\sigma$ level, depending on the redshift bins \citep{2021ApJ...908..181M}.

 In the previous subsection, we have given constraints on the $\Lambda$CDM model combining our two samples and the MD-LGRB sample \citep{2021prepWang}. The results are in line with the $\Lambda$CDM model at 2$\sigma$ confidence level. In this subsection, also utilizing these four samples, we constrain the $w(z)$ = $w_{0}$ model. Firstly, we test the $w(z)$ = $w_{0}$ model using MD-total sample and Pantheon sample. Then, we replace the sample with the GW-LGRBs sample and Pantheon sample.

\begin{figure}
	\centering
	\includegraphics[width=0.5\hsize]{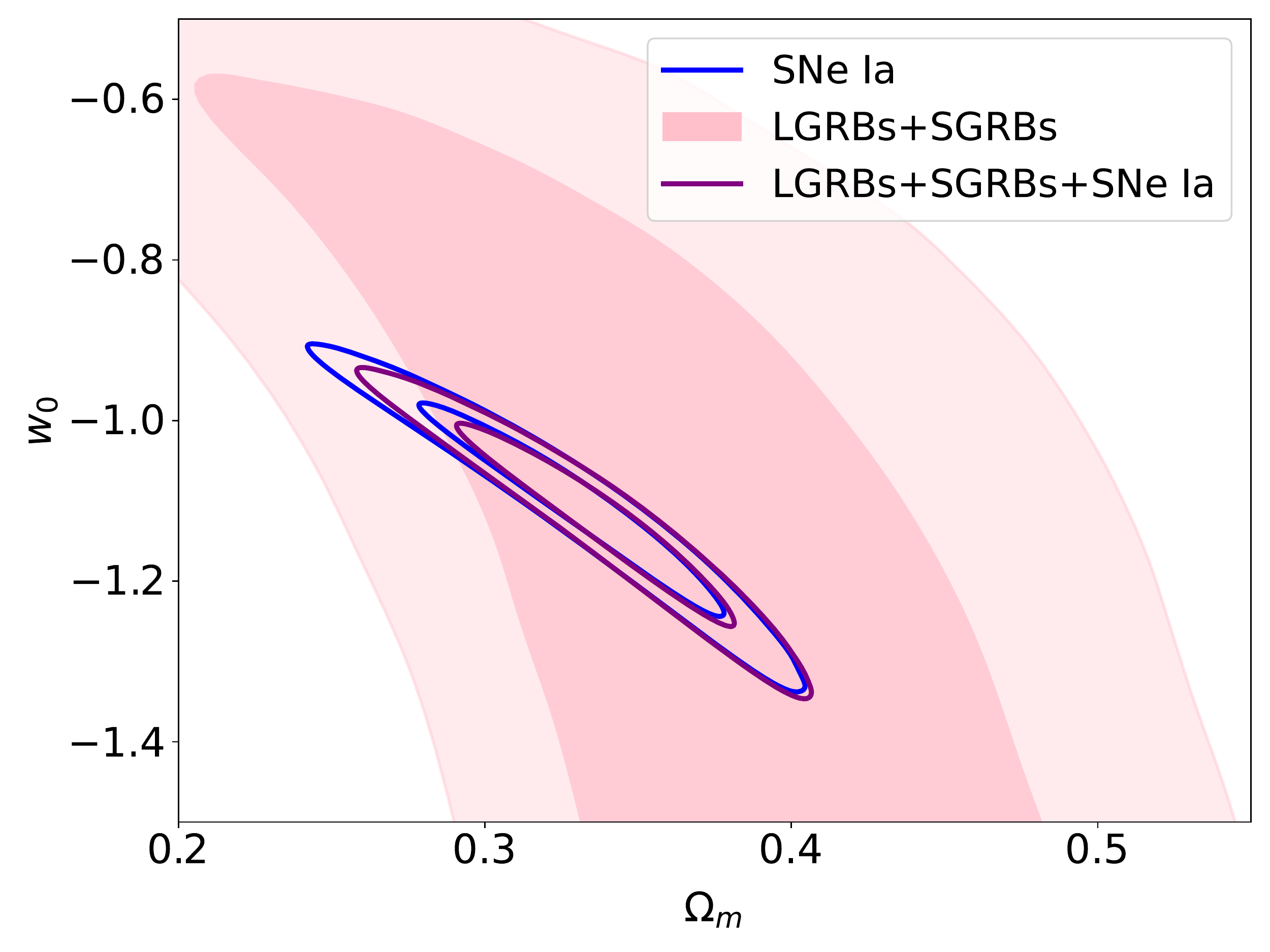}
	\caption{Confidence contours ($1\sigma$ and $2\sigma$) of the parameters $w_0$ and $\Omega_{m}$ in the $w(z)$ = $w_{0}$ model using the MD-LGRBs sample, MD-SGRBs sample, and Pantheon sample. It is worth noticing the confidence levels for GRBs do not include the systematic errors.}
	\label{F:S_w}       
\end{figure}

\begin{figure}
	\centering
	\includegraphics[width=0.5\hsize]{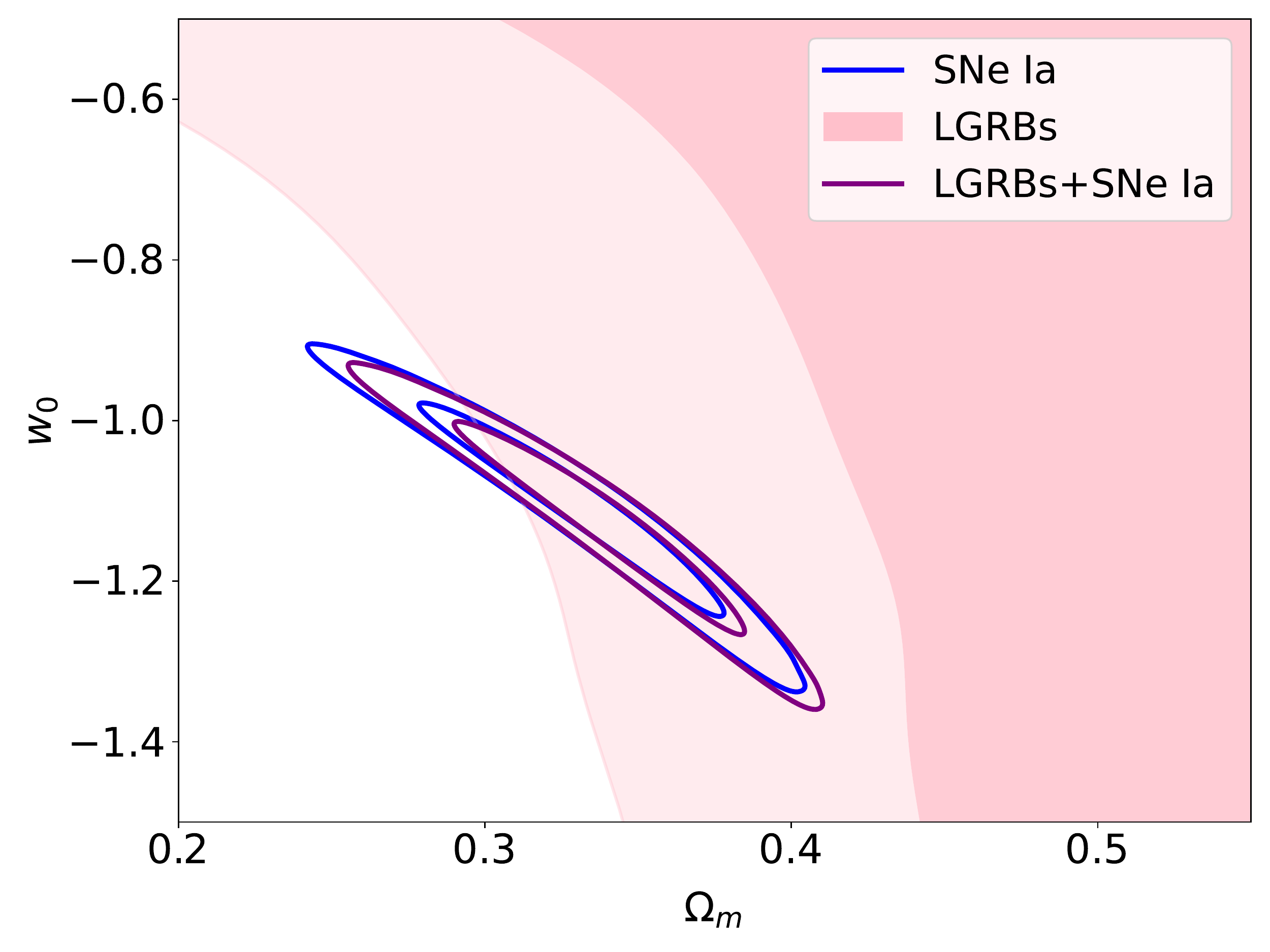}
	\caption{Confidence contours ($1\sigma$ and $2\sigma$) of the parameters $w_0$ and $\Omega_{m}$ in the $w(z)$ = $w_{0}$ model using the GW-LGRBs sample and Pantheon sample. It is worth noticing the confidence levels for GRBs do not include the systematic errors.}
	\label{F:L_w}       
\end{figure}

The luminosity distance in the flat $w(z) = w_{0}$ model can be written as
\begin{eqnarray}
	\label{eq:dl_w}
	d_{\rm L} = \frac{c(1+z)}{H_{0}} \int_{0}^{z} \frac{dz'}{\sqrt{\Omega_{m} (1+z')^{3} + (1 - \Omega_{m})(1+z')^{3(1+w_{0})}}}.
\end{eqnarray}
Taking equation (\ref{eq:dl_w}) into equation (\ref{eq:mu}), we obtain the theoretical distance modulus which used to calculate the corresponding $\chi^{2}$. From the MD-total sample, we obtain $\Omega_{m}$ = $0.40_{-0.19}^{+0.11}$, and the 1$\sigma$ upper limit of $w_{0}$ is -0.58. Then adding the Pantheon sample, the best fitting results become $\Omega_{m}$ = $0.34_{-0.04}^{+0.05}$ and $w_{0}$ = $-1.11_{-0.15}^{+0.11}$, as shown in figure \ref{F:S_w}. These results are consistent with the $\Lambda$CDM model at 2$\sigma$ confidence level. Figure \ref{F:L_w} sketches the constraints on $\Omega_{m}$ and $w_{0}$ in $w(z) = w_{0}$ model using the GW-LGRBs sample and Pantheon sample. For the GW-LGRBs sample, the best-fitting results are $\Omega_{m}$ = $0.63_{-0.52}^{+0.26}$ and the 1$\sigma$ upper limit of $w_{0}$ is -0.16. Combining the Pantheon sample, we obtain $\Omega_{m}$ = $0.34_{-0.04}^{+0.04}$ and $w_{0}$ = $-1.12_{-0.14}^{+0.11}$. It is still consistent with the $\Lambda$CDM model within 2$\sigma$ confidence level.

\section{Conclusions and discussions}
In this work, we employ the \ltb{} correlation to constrain the cosmological parameters. According to the five criteria provided at Section \ref{sec:selec}, we pick out two samples: MD-SGRBs sample and GW-LGRBs sample. Using the corresponding \ltb{} correlation, we standardize the light curves of 5 SGRBs in the MD-SGRBs sample and the light curves of 24 LGRBs in the GW-LGRBs sample, respectively. The light curves of short GRBs in the MD-SGRBs sample are well standardized to a universal behavior with a dispersion of 0.4 dex for luminosity. The dispersion of standardized result for GW-LGRBs sample is a little bit larger than that for the MD-SGRBs sample. Adopting the GP method with 37 $H(z)$ data, we calibrate the \ltb{} correlation. Employing the calibrated \ltb{} correlation, we constrain cosmological parameters using the MD-total sample, GW-LGRBs sample and Pantheon sample. From the MD-total sample, we measure $\Omega_{m}$ = 0.34$_{-0.07}^{+0.08}$ in a flat universe and $\Omega_{m}$ = 0.33$_{-0.09}^{+0.06}$ and $\Omega_{\Lambda}$ = 1.06$_{-0.34}^{+0.15}$ in a nonflat universe. For the GW-LGRBs sample, we obtain $\Omega_{m}$ = $0.69_{-0.18}^{+0.24}$ in a flat universe and $\Omega_{m} = 0.59_{-0.27}^{+0.30}$ and $\Omega_{\Lambda}$ = $1.38_{-0.52}^{+0.23}$ in a nonflat universe. The results obtained from the calibrated \ltb{} correlation are consistent with that from the Pantheon sample within $2\sigma$ level. The above results all support the $\Lambda$CDM model. Combining the Pantheon SNe Ia sample, we also constrain $\Omega_{m}$ and $w_{0}$ in the $w(z) = w_{0}$ model. The results still support the $\Lambda$CDM model using the high redshift GRBs. In addition, it is worth noticing the confidence levels for GRBs do not include the systematic errors.

Our analysis also poses some interesting points. First, from the fitting results of the \ltb{} correlation, it is easy to find that the standardized result obtained from the GW-LGRBs sample is not satisfactory. The reason is that a similar X-ray afterglow decaying index is expected for GW emission and external shock emission. Therefore, the GW-LGRBs sample may be contaminated. This dilemma will be improved by the third generation Einstein telescope which may detect the corresponding GW signal from newly born magnetars \citep{2010CQGra..27h4007P}. According to the stain of detected GW signal, the total energy converted into gravitational waves could be estimated \citep{2011gwpa.book.....C}, which can be used to obtain the luminosity dominated by the GW radiation. If this luminosity is larger than that of the plateau phase and the decay index is nearly $-1$, we believe that the rotational energy losing of this kind of GRBs may be dominated by GW emissions. On the contrary, if the corresponding luminosity is smaller than that of the plateau phase, this kind of GRBs must not be what we are trying to find. In short, GW observations can help us to screen a clean GW-LGRBs sample in the future. At the same time, the GW signals also provide a model-independent method to calibrate the \ltb{} correlation. These two scenarios go hand in hand with GW emissions. For the SGRBs whose rotational energy losing by DM radiation (MD-SGRBs), the GWs will be produced in the process of compact star merger. With the development of GW detection technology, if the corresponding GW signal and the associated electromagnetic counterparts produced by a newly born magnetar could be detected by the corresponding detectors in the future. This breakthrough can make sure that the decay of index -1 of GRBs afterglow powered by the GW emission.

Combining GW signal and corresponding electromagnetic counterpart, more detailed information about the universe can be discovered. The GW signal as a model-independent method can be employed to obtain the luminosity distance \citep{2019ApJ...873...39W}. The redshift can be obtained by analyzing the corresponding GRB. These two parameters are usually used to constrain the cosmological parameters \citep{Yu2021}. For $H_{0}$ tension \citep{2018MNRAS.480.3879A,2019ApJ...876...85R,2020A&A...641A...6P,2020MNRAS.498.1420W}, GRBs with GW signals would be helpful to confirm further and/or constrain the value of $H_{0}$, like GRB 170817A \citep{2017PhRvL.119p1101A,2017Natur.551...85A}. In addition, combining the fitting results of the \ltb{} correlation obtained from the MD-LGRBs sample, we find that the fitting results of different types with the same plateau are different. For example, the slope found from the MD-LGRBs sample is near $-1$ \citep{2021prepWang}, but that of the MD-SGRBs sample is not. The slope $k$ equals $-1$, indicating that the rotational energy could be a constant for MD-LGRBs. Different slope values hint that the property of a newly born magnetars produced by LGRBs and SGRBs is different. For the GW-LGRBs sample, the fitting slope also deviates $-1$. This result shows that although the light curves both include the plateau phase, we should not mix these two cases together. Decay behaviors of GRBs light curve that we see reflect different physical mechanisms. It is more meaningful to make independent research for the different decay indices. Perhaps among the several types of GRBs that include plateau phases, only one or a few types of GRBs can be regarded as standard candles. As expressed above, with considerably improved sensitivity, the investigation of GRB will be more refined. Our work demonstrates that GRBs are promising standard candles, especially at high redshifts.


\section*{Acknowledgements}
We thank the anonymous referee for valuable comments. This work was supported by the National Natural Science
Foundation of China (grant No. U1831207 and 11833003), the Fundamental Research Funds for the Central Universities (No. 0201-14380045), the National SKA Program of China (grant No. 2020SKA0120300), and the National Key Research
and Development Program of China (grant No. 2017YFA0402600). This work made use of data supplied by the UK Swift Science Data Centre at the University of
Leicester.

\section*{DATA AVAILABILITY}
The data used in the paper are available at the UK Swift Science Data Centre at the University of
Leicester (https://www.swift.ac.uk/).

\bibliographystyle{mnras}
\bibliography{mnras_mg} 

\begin{thebibliography}{}
\makeatletter
\relax
\def\mn@urlcharsother{\let\do\@makeother \do\$\do\&\do\#\do\^\do\_\do\%\do\~}
\def\mn@doi{\begingroup\mn@urlcharsother \@ifnextchar [ {\mn@doi@}
  {\mn@doi@[]}}
\def\mn@doi@[#1]#2{\def\@tempa{#1}\ifx\@tempa\@empty \href
  {http://dx.doi.org/#2} {doi:#2}\else \href {http://dx.doi.org/#2} {#1}\fi
  \endgroup}
\def\mn@eprint#1#2{\mn@eprint@#1:#2::\@nil}
\def\mn@eprint@arXiv#1{\href {http://arxiv.org/abs/#1} {{\tt arXiv:#1}}}
\def\mn@eprint@dblp#1{\href {http://dblp.uni-trier.de/rec/bibtex/#1.xml}
  {dblp:#1}}
\def\mn@eprint@#1:#2:#3:#4\@nil{\def\@tempa {#1}\def\@tempb {#2}\def\@tempc
  {#3}\ifx \@tempc \@empty \let \@tempc \@tempb \let \@tempb \@tempa \fi \ifx
  \@tempb \@empty \def\@tempb {arXiv}\fi \@ifundefined
  {mn@eprint@\@tempb}{\@tempb:\@tempc}{\expandafter \expandafter \csname
  mn@eprint@\@tempb\endcsname \expandafter{\@tempc}}}

\bibitem[\protect\citeauthoryear{{Abbott} et~al.,}{{Abbott}
  et~al.}{2017a}]{2017PhRvL.119p1101A}
{Abbott} B.~P.,  et~al., 2017a, \mn@doi [\prl]
  {10.1103/PhysRevLett.119.161101}, \href
  {https://ui.adsabs.harvard.edu/abs/2017PhRvL.119p1101A} {119, 161101}

\bibitem[\protect\citeauthoryear{{Abbott} et~al.,}{{Abbott}
  et~al.}{2017b}]{2017Natur.551...85A}
{Abbott} B.~P.,  et~al., 2017b, \mn@doi [\nat] {10.1038/nature24471}, \href
  {https://ui.adsabs.harvard.edu/abs/2017Natur.551...85A} {551, 85}

\bibitem[\protect\citeauthoryear{{Abbott} et~al.,}{{Abbott}
  et~al.}{2018}]{2018MNRAS.480.3879A}
{Abbott} T.~M.~C.,  et~al., 2018, \mn@doi [\mnras] {10.1093/mnras/sty1939},
  \href {https://ui.adsabs.harvard.edu/abs/2018MNRAS.480.3879A} {480, 3879}

\bibitem[\protect\citeauthoryear{{Amati} \& {Della Valle}}{{Amati} \& {Della
  Valle}}{2013}]{2013IJMPD..2230028A}
{Amati} L.,  {Della Valle} M.,  2013, \mn@doi [International Journal of Modern
  Physics D] {10.1142/S0218271813300280}, \href
  {https://ui.adsabs.harvard.edu/abs/2013IJMPD..2230028A} {22, 1330028}

\bibitem[\protect\citeauthoryear{{Amati} et~al.,}{{Amati}
  et~al.}{2002}]{2002A&A...390...81A}
{Amati} L.,  et~al., 2002, \mn@doi [\aap] {10.1051/0004-6361:20020722}, \href
  {https://ui.adsabs.harvard.edu/abs/2002A&A...390...81A} {390, 81}

\bibitem[\protect\citeauthoryear{{Amati}, {Guidorzi}, {Frontera}, {Della
  Valle}, {Finelli}, {Landi}  \& {Montanari}}{{Amati}
  et~al.}{2008}]{2008MNRAS.391..577A}
{Amati} L.,  {Guidorzi} C.,  {Frontera} F.,  {Della Valle} M.,  {Finelli} F.,
  {Landi} R.,   {Montanari} E.,  2008, \mn@doi [\mnras]
  {10.1111/j.1365-2966.2008.13943.x}, \href
  {https://ui.adsabs.harvard.edu/abs/2008MNRAS.391..577A} {391, 577}

\bibitem[\protect\citeauthoryear{{Amati}, {D'Agostino}, {Luongo}, {Muccino}  \&
  {Tantalo}}{{Amati} et~al.}{2019}]{2019MNRAS.486L..46A}
{Amati} L.,  {D'Agostino} R.,  {Luongo} O.,  {Muccino} M.,   {Tantalo} M.,
  2019, \mn@doi [\mnras] {10.1093/mnrasl/slz056}, \href
  {https://ui.adsabs.harvard.edu/abs/2019MNRAS.486L..46A} {486, L46}

\bibitem[\protect\citeauthoryear{{Barkana} \& {Loeb}}{{Barkana} \&
  {Loeb}}{2004}]{2004ApJ...601...64B}
{Barkana} R.,  {Loeb} A.,  2004, \mn@doi [\apj] {10.1086/380435}, \href
  {https://ui.adsabs.harvard.edu/abs/2004ApJ...601...64B} {601, 64}

\bibitem[\protect\citeauthoryear{{Berger}}{{Berger}}{2007}]{2007ApJ...670.1254B}
{Berger} E.,  2007, \mn@doi [\apj] {10.1086/522195}, \href
  {https://ui.adsabs.harvard.edu/abs/2007ApJ...670.1254B} {670, 1254}

\bibitem[\protect\citeauthoryear{{Bloom}, {Frail}  \& {Sari}}{{Bloom}
  et~al.}{2001}]{2001AJ....121.2879B}
{Bloom} J.~S.,  {Frail} D.~A.,   {Sari} R.,  2001, \mn@doi [\aj]
  {10.1086/321093}, \href
  {https://ui.adsabs.harvard.edu/abs/2001AJ....121.2879B} {121, 2879}

\bibitem[\protect\citeauthoryear{{Bromm} \& {Loeb}}{{Bromm} \&
  {Loeb}}{2007}]{2007AIPC..937..532B}
{Bromm} V.,  {Loeb} A.,  2007, in {Immler} S.,  {Weiler} K.,   {McCray} R.,
  eds,  American Institute of Physics Conference Series Vol. 937, Supernova
  1987A: 20 Years After: Supernovae and Gamma-Ray Bursters. pp 532--541,
  \mn@doi{10.1063/1.3682957}

\bibitem[\protect\citeauthoryear{{Bromm}, {Coppi}  \& {Larson}}{{Bromm}
  et~al.}{2002}]{2002ApJ...564...23B}
{Bromm} V.,  {Coppi} P.~S.,   {Larson} R.~B.,  2002, \mn@doi [\apj]
  {10.1086/323947}, \href
  {https://ui.adsabs.harvard.edu/abs/2002ApJ...564...23B} {564, 23}

\bibitem[\protect\citeauthoryear{{Capozziello} \& {Izzo}}{{Capozziello} \&
  {Izzo}}{2008}]{2008A&A...490...31C}
{Capozziello} S.,  {Izzo} L.,  2008, \mn@doi [\aap]
  {10.1051/0004-6361:200810337}, \href
  {https://ui.adsabs.harvard.edu/abs/2008A&A...490...31C} {490, 31}

\bibitem[\protect\citeauthoryear{{Creighton} \& {Anderson}}{{Creighton} \&
  {Anderson}}{2011}]{2011gwpa.book.....C}
{Creighton} J.,  {Anderson} W.,  2011, {Gravitational-Wave Physics and
  Astronomy: An Introduction to Theory, Experiment and Data Analysis.}

\bibitem[\protect\citeauthoryear{{Cucchiara}, {Levan}, {Fox}, {Tanvir}  \&
  {Ukwatta}}{{Cucchiara} et~al.}{2011}]{2011ApJ...736....7C}
{Cucchiara} A.,  {Levan} A.~J.,  {Fox} D.~B.,  {Tanvir} N.~R.,   {Ukwatta}
  T.~N.,  2011, \mn@doi [\apj] {10.1088/0004-637X/736/1/7}, \href
  {https://ui.adsabs.harvard.edu/abs/2011ApJ...736....7C} {736, 7}

\bibitem[\protect\citeauthoryear{{D'Agostini}}{{D'Agostini}}{2005}]{2005physics..11182D}
{D'Agostini} G.,  2005, arXiv e-prints, \href
  {https://ui.adsabs.harvard.edu/abs/2005physics..11182D} {p. physics/0511182}

\bibitem[\protect\citeauthoryear{{Dai} \& {Lu}}{{Dai} \&
  {Lu}}{1998}]{1998A&A...333L..87D}
{Dai} Z.~G.,  {Lu} T.,  1998, \aap, \href
  {https://ui.adsabs.harvard.edu/abs/1998A&A...333L..87D} {333, L87}

\bibitem[\protect\citeauthoryear{{Dai}, {Liang}  \& {Xu}}{{Dai}
  et~al.}{2004}]{2004ApJ...612L.101D}
{Dai} Z.~G.,  {Liang} E.~W.,   {Xu} D.,  2004, \mn@doi [\apjl]
  {10.1086/424694}, \href
  {https://ui.adsabs.harvard.edu/abs/2004ApJ...612L.101D} {612, L101}

\bibitem[\protect\citeauthoryear{{Dainotti} \& {Amati}}{{Dainotti} \&
  {Amati}}{2018}]{2018PASP..130e1001D}
{Dainotti} M.~G.,  {Amati} L.,  2018, \mn@doi [\pasp]
  {10.1088/1538-3873/aaa8d7}, \href
  {https://ui.adsabs.harvard.edu/abs/2018PASP..130e1001D} {130, 051001}

\bibitem[\protect\citeauthoryear{{Dainotti} \& {Del Vecchio}}{{Dainotti} \&
  {Del Vecchio}}{2017}]{2017NewAR..77...23D}
{Dainotti} M.~G.,  {Del Vecchio} R.,  2017, \mn@doi [\nar]
  {10.1016/j.newar.2017.04.001}, \href
  {https://ui.adsabs.harvard.edu/abs/2017NewAR..77...23D} {77, 23}

\bibitem[\protect\citeauthoryear{{Dainotti}, {Cardone}  \&
  {Capozziello}}{{Dainotti} et~al.}{2008}]{2008MNRAS.391L..79D}
{Dainotti} M.~G.,  {Cardone} V.~F.,   {Capozziello} S.,  2008, \mn@doi [\mnras]
  {10.1111/j.1745-3933.2008.00560.x}, \href
  {https://ui.adsabs.harvard.edu/abs/2008MNRAS.391L..79D} {391, L79}

\bibitem[\protect\citeauthoryear{{Dainotti}, {Willingale}, {Capozziello},
  {Fabrizio Cardone}  \& {Ostrowski}}{{Dainotti}
  et~al.}{2010}]{2010ApJ...722L.215D}
{Dainotti} M.~G.,  {Willingale} R.,  {Capozziello} S.,  {Fabrizio Cardone} V.,
   {Ostrowski} M.,  2010, \mn@doi [\apjl] {10.1088/2041-8205/722/2/L215}, \href
  {https://ui.adsabs.harvard.edu/abs/2010ApJ...722L.215D} {722, L215}

\bibitem[\protect\citeauthoryear{{Dainotti}, {Ostrowski}  \&
  {Willingale}}{{Dainotti} et~al.}{2011}]{2011MNRAS.418.2202D}
{Dainotti} M.~G.,  {Ostrowski} M.,   {Willingale} R.,  2011, \mn@doi [\mnras]
  {10.1111/j.1365-2966.2011.19433.x}, \href
  {https://ui.adsabs.harvard.edu/abs/2011MNRAS.418.2202D} {418, 2202}

\bibitem[\protect\citeauthoryear{{Dainotti}, {Cardone}, {Piedipalumbo}  \&
  {Capozziello}}{{Dainotti} et~al.}{2013a}]{2013MNRAS.436...82D}
{Dainotti} M.~G.,  {Cardone} V.~F.,  {Piedipalumbo} E.,   {Capozziello} S.,
  2013a, \mn@doi [\mnras] {10.1093/mnras/stt1516}, \href
  {https://ui.adsabs.harvard.edu/abs/2013MNRAS.436...82D} {436, 82}

\bibitem[\protect\citeauthoryear{{Dainotti}, {Petrosian}, {Singal}  \&
  {Ostrowski}}{{Dainotti} et~al.}{2013b}]{2013ApJ...774..157D}
{Dainotti} M.~G.,  {Petrosian} V.,  {Singal} J.,   {Ostrowski} M.,  2013b,
  \mn@doi [\apj] {10.1088/0004-637X/774/2/157}, \href
  {https://ui.adsabs.harvard.edu/abs/2013ApJ...774..157D} {774, 157}

\bibitem[\protect\citeauthoryear{{Dall'Osso}, {Giacomazzo}, {Perna}  \&
  {Stella}}{{Dall'Osso} et~al.}{2015}]{2015ApJ...798...25D}
{Dall'Osso} S.,  {Giacomazzo} B.,  {Perna} R.,   {Stella} L.,  2015, \mn@doi
  [\apj] {10.1088/0004-637X/798/1/25}, \href
  {https://ui.adsabs.harvard.edu/abs/2015ApJ...798...25D} {798, 25}

\bibitem[\protect\citeauthoryear{{Demianski}, {Piedipalumbo}, {Sawant}  \&
  {Amati}}{{Demianski} et~al.}{2017}]{2017A&A...598A.112D}
{Demianski} M.,  {Piedipalumbo} E.,  {Sawant} D.,   {Amati} L.,  2017, \mn@doi
  [\aap] {10.1051/0004-6361/201628909}, \href
  {https://ui.adsabs.harvard.edu/abs/2017A&A...598A.112D} {598, A112}

\bibitem[\protect\citeauthoryear{{Dirirsa} \& {Razzaque}}{{Dirirsa} \&
  {Razzaque}}{2017}]{2017heas.confE...2D}
{Dirirsa} F.~F.,  {Razzaque} S.,  2017, in 5th Annual Conference on High Energy
  Astrophysics in Southern Africa. p.~2

\bibitem[\protect\citeauthoryear{{Fan}, {Wu}  \& {Wei}}{{Fan}
  et~al.}{2013}]{2013PhRvD..88f7304F}
{Fan} Y.-Z.,  {Wu} X.-F.,   {Wei} D.-M.,  2013, \mn@doi [\prd]
  {10.1103/PhysRevD.88.067304}, \href
  {https://ui.adsabs.harvard.edu/abs/2013PhRvD..88f7304F} {88, 067304}

\bibitem[\protect\citeauthoryear{{Foreman-Mackey}, {Hogg}, {Lang}  \&
  {Goodman}}{{Foreman-Mackey} et~al.}{2013}]{2013PASP..125..306F}
{Foreman-Mackey} D.,  {Hogg} D.~W.,  {Lang} D.,   {Goodman} J.,  2013, \mn@doi
  [\pasp] {10.1086/670067}, \href
  {https://ui.adsabs.harvard.edu/abs/2013PASP..125..306F} {125, 306}

\bibitem[\protect\citeauthoryear{{Ghirlanda}, {Ghisellini}, {Lazzati}  \&
  {Firmani}}{{Ghirlanda} et~al.}{2004a}]{2004ApJ...613L..13G}
{Ghirlanda} G.,  {Ghisellini} G.,  {Lazzati} D.,   {Firmani} C.,  2004a,
  \mn@doi [\apjl] {10.1086/424915}, \href
  {https://ui.adsabs.harvard.edu/abs/2004ApJ...613L..13G} {613, L13}

\bibitem[\protect\citeauthoryear{{Ghirlanda}, {Ghisellini}  \&
  {Lazzati}}{{Ghirlanda} et~al.}{2004b}]{2004ApJ...616..331G}
{Ghirlanda} G.,  {Ghisellini} G.,   {Lazzati} D.,  2004b, \mn@doi [\apj]
  {10.1086/424913}, \href
  {https://ui.adsabs.harvard.edu/abs/2004ApJ...616..331G} {616, 331}

\bibitem[\protect\citeauthoryear{{Ghisellini}, {Nardini}, {Ghirlanda}  \&
  {Celotti}}{{Ghisellini} et~al.}{2009}]{2009MNRAS.393..253G}
{Ghisellini} G.,  {Nardini} M.,  {Ghirlanda} G.,   {Celotti} A.,  2009, \mn@doi
  [\mnras] {10.1111/j.1365-2966.2008.14214.x}, \href
  {https://ui.adsabs.harvard.edu/abs/2009MNRAS.393..253G} {393, 253}

\bibitem[\protect\citeauthoryear{{Hartoog} et~al.,}{{Hartoog}
  et~al.}{2015}]{2015A&A...580A.139H}
{Hartoog} O.~E.,  et~al., 2015, \mn@doi [\aap] {10.1051/0004-6361/201425001},
  \href {https://ui.adsabs.harvard.edu/abs/2015A&A...580A.139H} {580, A139}

\bibitem[\protect\citeauthoryear{{Izzo}, {Muccino}, {Zaninoni}, {Amati}  \&
  {Della Valle}}{{Izzo} et~al.}{2015}]{2015A&A...582A.115I}
{Izzo} L.,  {Muccino} M.,  {Zaninoni} E.,  {Amati} L.,   {Della Valle} M.,
  2015, \mn@doi [\aap] {10.1051/0004-6361/201526461}, \href
  {https://ui.adsabs.harvard.edu/abs/2015A&A...582A.115I} {582, A115}

\bibitem[\protect\citeauthoryear{{Khadka} \& {Ratra}}{{Khadka} \&
  {Ratra}}{2020a}]{2020MNRAS.492.4456K}
{Khadka} N.,  {Ratra} B.,  2020a, \mn@doi [\mnras] {10.1093/mnras/staa101},
  \href {https://ui.adsabs.harvard.edu/abs/2020MNRAS.492.4456K} {492, 4456}

\bibitem[\protect\citeauthoryear{{Khadka} \& {Ratra}}{{Khadka} \&
  {Ratra}}{2020b}]{2020MNRAS.497..263K}
{Khadka} N.,  {Ratra} B.,  2020b, \mn@doi [\mnras] {10.1093/mnras/staa1855},
  \href {https://ui.adsabs.harvard.edu/abs/2020MNRAS.497..263K} {497, 263}

\bibitem[\protect\citeauthoryear{{Khadka} \& {Ratra}}{{Khadka} \&
  {Ratra}}{2020c}]{2020MNRAS.499..391K}
{Khadka} N.,  {Ratra} B.,  2020c, \mn@doi [\mnras] {10.1093/mnras/staa2779},
  \href {https://ui.adsabs.harvard.edu/abs/2020MNRAS.499..391K} {499, 391}

\bibitem[\protect\citeauthoryear{{Khadka}, {Luongo}, {Muccino}  \&
  {Ratra}}{{Khadka} et~al.}{2021}]{Khadka21}
{Khadka} N.,  {Luongo} O.,  {Muccino} M.,   {Ratra} B.,  2021, arXiv e-prints,
  \href {https://ui.adsabs.harvard.edu/abs/arXiv210512692} {p.
  arXiv:2105.12692}

\bibitem[\protect\citeauthoryear{{Kodama}, {Yonetoku}, {Murakami}, {Tanabe},
  {Tsutsui}  \& {Nakamura}}{{Kodama} et~al.}{2008}]{2008MNRAS.391L...1K}
{Kodama} Y.,  {Yonetoku} D.,  {Murakami} T.,  {Tanabe} S.,  {Tsutsui} R.,
  {Nakamura} T.,  2008, \mn@doi [\mnras] {10.1111/j.1745-3933.2008.00508.x},
  \href {https://ui.adsabs.harvard.edu/abs/2008MNRAS.391L...1K} {391, L1}

\bibitem[\protect\citeauthoryear{{Kouveliotou}, {Meegan}, {Fishman}, {Bhat},
  {Briggs}, {Koshut}, {Paciesas}  \& {Pendleton}}{{Kouveliotou}
  et~al.}{1993}]{1993ApJ...413L.101K}
{Kouveliotou} C.,  {Meegan} C.~A.,  {Fishman} G.~J.,  {Bhat} N.~P.,  {Briggs}
  M.~S.,  {Koshut} T.~M.,  {Paciesas} W.~S.,   {Pendleton} G.~N.,  1993,
  \mn@doi [\apjl] {10.1086/186969}, \href
  {https://ui.adsabs.harvard.edu/abs/1993ApJ...413L.101K} {413, L101}

\bibitem[\protect\citeauthoryear{{Kumar} \& {Zhang}}{{Kumar} \&
  {Zhang}}{2015}]{2015PhR...561....1K}
{Kumar} P.,  {Zhang} B.,  2015, \mn@doi [\physrep]
  {10.1016/j.physrep.2014.09.008}, \href
  {https://ui.adsabs.harvard.edu/abs/2015PhR...561....1K} {561, 1}

\bibitem[\protect\citeauthoryear{{Lasky} \& {Glampedakis}}{{Lasky} \&
  {Glampedakis}}{2016}]{2016MNRAS.458.1660L}
{Lasky} P.~D.,  {Glampedakis} K.,  2016, \mn@doi [\mnras]
  {10.1093/mnras/stw435}, \href
  {https://ui.adsabs.harvard.edu/abs/2016MNRAS.458.1660L} {458, 1660}

\bibitem[\protect\citeauthoryear{{Li}, {Xia}, {Liu}, {Zhao}, {Fan}  \&
  {Zhang}}{{Li} et~al.}{2008}]{2008ApJ...680...92L}
{Li} H.,  {Xia} J.-Q.,  {Liu} J.,  {Zhao} G.-B.,  {Fan} Z.-H.,   {Zhang} X.,
  2008, \mn@doi [\apj] {10.1086/529582}, \href
  {https://ui.adsabs.harvard.edu/abs/2008ApJ...680...92L} {680, 92}

\bibitem[\protect\citeauthoryear{{Liang} \& {Zhang}}{{Liang} \&
  {Zhang}}{2005}]{2005ApJ...633..611L}
{Liang} E.,  {Zhang} B.,  2005, \mn@doi [\apj] {10.1086/491594}, \href
  {https://ui.adsabs.harvard.edu/abs/2005ApJ...633..611L} {633, 611}

\bibitem[\protect\citeauthoryear{{Liang} \& {Zhang}}{{Liang} \&
  {Zhang}}{2006}]{2006MNRAS.369L..37L}
{Liang} E.,  {Zhang} B.,  2006, \mn@doi [\mnras]
  {10.1111/j.1745-3933.2006.00169.x}, \href
  {https://ui.adsabs.harvard.edu/abs/2006MNRAS.369L..37L} {369, L37}

\bibitem[\protect\citeauthoryear{{Liang}, {Zhang}  \& {Zhang}}{{Liang}
  et~al.}{2007}]{2007ApJ...670..565L}
{Liang} E.-W.,  {Zhang} B.-B.,   {Zhang} B.,  2007, \mn@doi [\apj]
  {10.1086/521870}, \href
  {https://ui.adsabs.harvard.edu/abs/2007ApJ...670..565L} {670, 565}

\bibitem[\protect\citeauthoryear{{Liang}, {Xiao}, {Liu}  \& {Zhang}}{{Liang}
  et~al.}{2008}]{2008ApJ...685..354L}
{Liang} N.,  {Xiao} W.~K.,  {Liu} Y.,   {Zhang} S.~N.,  2008, \mn@doi [\apj]
  {10.1086/590903}, \href
  {https://ui.adsabs.harvard.edu/abs/2008ApJ...685..354L} {685, 354}

\bibitem[\protect\citeauthoryear{{Lloyd-Ronning}, {Johnson}  \&
  {Aykutalp}}{{Lloyd-Ronning} et~al.}{2020}]{2020MNRAS.498.5041L}
{Lloyd-Ronning} N.~M.,  {Johnson} J.~L.,   {Aykutalp} A.,  2020, \mn@doi
  [\mnras] {10.1093/mnras/staa2787}, \href
  {https://ui.adsabs.harvard.edu/abs/2020MNRAS.498.5041L} {498, 5041}

\bibitem[\protect\citeauthoryear{{L{\"u}} et~al.,}{{L{\"u}}
  et~al.}{2020}]{2020ApJ...898L...6L}
{L{\"u}} H.-J.,  et~al., 2020, \mn@doi [\apjl] {10.3847/2041-8213/aba1ed},
  \href {https://ui.adsabs.harvard.edu/abs/2020ApJ...898L...6L} {898, L6}

\bibitem[\protect\citeauthoryear{{Mehrabi} \& {Basilakos}}{{Mehrabi} \&
  {Basilakos}}{2020}]{2020EPJC...80..632M}
{Mehrabi} A.,  {Basilakos} S.,  2020, \mn@doi [European Physical Journal C]
  {10.1140/epjc/s10052-020-8221-2}, \href
  {https://ui.adsabs.harvard.edu/abs/2020EPJC...80..632M} {80, 632}

\bibitem[\protect\citeauthoryear{{M{\'e}sz{\'a}ros}}{{M{\'e}sz{\'a}ros}}{2006}]{Meszaros2006}
{M{\'e}sz{\'a}ros} P.,  2006, \mn@doi [Reports on Progress in Physics]
  {10.1088/0034-4885/69/8/R01}, \href
  {https://ui.adsabs.harvard.edu/abs/2006RPPh...69.2259M} {69, 2259}

\bibitem[\protect\citeauthoryear{{Montiel}, {Cabrera}  \& {Hidalgo}}{{Montiel}
  et~al.}{2021}]{2021MNRAS.501.3515M}
{Montiel} A.,  {Cabrera} J.~I.,   {Hidalgo} J.~C.,  2021, \mn@doi [\mnras]
  {10.1093/mnras/staa3926}, \href
  {https://ui.adsabs.harvard.edu/abs/2021MNRAS.501.3515M} {501, 3515}

\bibitem[\protect\citeauthoryear{{Muccino}, {Izzo}, {Luongo}, {Boshkayev},
  {Amati}, {Della Valle}, {Pisani}  \& {Zaninoni}}{{Muccino}
  et~al.}{2021}]{2021ApJ...908..181M}
{Muccino} M.,  {Izzo} L.,  {Luongo} O.,  {Boshkayev} K.,  {Amati} L.,  {Della
  Valle} M.,  {Pisani} G.~B.,   {Zaninoni} E.,  2021, \mn@doi [\apj]
  {10.3847/1538-4357/abd254}, \href
  {https://ui.adsabs.harvard.edu/abs/2021ApJ...908..181M} {908, 181}

\bibitem[\protect\citeauthoryear{{Norris}, {Marani}  \& {Bonnell}}{{Norris}
  et~al.}{2000}]{2000ApJ...534..248N}
{Norris} J.~P.,  {Marani} G.~F.,   {Bonnell} J.~T.,  2000, \mn@doi [\apj]
  {10.1086/308725}, \href
  {https://ui.adsabs.harvard.edu/abs/2000ApJ...534..248N} {534, 248}

\bibitem[\protect\citeauthoryear{{Nousek} et~al.,}{{Nousek}
  et~al.}{2006}]{2006ApJ...642..389N}
{Nousek} J.~A.,  et~al., 2006, \mn@doi [\apj] {10.1086/500724}, \href
  {https://ui.adsabs.harvard.edu/abs/2006ApJ...642..389N} {642, 389}

\bibitem[\protect\citeauthoryear{{O'Brien} et~al.,}{{O'Brien}
  et~al.}{2006}]{2006ApJ...647.1213O}
{O'Brien} P.~T.,  et~al., 2006, \mn@doi [\apj] {10.1086/505457}, \href
  {https://ui.adsabs.harvard.edu/abs/2006ApJ...647.1213O} {647, 1213}

\bibitem[\protect\citeauthoryear{{Oates}, {Page}, {De Pasquale}, {Schady},
  {Breeveld}, {Holland}, {Kuin}  \& {Marshall}}{{Oates}
  et~al.}{2012}]{2012MNRAS.426L..86O}
{Oates} S.~R.,  {Page} M.~J.,  {De Pasquale} M.,  {Schady} P.,  {Breeveld}
  A.~A.,  {Holland} S.~T.,  {Kuin} N.~P.~M.,   {Marshall} F.~E.,  2012, \mn@doi
  [\mnras] {10.1111/j.1745-3933.2012.01331.x}, \href
  {https://ui.adsabs.harvard.edu/abs/2012MNRAS.426L..86O} {426, L86}

\bibitem[\protect\citeauthoryear{{Phillips}}{{Phillips}}{1993}]{1993ApJ...413L.105P}
{Phillips} M.~M.,  1993, \mn@doi [\apjl] {10.1086/186970}, \href
  {https://ui.adsabs.harvard.edu/abs/1993ApJ...413L.105P} {413, L105}

\bibitem[\protect\citeauthoryear{{Planck Collaboration} et~al.,}{{Planck
  Collaboration} et~al.}{2020}]{2020A&A...641A...6P}
{Planck Collaboration} et~al., 2020, \mn@doi [\aap]
  {10.1051/0004-6361/201833910}, \href
  {https://ui.adsabs.harvard.edu/abs/2020A&A...641A...6P} {641, A6}

\bibitem[\protect\citeauthoryear{{Punturo} et~al.,}{{Punturo}
  et~al.}{2010}]{2010CQGra..27h4007P}
{Punturo} M.,  et~al., 2010, \mn@doi [Classical and Quantum Gravity]
  {10.1088/0264-9381/27/8/084007}, \href
  {https://ui.adsabs.harvard.edu/abs/2010CQGra..27h4007P} {27, 084007}

\bibitem[\protect\citeauthoryear{{Riess}, {Casertano}, {Yuan}, {Macri}  \&
  {Scolnic}}{{Riess} et~al.}{2019}]{2019ApJ...876...85R}
{Riess} A.~G.,  {Casertano} S.,  {Yuan} W.,  {Macri} L.~M.,   {Scolnic} D.,
  2019, \mn@doi [\apj] {10.3847/1538-4357/ab1422}, \href
  {https://ui.adsabs.harvard.edu/abs/2019ApJ...876...85R} {876, 85}

\bibitem[\protect\citeauthoryear{{Risaliti} \& {Lusso}}{{Risaliti} \&
  {Lusso}}{2019}]{2019NatAs...3..272R}
{Risaliti} G.,  {Lusso} E.,  2019, \mn@doi [Nature Astronomy]
  {10.1038/s41550-018-0657-z}, \href
  {https://ui.adsabs.harvard.edu/abs/2019NatAs...3..272R} {3, 272}

\bibitem[\protect\citeauthoryear{{Rowlinson} et~al.,}{{Rowlinson}
  et~al.}{2010}]{2010MNRAS.409..531R}
{Rowlinson} A.,  et~al., 2010, \mn@doi [\mnras]
  {10.1111/j.1365-2966.2010.17354.x}, \href
  {https://ui.adsabs.harvard.edu/abs/2010MNRAS.409..531R} {409, 531}

\bibitem[\protect\citeauthoryear{{Sari}, {Piran}  \& {Narayan}}{{Sari}
  et~al.}{1998}]{Sari1998}
{Sari} R.,  {Piran} T.,   {Narayan} R.,  1998, \mn@doi [\apjl]
  {10.1086/311269}, \href
  {https://ui.adsabs.harvard.edu/abs/1998ApJ...497L..17S} {497, L17}

\bibitem[\protect\citeauthoryear{{Schaefer}}{{Schaefer}}{2007}]{2007ApJ...660...16S}
{Schaefer} B.~E.,  2007, \mn@doi [\apj] {10.1086/511742}, \href
  {https://ui.adsabs.harvard.edu/abs/2007ApJ...660...16S} {660, 16}

\bibitem[\protect\citeauthoryear{{Seikel}, {Clarkson}  \& {Smith}}{{Seikel}
  et~al.}{2012}]{2012JCAP...06..036S}
{Seikel} M.,  {Clarkson} C.,   {Smith} M.,  2012, \mn@doi [\jcap]
  {10.1088/1475-7516/2012/06/036}, \href
  {https://ui.adsabs.harvard.edu/abs/2012JCAP...06..036S} {2012, 036}

\bibitem[\protect\citeauthoryear{{Shapiro} \& {Teukolsky}}{{Shapiro} \&
  {Teukolsky}}{1983}]{1983bhwd.book.....S}
{Shapiro} S.~L.,  {Teukolsky} S.~A.,  1983, {Black holes, white dwarfs, and
  neutron stars : the physics of compact objects}

\bibitem[\protect\citeauthoryear{{Tang}, {Huang}, {Geng}  \& {Zhang}}{{Tang}
  et~al.}{2019}]{2019ApJS..245....1T}
{Tang} C.-H.,  {Huang} Y.-F.,  {Geng} J.-J.,   {Zhang} Z.-B.,  2019, \mn@doi
  [\apjs] {10.3847/1538-4365/ab4711}, \href
  {https://ui.adsabs.harvard.edu/abs/2019ApJS..245....1T} {245, 1}

\bibitem[\protect\citeauthoryear{{Totani}}{{Totani}}{1997}]{1997ApJ...486L..71T}
{Totani} T.,  1997, \mn@doi [\apjl] {10.1086/310853}, \href
  {https://ui.adsabs.harvard.edu/abs/1997ApJ...486L..71T} {486, L71}

\bibitem[\protect\citeauthoryear{{Totani}, {Kawai}, {Kosugi}, {Aoki}, {Yamada},
  {Iye}, {Ohta}  \& {Hattori}}{{Totani} et~al.}{2006}]{2006PASJ...58..485T}
{Totani} T.,  {Kawai} N.,  {Kosugi} G.,  {Aoki} K.,  {Yamada} T.,  {Iye} M.,
  {Ohta} K.,   {Hattori} T.,  2006, \mn@doi [\pasj] {10.1093/pasj/58.3.485},
  \href {https://ui.adsabs.harvard.edu/abs/2006PASJ...58..485T} {58, 485}

\bibitem[\protect\citeauthoryear{{Troja} et~al.,}{{Troja}
  et~al.}{2007}]{2007ApJ...665..599T}
{Troja} E.,  et~al., 2007, \mn@doi [\apj] {10.1086/519450}, \href
  {https://ui.adsabs.harvard.edu/abs/2007ApJ...665..599T} {665, 599}

\bibitem[\protect\citeauthoryear{{Tsvetkova} et~al.,}{{Tsvetkova}
  et~al.}{2017}]{2017ApJ...850..161T}
{Tsvetkova} A.,  et~al., 2017, \mn@doi [\apj] {10.3847/1538-4357/aa96af}, \href
  {https://ui.adsabs.harvard.edu/abs/2017ApJ...850..161T} {850, 161}

\bibitem[\protect\citeauthoryear{{Vangioni}, {Olive}, {Prestegard}, {Silk},
  {Petitjean}  \& {Mandic}}{{Vangioni} et~al.}{2015}]{2015MNRAS.447.2575V}
{Vangioni} E.,  {Olive} K.~A.,  {Prestegard} T.,  {Silk} J.,  {Petitjean} P.,
  {Mandic} V.,  2015, \mn@doi [\mnras] {10.1093/mnras/stu2600}, \href
  {https://ui.adsabs.harvard.edu/abs/2015MNRAS.447.2575V} {447, 2575}

\bibitem[\protect\citeauthoryear{{Wang}}{{Wang}}{2008}]{2008PhRvD..78l3532W}
{Wang} Y.,  2008, \mn@doi [\prd] {10.1103/PhysRevD.78.123532}, \href
  {https://ui.adsabs.harvard.edu/abs/2008PhRvD..78l3532W} {78, 123532}

\bibitem[\protect\citeauthoryear{{Wang}}{{Wang}}{2013}]{2013A&A...556A..90W}
{Wang} F.~Y.,  2013, \mn@doi [\aap] {10.1051/0004-6361/201321623}, \href
  {https://ui.adsabs.harvard.edu/abs/2013A&A...556A..90W} {556, A90}

\bibitem[\protect\citeauthoryear{{Wang} \& {Dai}}{{Wang} \&
  {Dai}}{2011}]{2011A&A...536A..96W}
{Wang} F.~Y.,  {Dai} Z.~G.,  2011, \mn@doi [\aap]
  {10.1051/0004-6361/201117517}, \href
  {https://ui.adsabs.harvard.edu/abs/2011A&A...536A..96W} {536, A96}

\bibitem[\protect\citeauthoryear{{Wang} \& {Wang}}{{Wang} \&
  {Wang}}{2019}]{2019ApJ...873...39W}
{Wang} Y.~Y.,  {Wang} F.~Y.,  2019, \mn@doi [\apj] {10.3847/1538-4357/ab037b},
  \href {https://ui.adsabs.harvard.edu/abs/2019ApJ...873...39W} {873, 39}

\bibitem[\protect\citeauthoryear{{Wang}, {Qi}  \& {Dai}}{{Wang}
  et~al.}{2011}]{2011MNRAS.415.3423W}
{Wang} F.-Y.,  {Qi} S.,   {Dai} Z.-G.,  2011, \mn@doi [\mnras]
  {10.1111/j.1365-2966.2011.18961.x}, \href
  {https://ui.adsabs.harvard.edu/abs/2011MNRAS.415.3423W} {415, 3423}

\bibitem[\protect\citeauthoryear{{Wang}, {Bromm}, {Greif}, {Stacy}, {Dai},
  {Loeb}  \& {Cheng}}{{Wang} et~al.}{2012}]{Wang2012}
{Wang} F.~Y.,  {Bromm} V.,  {Greif} T.~H.,  {Stacy} A.,  {Dai} Z.~G.,  {Loeb}
  A.,   {Cheng} K.~S.,  2012, \mn@doi [\apj] {10.1088/0004-637X/760/1/27},
  \href {https://ui.adsabs.harvard.edu/abs/2012ApJ...760...27W} {760, 27}

\bibitem[\protect\citeauthoryear{{Wang}, {Dai}  \& {Liang}}{{Wang}
  et~al.}{2015}]{2015NewAR..67....1W}
{Wang} F.~Y.,  {Dai} Z.~G.,   {Liang} E.~W.,  2015, \mn@doi [\nar]
  {10.1016/j.newar.2015.03.001}, \href
  {https://ui.adsabs.harvard.edu/abs/2015NewAR..67....1W} {67, 1}

\bibitem[\protect\citeauthoryear{{Wang}, {Wang}, {Cheng}  \& {Dai}}{{Wang}
  et~al.}{2016}]{2016A&A...585A..68W}
{Wang} J.~S.,  {Wang} F.~Y.,  {Cheng} K.~S.,   {Dai} Z.~G.,  2016, \mn@doi
  [\aap] {10.1051/0004-6361/201526485}, \href
  {https://ui.adsabs.harvard.edu/abs/2016A&A...585A..68W} {585, A68}

\bibitem[\protect\citeauthoryear{{Wang}, {Hu}, {Zhang}  \& {Dai}}{{Wang}
  et~al.}{2021}]{2021prepWang}
{Wang} F.~Y.,  {Hu} J.~P.,  {Zhang} G.~Q.,   {Dai} Z.~G.,  2021, arXiv
  e-prints, \href {https://ui.adsabs.harvard.edu/abs/2021arXiv210614155W} {p.
  arXiv:2106.14155}

\bibitem[\protect\citeauthoryear{{Willingale} et~al.,}{{Willingale}
  et~al.}{2007}]{2007ApJ...662.1093W}
{Willingale} R.,  et~al., 2007, \mn@doi [\apj] {10.1086/517989}, \href
  {https://ui.adsabs.harvard.edu/abs/2007ApJ...662.1093W} {662, 1093}

\bibitem[\protect\citeauthoryear{{Wong} et~al.,}{{Wong}
  et~al.}{2020}]{2020MNRAS.498.1420W}
{Wong} K.~C.,  et~al., 2020, \mn@doi [\mnras] {10.1093/mnras/stz3094}, \href
  {https://ui.adsabs.harvard.edu/abs/2020MNRAS.498.1420W} {498, 1420}

\bibitem[\protect\citeauthoryear{{Woosley} \& {Bloom}}{{Woosley} \&
  {Bloom}}{2006}]{2006ARA&A..44..507W}
{Woosley} S.~E.,  {Bloom} J.~S.,  2006, \mn@doi [\araa]
  {10.1146/annurev.astro.43.072103.150558}, \href
  {https://ui.adsabs.harvard.edu/abs/2006ARA&A..44..507W} {44, 507}

\bibitem[\protect\citeauthoryear{{Yonetoku}, {Murakami}, {Nakamura},
  {Yamazaki}, {Inoue}  \& {Ioka}}{{Yonetoku}
  et~al.}{2004}]{2004ApJ...609..935Y}
{Yonetoku} D.,  {Murakami} T.,  {Nakamura} T.,  {Yamazaki} R.,  {Inoue} A.~K.,
   {Ioka} K.,  2004, \mn@doi [\apj] {10.1086/421285}, \href
  {https://ui.adsabs.harvard.edu/abs/2004ApJ...609..935Y} {609, 935}

\bibitem[\protect\citeauthoryear{{Yu}, {Ratra}  \& {Wang}}{{Yu}
  et~al.}{2018}]{2018ApJ...856....3Y}
{Yu} H.,  {Ratra} B.,   {Wang} F.-Y.,  2018, \mn@doi [\apj]
  {10.3847/1538-4357/aab0a2}, \href
  {https://ui.adsabs.harvard.edu/abs/2018ApJ...856....3Y} {856, 3}

\bibitem[\protect\citeauthoryear{{Yu} et~al.,}{{Yu} et~al.}{2021}]{Yu2021}
{Yu} J.,  et~al., 2021, arXiv e-prints, \href
  {https://ui.adsabs.harvard.edu/abs/2021arXiv210412374Y} {p. arXiv:2104.12374}

\bibitem[\protect\citeauthoryear{{Y{\"u}ksel}, {Kistler}, {Beacom}  \&
  {Hopkins}}{{Y{\"u}ksel} et~al.}{2008}]{2008ApJ...683L...5Y}
{Y{\"u}ksel} H.,  {Kistler} M.~D.,  {Beacom} J.~F.,   {Hopkins} A.~M.,  2008,
  \mn@doi [\apjl] {10.1086/591449}, \href
  {https://ui.adsabs.harvard.edu/abs/2008ApJ...683L...5Y} {683, L5}

\bibitem[\protect\citeauthoryear{{Zhang} \& {M{\'e}sz{\'a}ros}}{{Zhang} \&
  {M{\'e}sz{\'a}ros}}{2001}]{2001ApJ...552L..35Z}
{Zhang} B.,  {M{\'e}sz{\'a}ros} P.,  2001, \mn@doi [\apjl] {10.1086/320255},
  \href {https://ui.adsabs.harvard.edu/abs/2001ApJ...552L..35Z} {552, L35}

\bibitem[\protect\citeauthoryear{{Zhang} \& {Wang}}{{Zhang} \&
  {Wang}}{2018}]{Zhang2018}
{Zhang} G.~Q.,  {Wang} F.~Y.,  2018, \mn@doi [\apj] {10.3847/1538-4357/aa9ce5},
  \href {https://ui.adsabs.harvard.edu/abs/2018ApJ...852....1Z} {852, 1}

\bibitem[\protect\citeauthoryear{{Zhang}, {Fan}, {Dyks}, {Kobayashi},
  {M{\'e}sz{\'a}ros}, {Burrows}, {Nousek}  \& {Gehrels}}{{Zhang}
  et~al.}{2006}]{2006ApJ...642..354Z}
{Zhang} B.,  {Fan} Y.~Z.,  {Dyks} J.,  {Kobayashi} S.,  {M{\'e}sz{\'a}ros} P.,
  {Burrows} D.~N.,  {Nousek} J.~A.,   {Gehrels} N.,  2006, \mn@doi [\apj]
  {10.1086/500723}, \href
  {https://ui.adsabs.harvard.edu/abs/2006ApJ...642..354Z} {642, 354}

\bibitem[\protect\citeauthoryear{{Zhao}, {Zhang}, {Gao}, {Lan}, {L{\"u}}  \&
  {Zhang}}{{Zhao} et~al.}{2019}]{2019ApJ...883...97Z}
{Zhao} L.,  {Zhang} B.,  {Gao} H.,  {Lan} L.,  {L{\"u}} H.,   {Zhang} B.,
  2019, \mn@doi [\apj] {10.3847/1538-4357/ab38c4}, \href
  {https://ui.adsabs.harvard.edu/abs/2019ApJ...883...97Z} {883, 97}

\makeatother
\end{thebibliography}
\bsp	

\clearpage
\onecolumn


\linespread{1.05}
\begin{table} \footnotesize	
	\centering
	\caption{Fitting results and some key parameters of the MD-SGRBs and GW-LGRBs samples. }		\label{T1}
	\begin{threeparttable}
		\begin{tabular}{p{3cm}p{1cm}p{1cm}p{3.0cm}p{2.5cm}p{1cm}p{1cm}p{1.7cm}} 
			\hline \hline
			Name&$z$\tnote{a}&$T_{90}$\tnote{b}&$F_0$\tnote{c} &$t_{\rm b}$\tnote{c}& {$\chi^{2}_{\rm r}$}&$\gamma$\tnote{d} & $\mu_{\rm obs}$\tnote{e} \\ \hline	
			&&(s)&10$^{-11}$(erg/cm$^{2}$/s) &10$^{3}$(s)&&&\\ \hline
			\textbf{MD-SGRBs sample} & & & & & & & \\
			\hline					
			GRB 051221A&0.55&1.40&0.24$_{-0.02}^{+0.02}$&45.30$_{-3.27}^{+3.50 }$&1.02&2.06&42.15$\pm$1.19\\
			GRB 090426&2.60&1.20&3.05$_{-0.46}^{+0.53}$&1.29$_{-0.20 }^{+0.26 }$&1.66&2.04&45.51$\pm$0.89\\
			GRB 090510&0.90&0.30&36.00$_{-1.73}^{+1.80}$&0.55$_{-0.03 }^{+0.03 }$&1.86&1.60\tnote{(1)}&43.45$\pm$0.86\\
			GRB 130603B&0.36&0.18&8.40$_{-0.40}^{+0.41}$&2.77$_{-0.12 }^{+0.13 }$&1.49&1.75\tnote{(1)}&42.14$\pm$0.82\\
			GRB 140903A&0.35&0.30&1.40$_{-0.01}^{+0.01}$&18.90$_{-1.41 }^{+1.50 }$&0.84&1.63&41.47$\pm$1.06\\
			\hline \hline
			\textbf{GW-LGRBs sample}& & & & & &  &\\
			\hline						
			GRB 050822&1.43&103.40&1.53$_{-0.22}^{+0.29}$&4.37$_{-0.77}^{+0.87 }$&1.22&2.11\tnote{(2)}&44.35$\pm$1.21\\
			GRB 051016B&0.94&4.00&0.79$_{-0.07}^{+0.07}$&4.48$_{-0.51 }^{+0.59 }$&1.16&1.86\tnote{(2)}&44.96$\pm$1.19\\
			GRB 060206&4.05&7.60&8.32$_{-0.80}^{+0.81}$&1.11$_{-0.11 }^{+0.13 }$&2.05&1.78\tnote{(2)}&46.75$\pm$1.23\\
			GRB 060502A&1.51&28.40&2.58$_{-0.36}^{+0.48}$&2.56$_{-0.48 }^{+0.51 }$&1.02&2.21\tnote{(2)}&44.78$\pm$1.20\\
			GRB 060719&1.53&66.90&1.87$_{-0.20}^{+0.22}$&1.54$_{-0.20 }^{+0.23 }$&1.32&2.36\tnote{(2)}&45.87$\pm$1.18\\
			GRB 070129&2.34&460.60&0.69$_{-0.05}^{+0.06}$&7.46$_{-0.75 }^{+0.81 }$&1.12&2.21\tnote{(1)}&44.79$\pm$1.18\\
			GRB 070802&2.45&16.40&0.50$_{-0.07}^{+0.08}$&2.52$_{-0.45 }^{+0.53 }$&2.84&1.99&47.37$\pm$1.20\\
			GRB 080516&3.20&5.80&1.38$_{-0.15}^{+0.16}$&1.37$_{-0.19 }^{+0.22 }$&1.02&2.69\tnote{(1)}&46.57$\pm$1.21\\
			GRB 080707&1.23&27.10&0.66$_{-0.09}^{+0.10}$&3.76$_{-0.70}^{+0.96 }$&0.60&2.19\tnote{(1)}&45.45$\pm$1.22\\
			GRB 090530&1.27&48.00&1.14$_{-0.25}^{+0.42}$&2.84$_{-0.86 }^{+0.97 }$&1.38&1.97\tnote{(1)}&45.55$\pm$1.31\\
			GRB 090927&1.37&2.20&0.74$_{-0.12}^{+0.17}$&2.91$_{-0.63 }^{+0.74 }$&2.44&2.07&45.97$\pm$1.23\\
			GRB 091029&2.75&39.20&1.33$_{-0.07}^{+0.07}$&5.93$_{-0.43 }^{+0.48 }$&1.06&2.10\tnote{(1)}&44.81$\pm$1.16\\
			GRB 100302A&4.81&17.90&0.18$_{-0.03}^{+0.03}$&10.10$_{-1.98 }^{+2.44 }$&0.69&1.80\tnote{(1)}&47.35$\pm$1.22\\
			GRB 100425A&1.76&37.00&0.58$_{-0.09}^{+0.09}$&2.13$_{-0.39 }^{+0.49 }$&1.75&2.24\tnote{(1)}&46.82$\pm$1.21\\
			GRB 100615A&1.40&39.00&7.16$_{-0.41}^{+0.43}$&3.20$_{-0.26 }^{+0.28 }$&0.86&2.38&43.06$\pm$1.15\\
			GRB 101219B&0.55&34.00&0.12$_{-0.02}^{+0.02}$&37.30$_{-7.17 }^{+9.09 }$&0.93&1.86& 42.93$\pm$1.49\\
			GRB 110808A&1.35&48.00&0.15$_{-0.03}^{+0.03}$&12.70$_{-2.86 }^{+3.60 }$&0.80&2.32\tnote{(1)}&44.94$\pm$1.32\\
			GRB 120118B&2.94&23.30&0.95$_{-0.07}^{+0.07}$&3.85$_{-0.42 }^{+0.47 }$&1.79&2.21& 45.84$\pm$1.16\\
			GRB 131105A&1.69&112.30&3.95$_{-0.32}^{+0.35}$&1.30$_{-0.14 }^{+0.15 }$&1.56&1.92\tnote{(1)}& 45.90$\pm$1.18\\
			GRB 151027A&4.06&80.00&1.51$_{-0.23}^{+0.29}$&4.00$_{-0.85 }^{+0.99 }$&0.75&1.81&  46.34$\pm$1.22\\
			GRB 160804A&0.74&144.20&0.71$_{-0.07}^{+0.08}$&8.74$_{-1.32 }^{+1.51 }$&1.46&1.97\tnote{(1)}&43.65$\pm$1.26\\
			GRB 170202A&3.65&46.20&5.52$_{-0.47}^{+0.52}$&1.04$_{-0.11 }^{+0.12 }$&2.04&2.12\tnote{(1)}& 46.58$\pm$1.23\\
			GRB 170607A&0.56&23.00&4.16$_{-0.19}^{+0.20}$&5.38$_{-0.33 }^{+0.36 }$&1.41&2.08\tnote{(1)}& 42.33$\pm$1.20\\
			GRB 170705A&2.01&217.30&13.30$_{-0.81}^{+0.87}$&2.11$_{-0.15 }^{+0.16 }$&2.13&1.89&  43.99$\pm$1.15\\  \hline		
		\end{tabular}
		\begin{tablenotes}
			\footnotesize
			\item[] \textbf{Reference.} (1) \cite{2019ApJ...883...97Z}, (2) \cite{2007ApJ...662.1093W}.
			\item[a] The measured redshifts are adopted from the published papers and GCNs.
			\item[b] The duration of GRBs are obtained from the Swift GRB table at https://swift.gsfc.nasa.gov/archive/grb$_{-}$table.html/
			\item[c] The parameters $F_0$ and $t_b$ are derived from equation (\ref{q_rev}). The corresponding 1$\sigma$ errors, $\sigma_{\rm F_{0}}$ and $\sigma_{t_{\rm b}}$, are derived from the function $\sqrt{({\sigma_{\rm u}}^2+{\sigma_{\rm d}}^2)/2}$. Here, $\sigma_{\rm u}$ and $\sigma_{\rm d}$ are the upper error and the lower error of the parameters, respectively.
			\item[d] $\gamma$ is the photon index of the plateau phase. When $\gamma$ is not given in published papers, the value from the Swift GRB table is used.
			\item[e]  $\mu_{\rm obs}$ is distance modulus derived from equation (\ref{eq:muobs}) by using the calibrated \ltb{} correlation.
		\end{tablenotes}
	\end{threeparttable}
\end{table}

\clearpage
\appendix
\section{XRT light curves (0.3-10 keV) of all GRBs used in this paper}

\clearpage

\begin{figure}
	\centering
	\includegraphics[width=0.31\hsize]{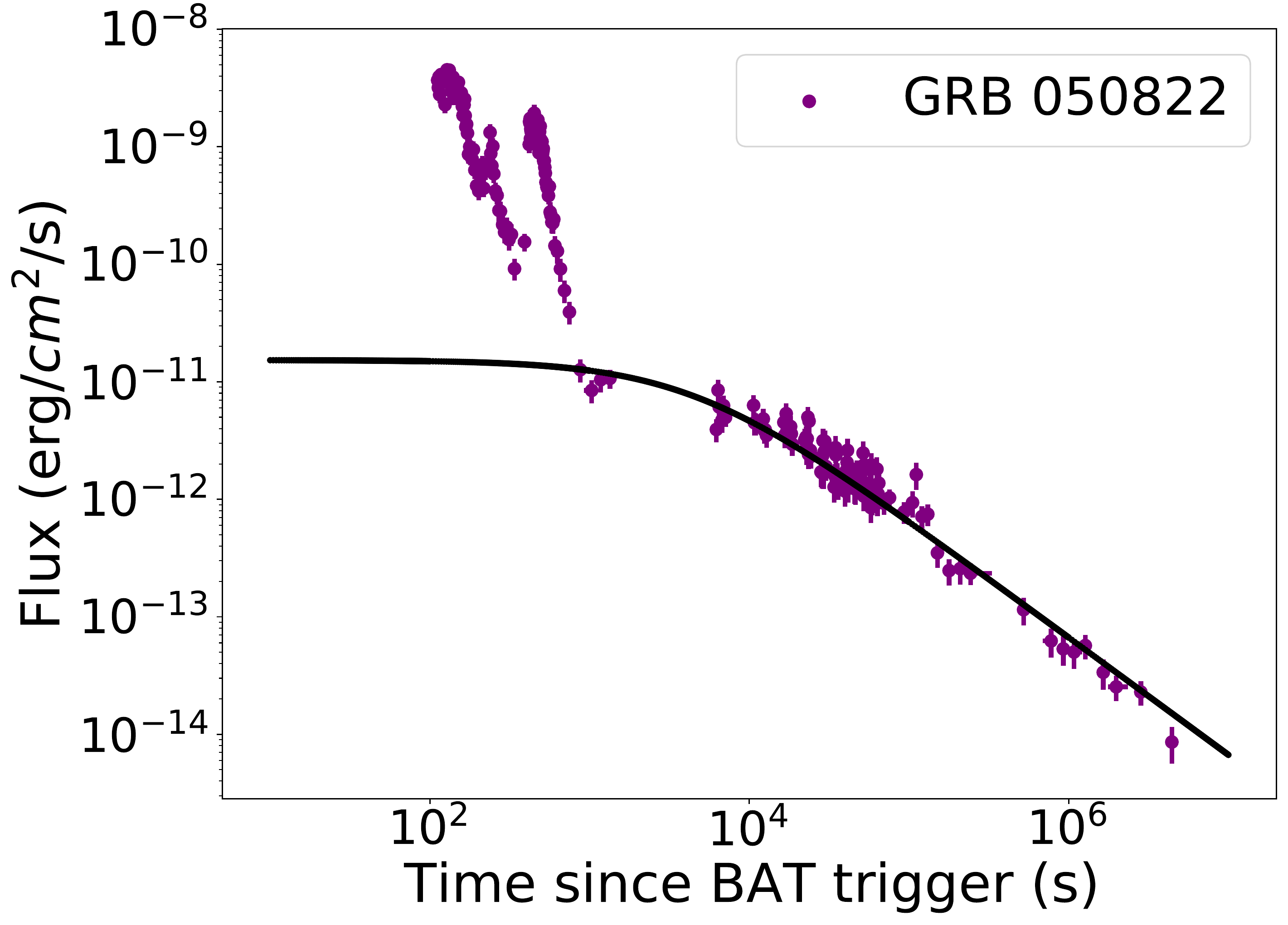}
	\includegraphics[width=0.31\hsize]{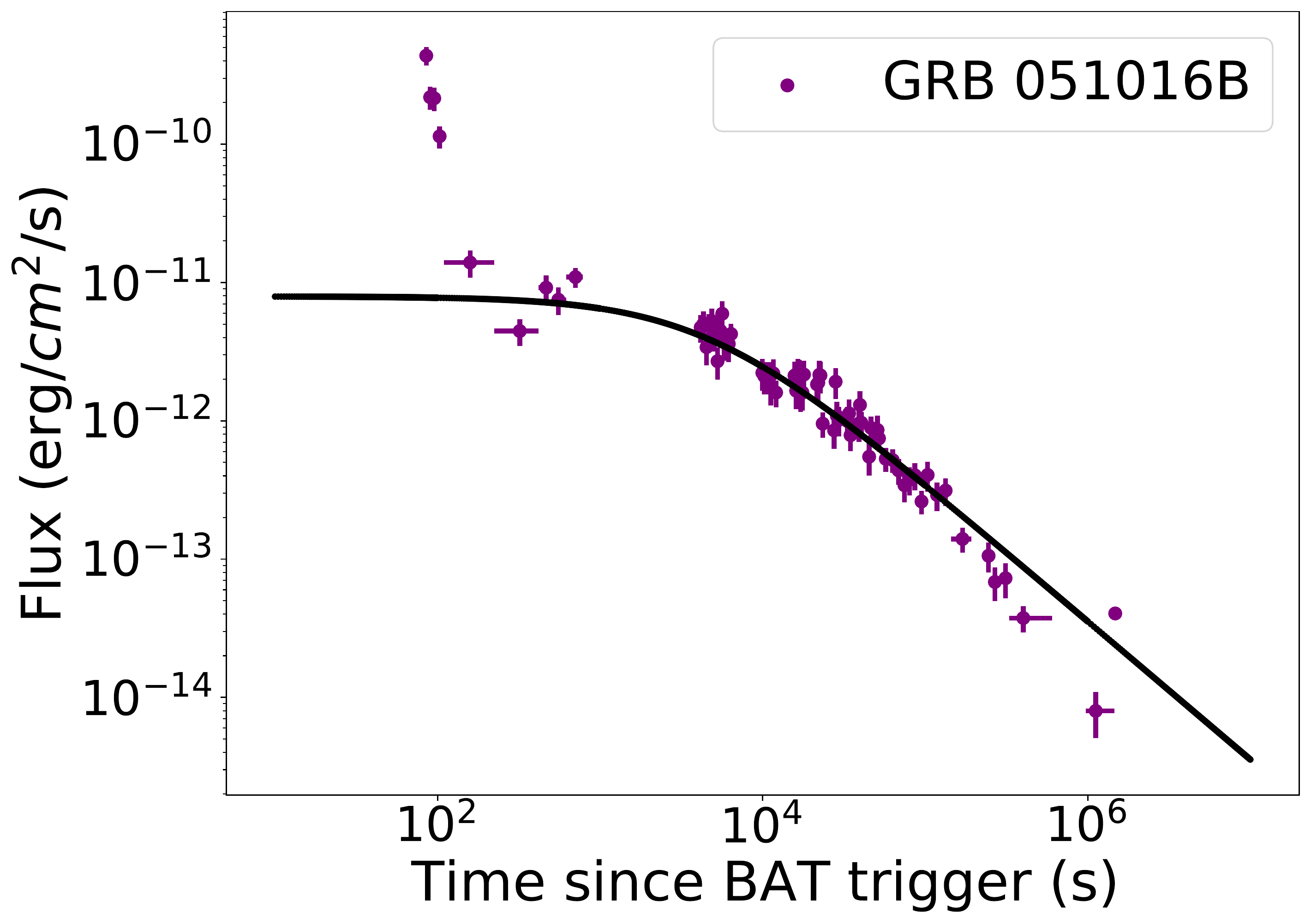}
	\includegraphics[width=0.31\hsize]{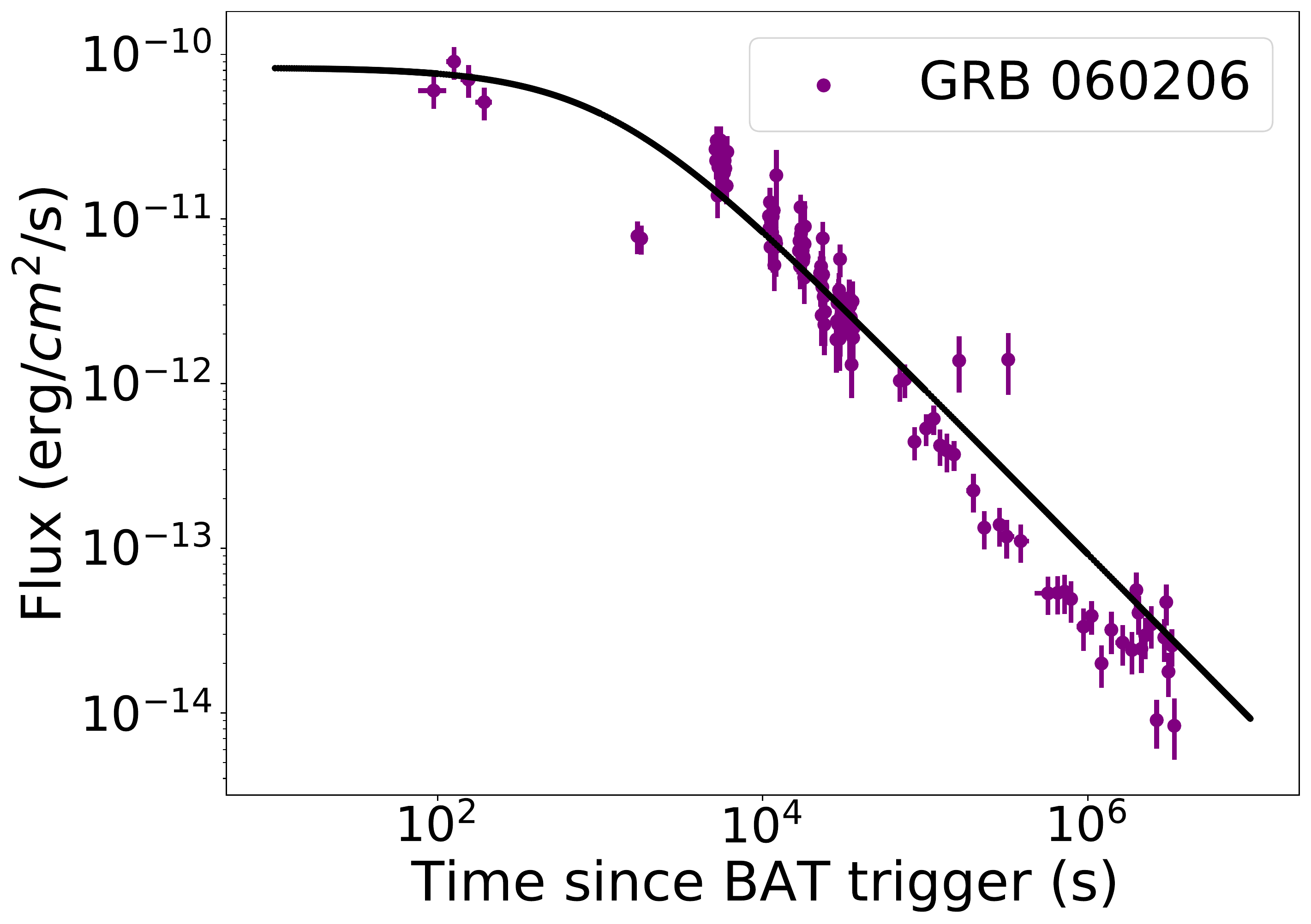}
	\includegraphics[width=0.31\hsize]{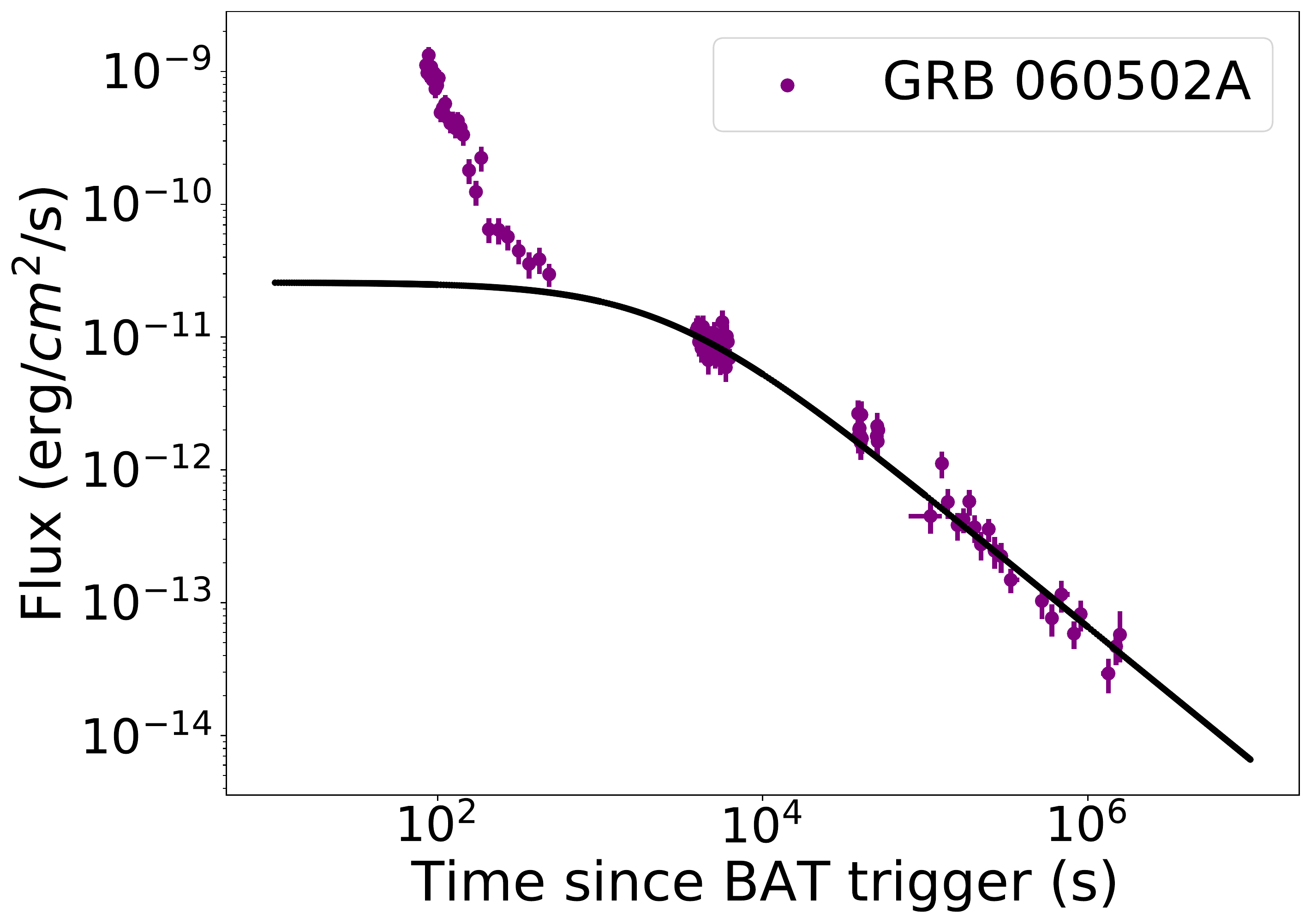}
	\includegraphics[width=0.31\hsize]{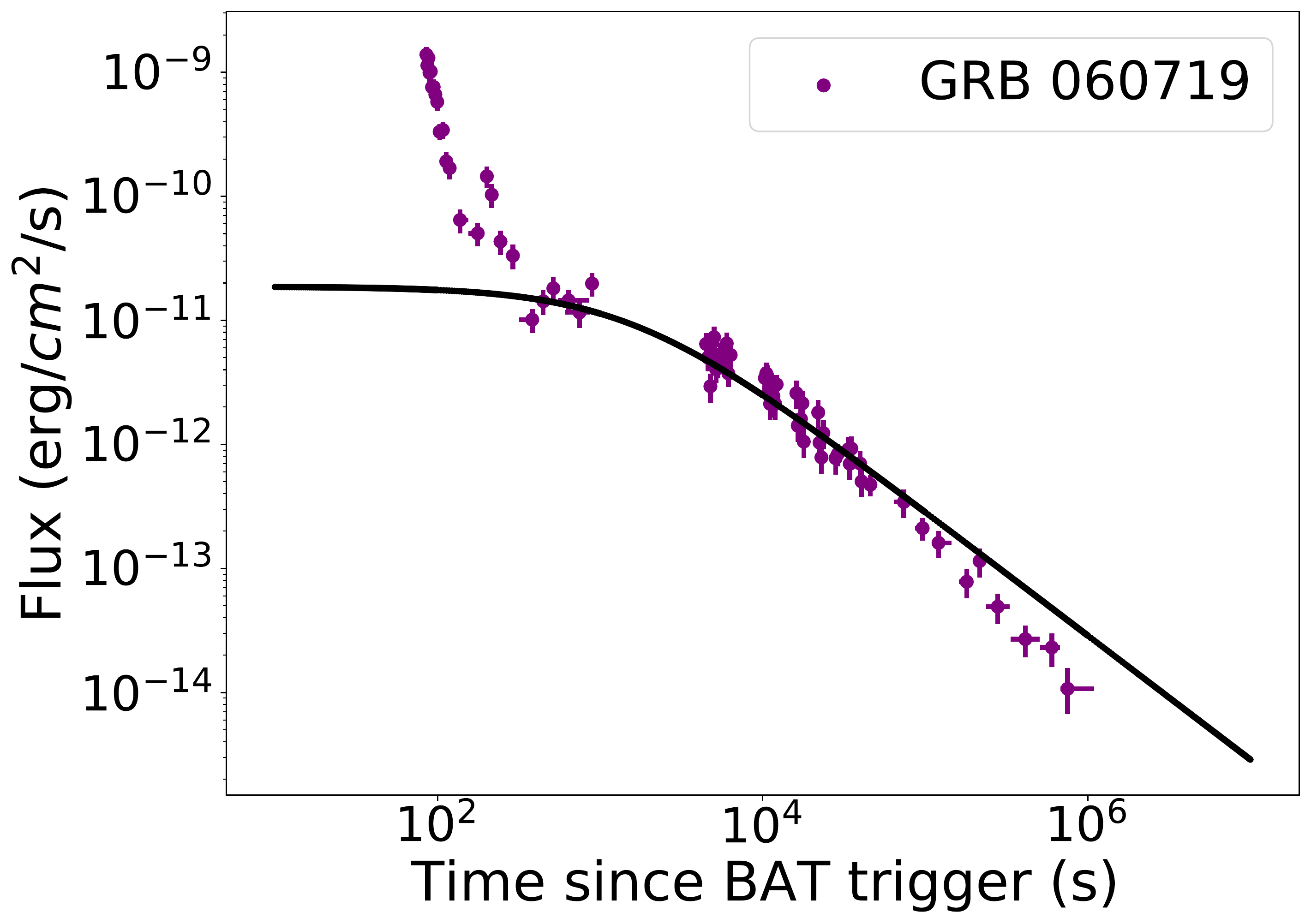}
	\includegraphics[width=0.31\hsize]{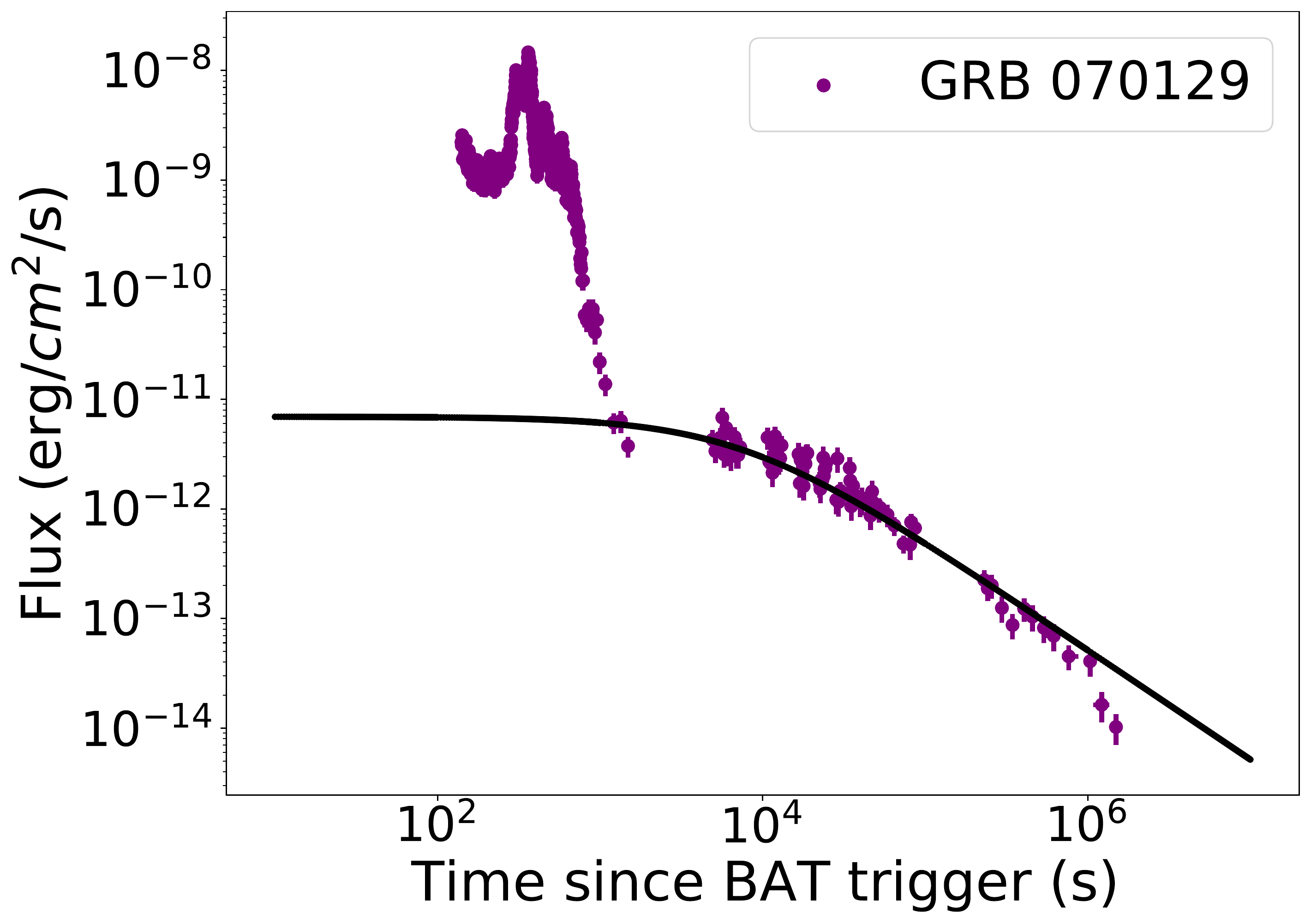}
	\includegraphics[width=0.31\hsize]{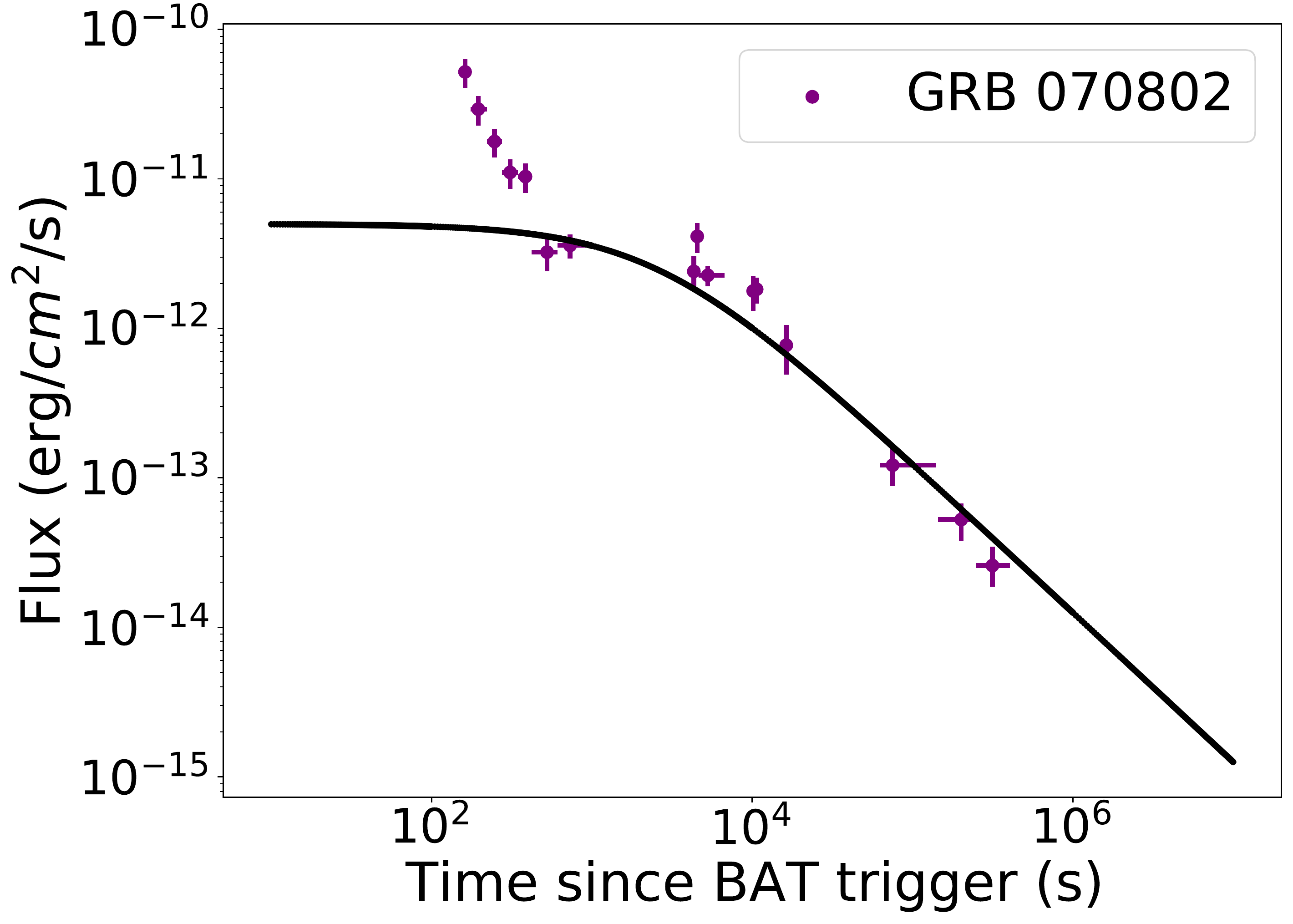}
	\includegraphics[width=0.31\hsize]{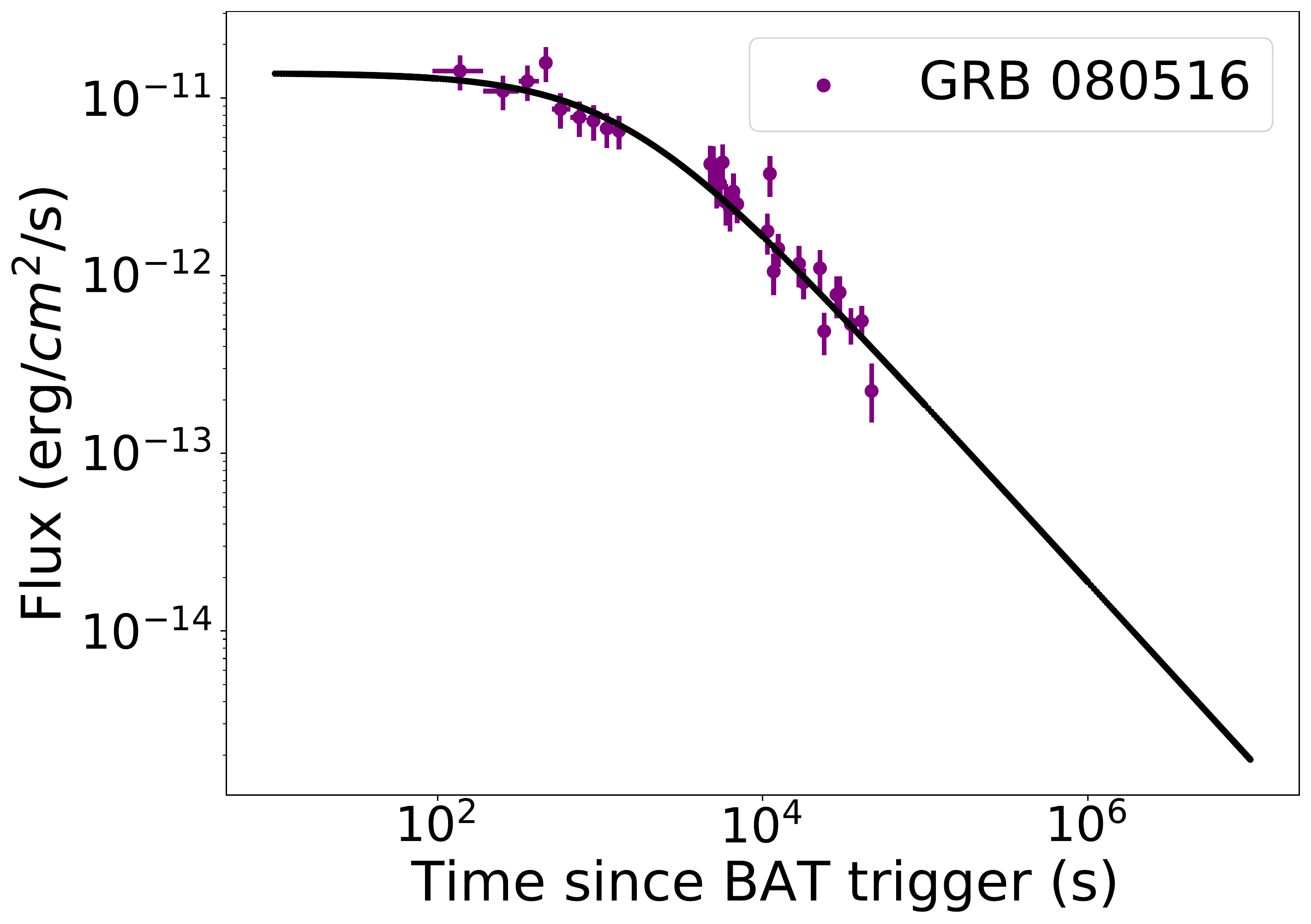}
	\includegraphics[width=0.31\hsize]{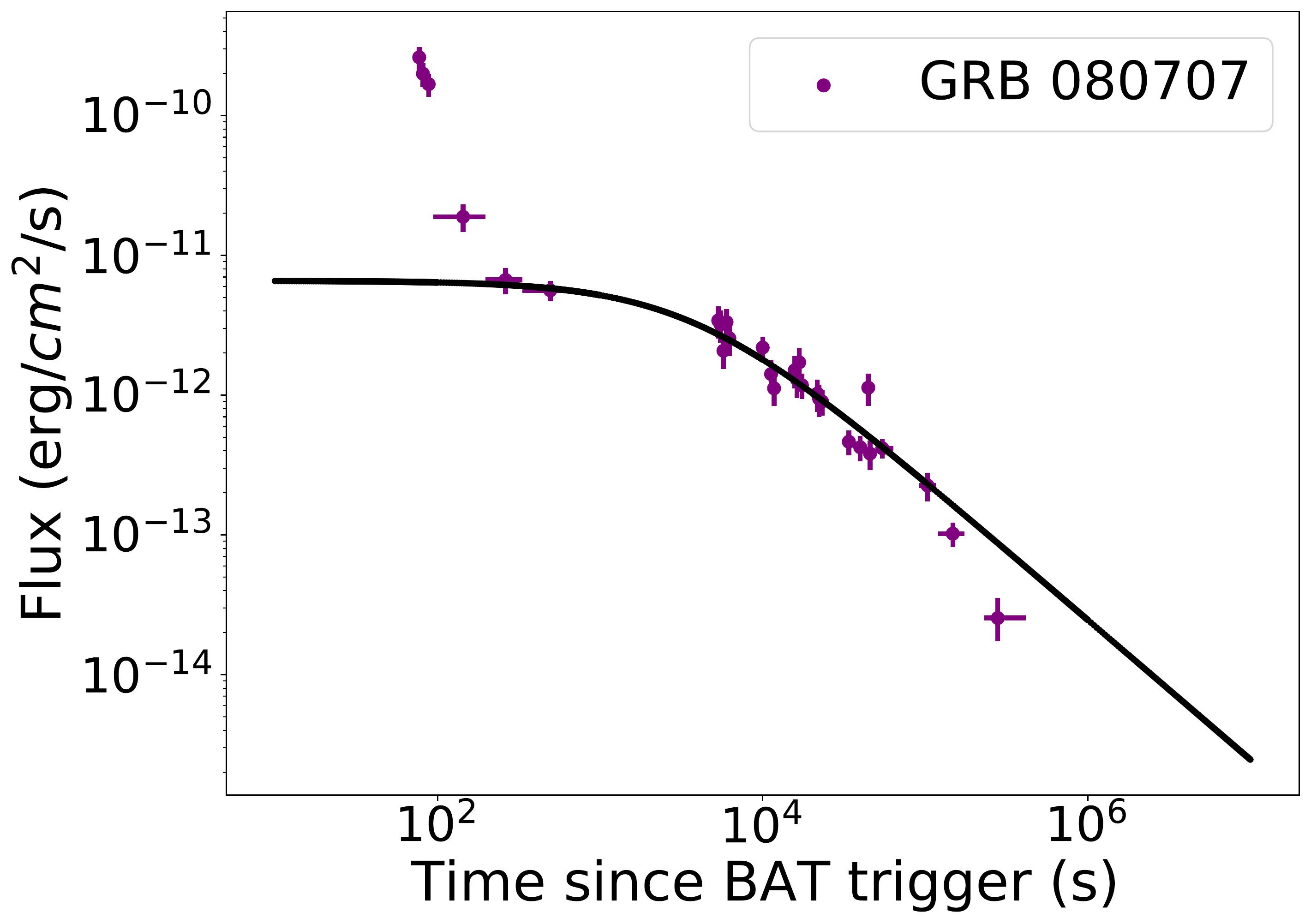}
	\includegraphics[width=0.31\hsize]{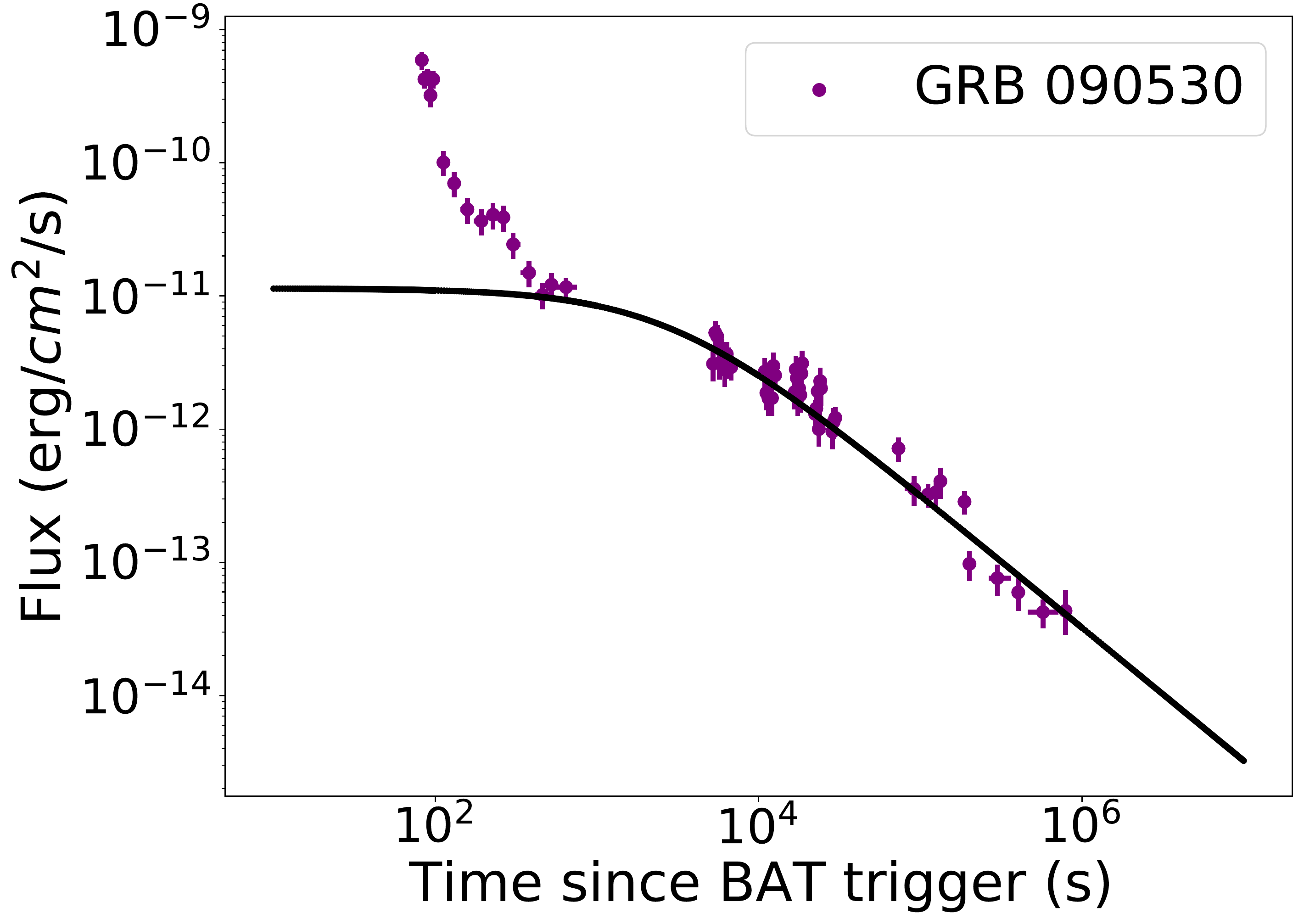}
	\includegraphics[width=0.31\hsize]{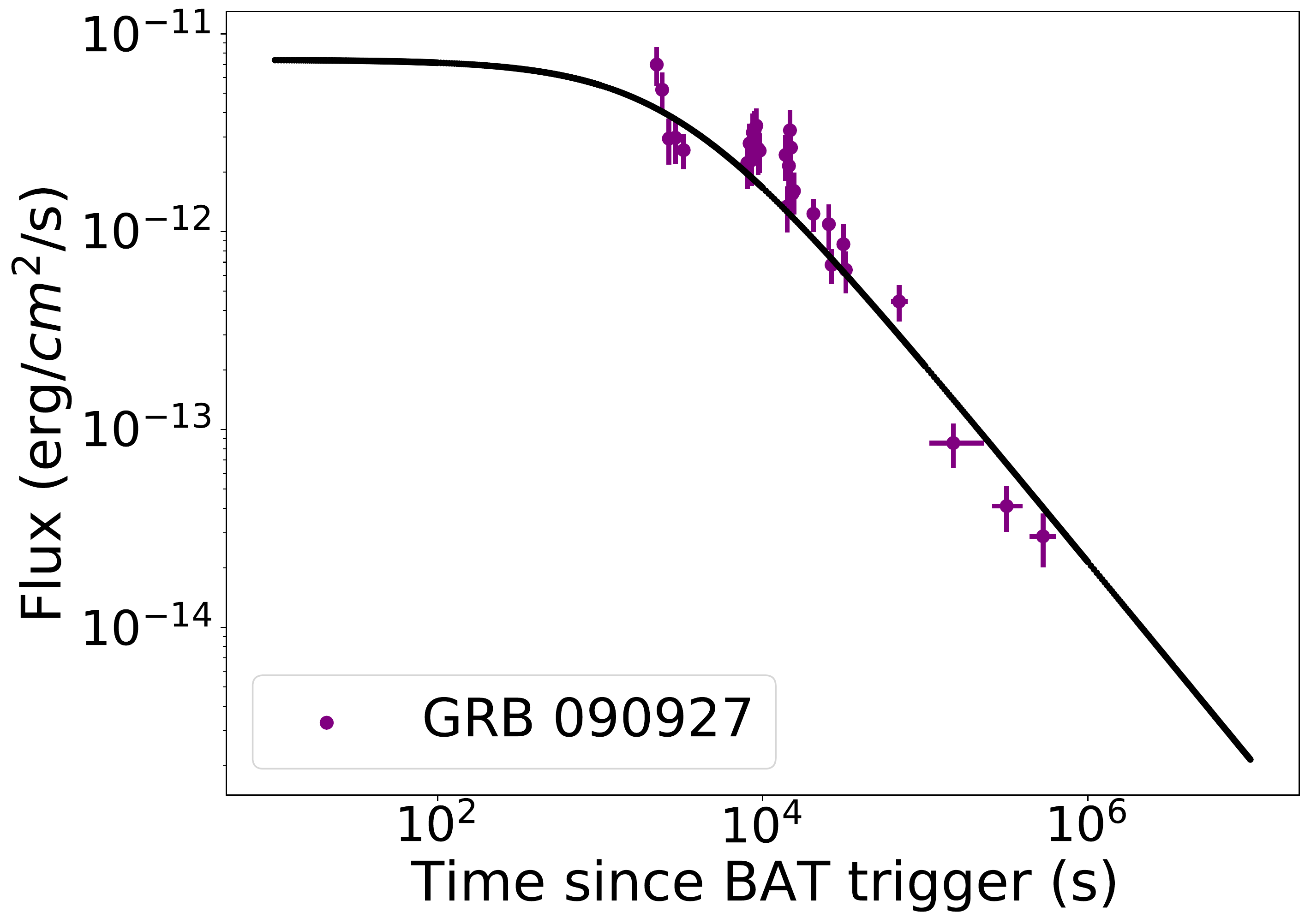}
	\includegraphics[width=0.31\hsize]{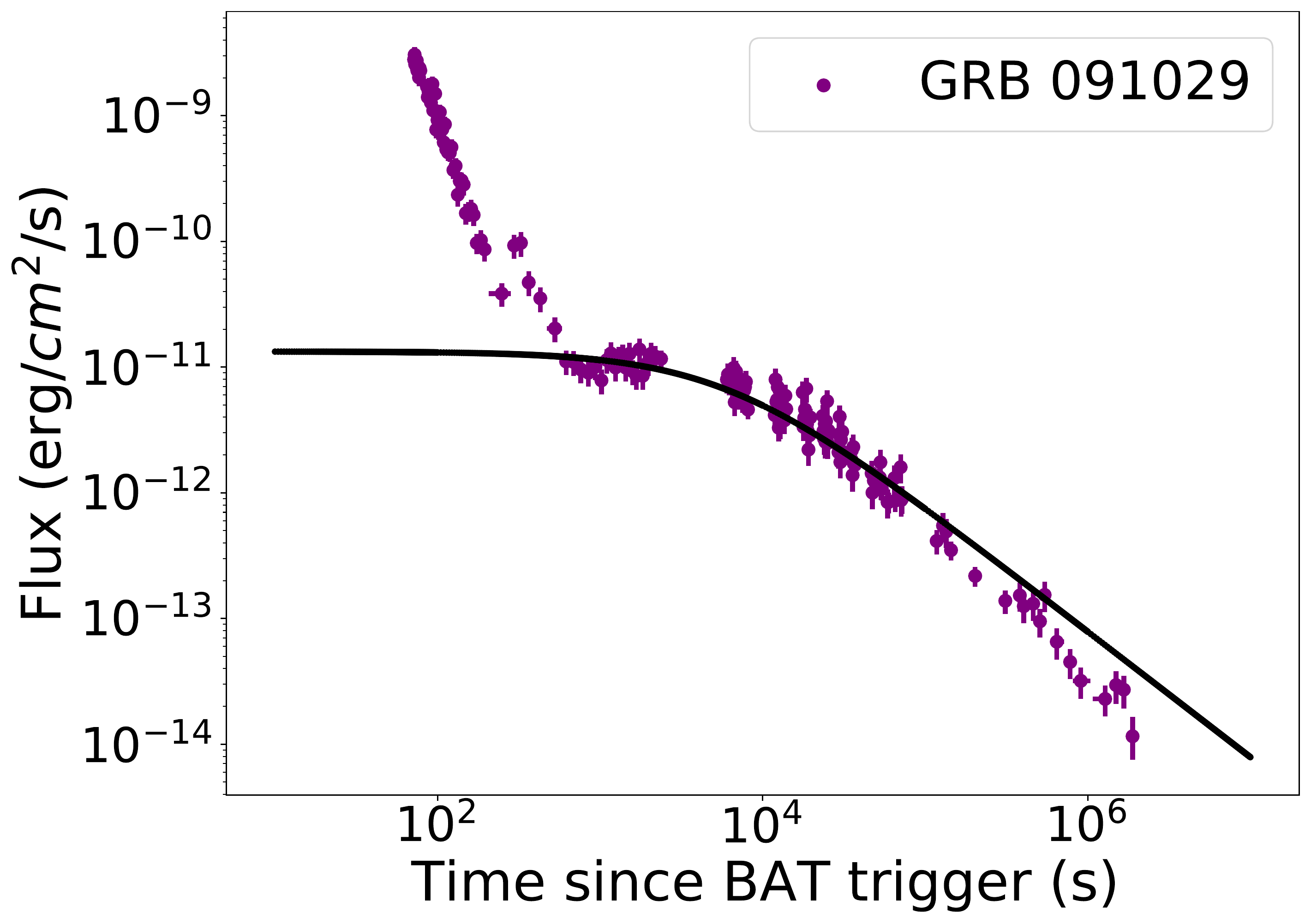}
	\includegraphics[width=0.31\hsize]{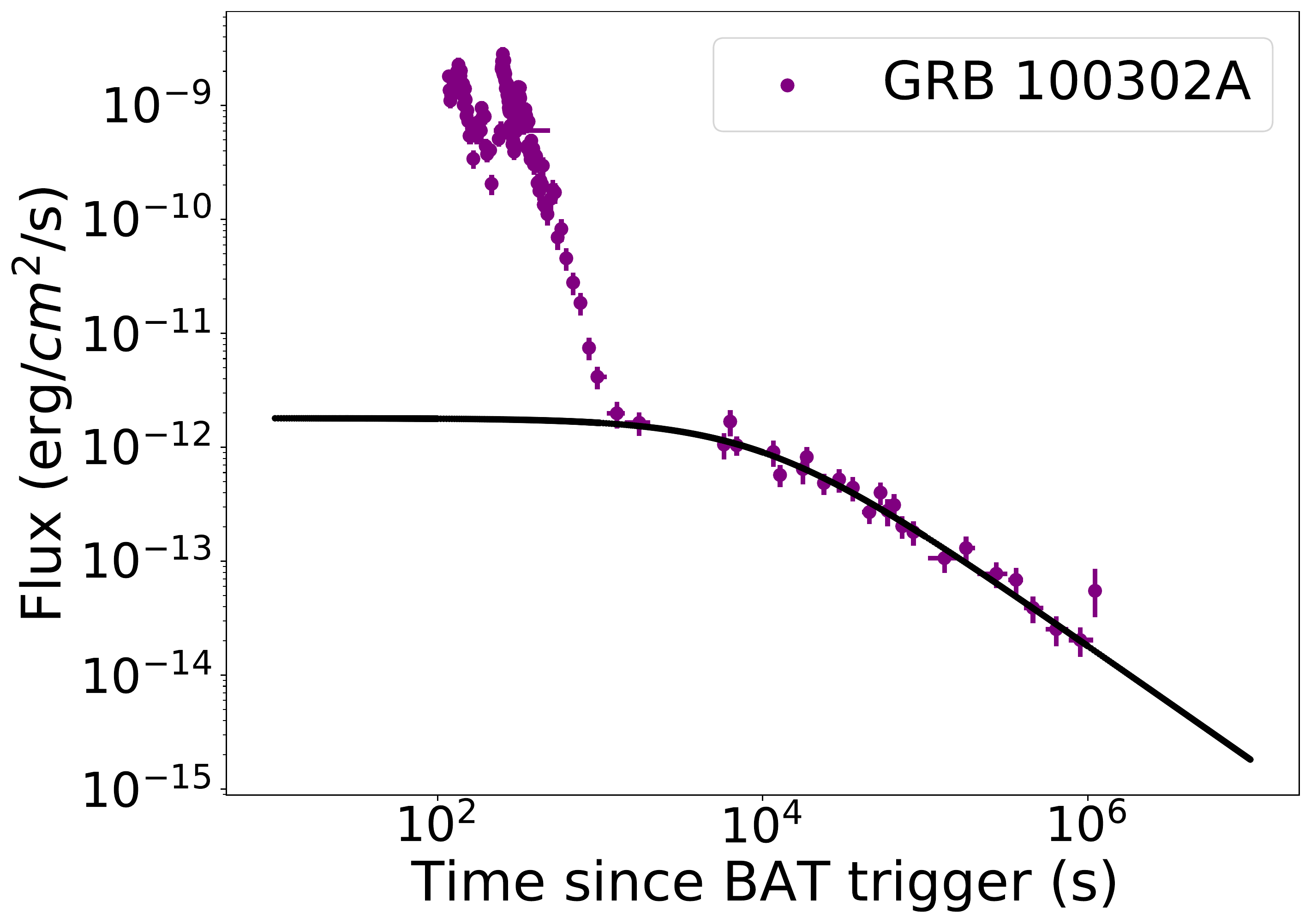}
	\includegraphics[width=0.31\hsize]{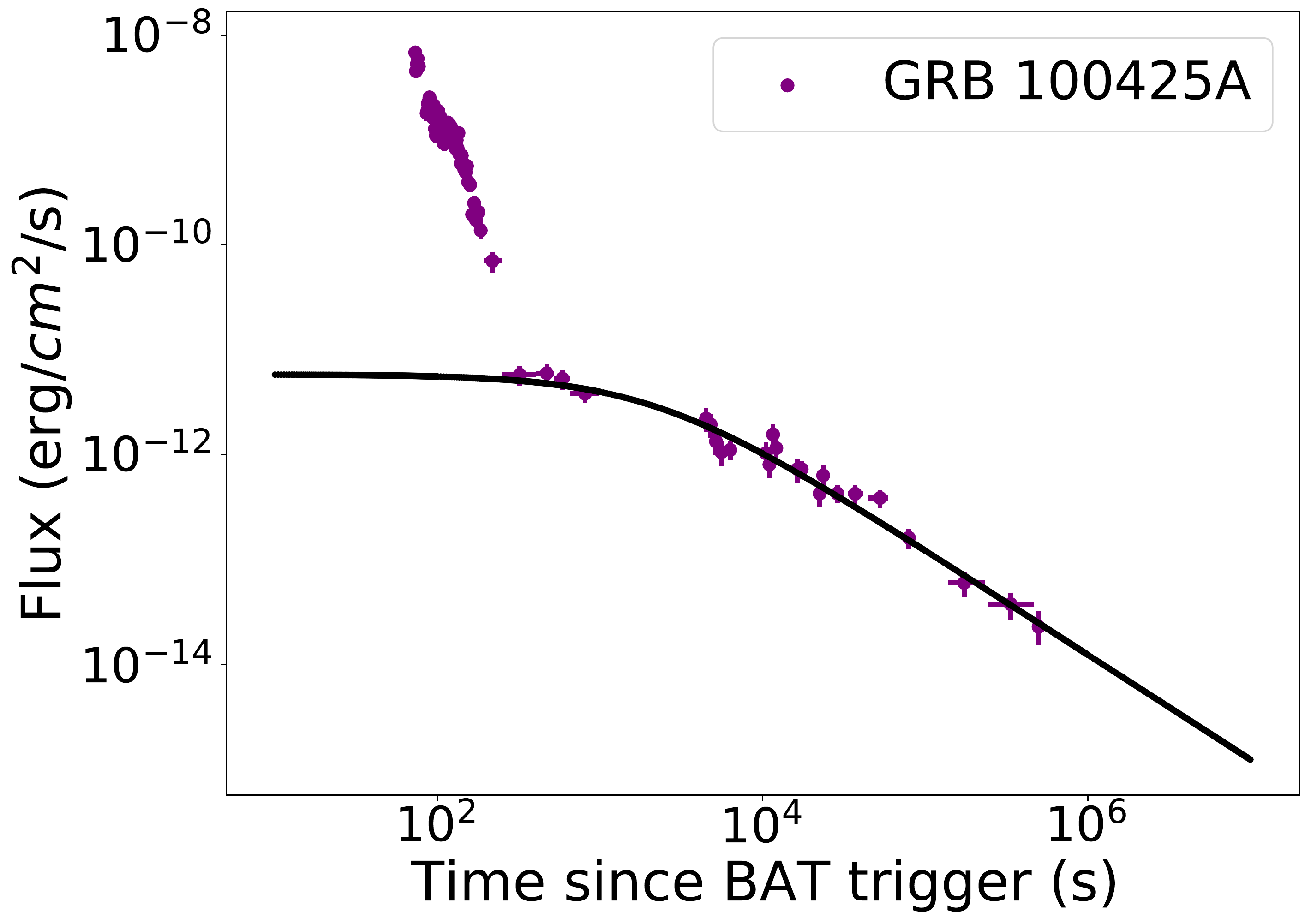}
	\includegraphics[width=0.31\hsize]{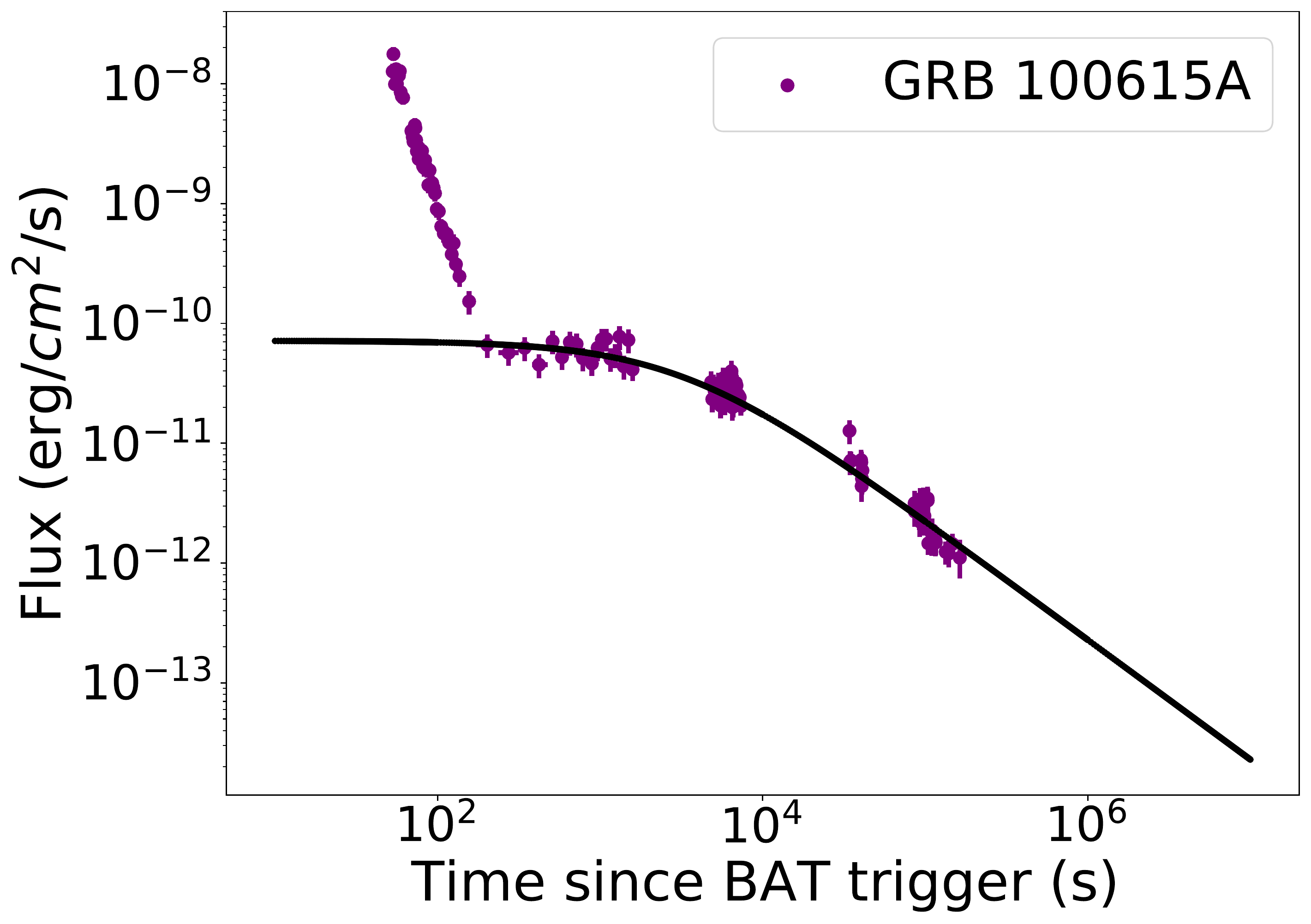}

	\caption{XRT light curves (0.3-10 keV) of long GRBs in GW-LGRBs sample. The black solid curves are the best fits with a
		smooth power-law model to the data (purple points).}
	\label{F:LGRB1}       
\end{figure}

\addtocounter{figure}{-1}
\begin{figure}
	\centering
	\includegraphics[width=0.31\hsize]{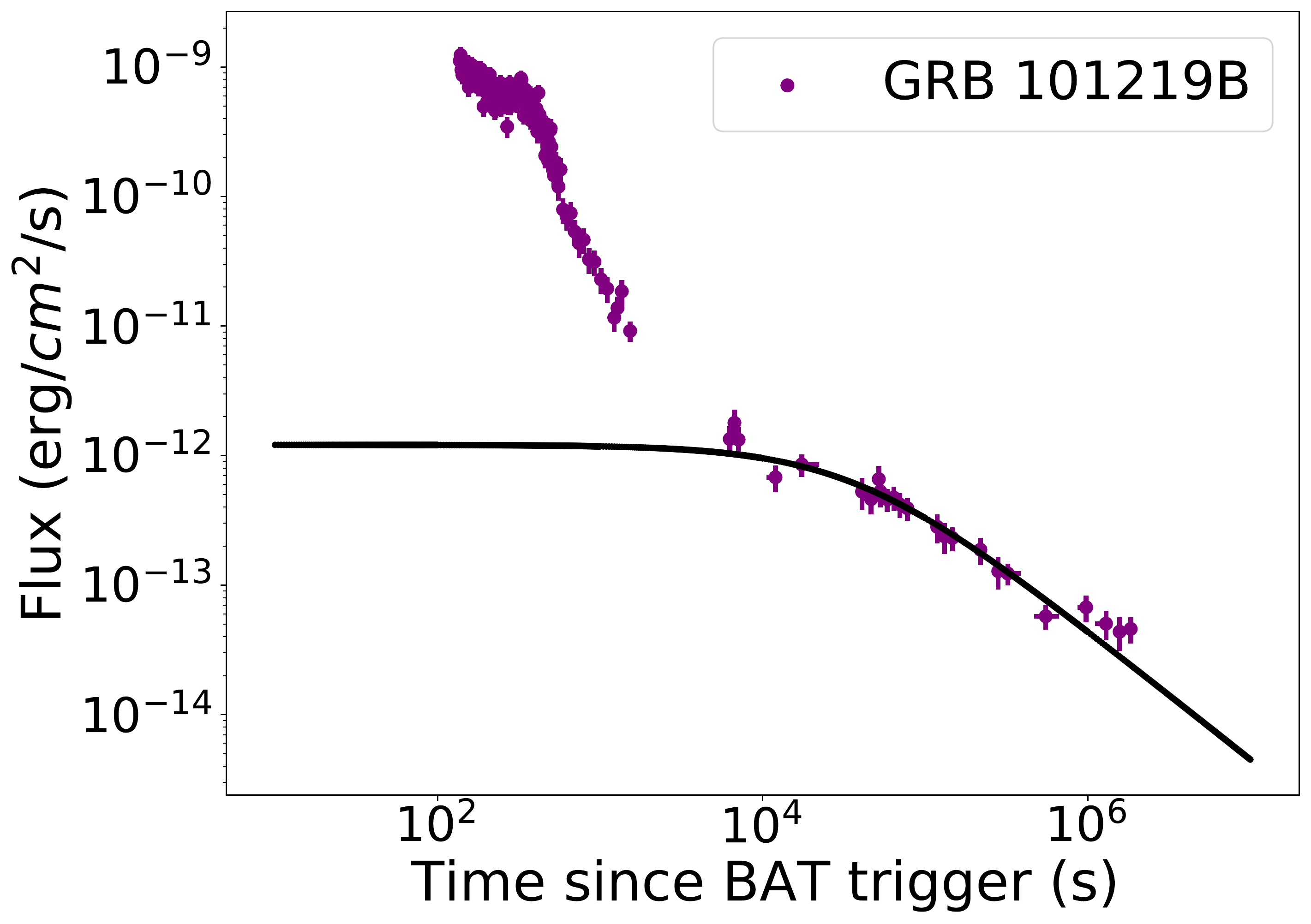}
	\includegraphics[width=0.31\hsize]{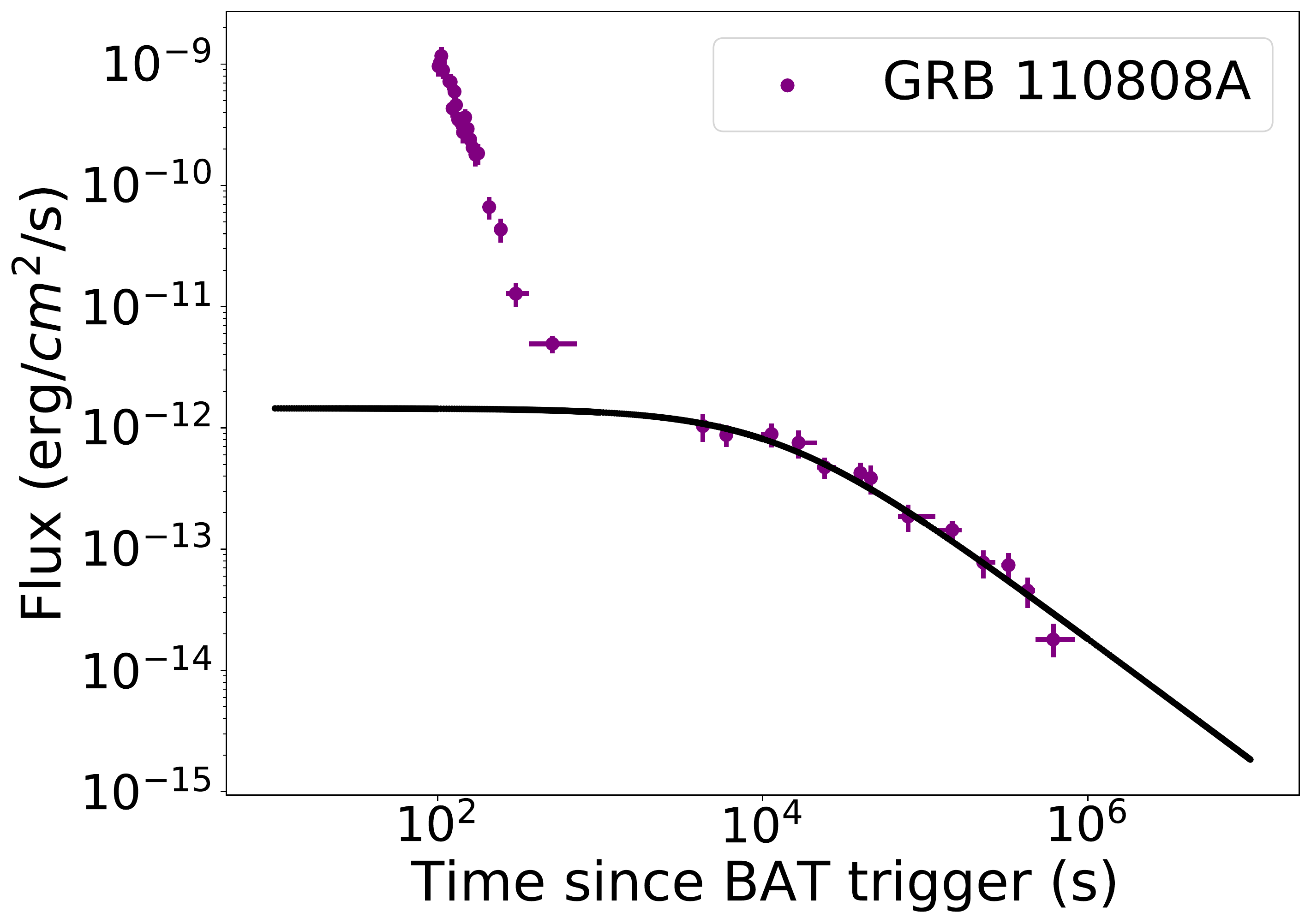}
	\includegraphics[width=0.31\hsize]{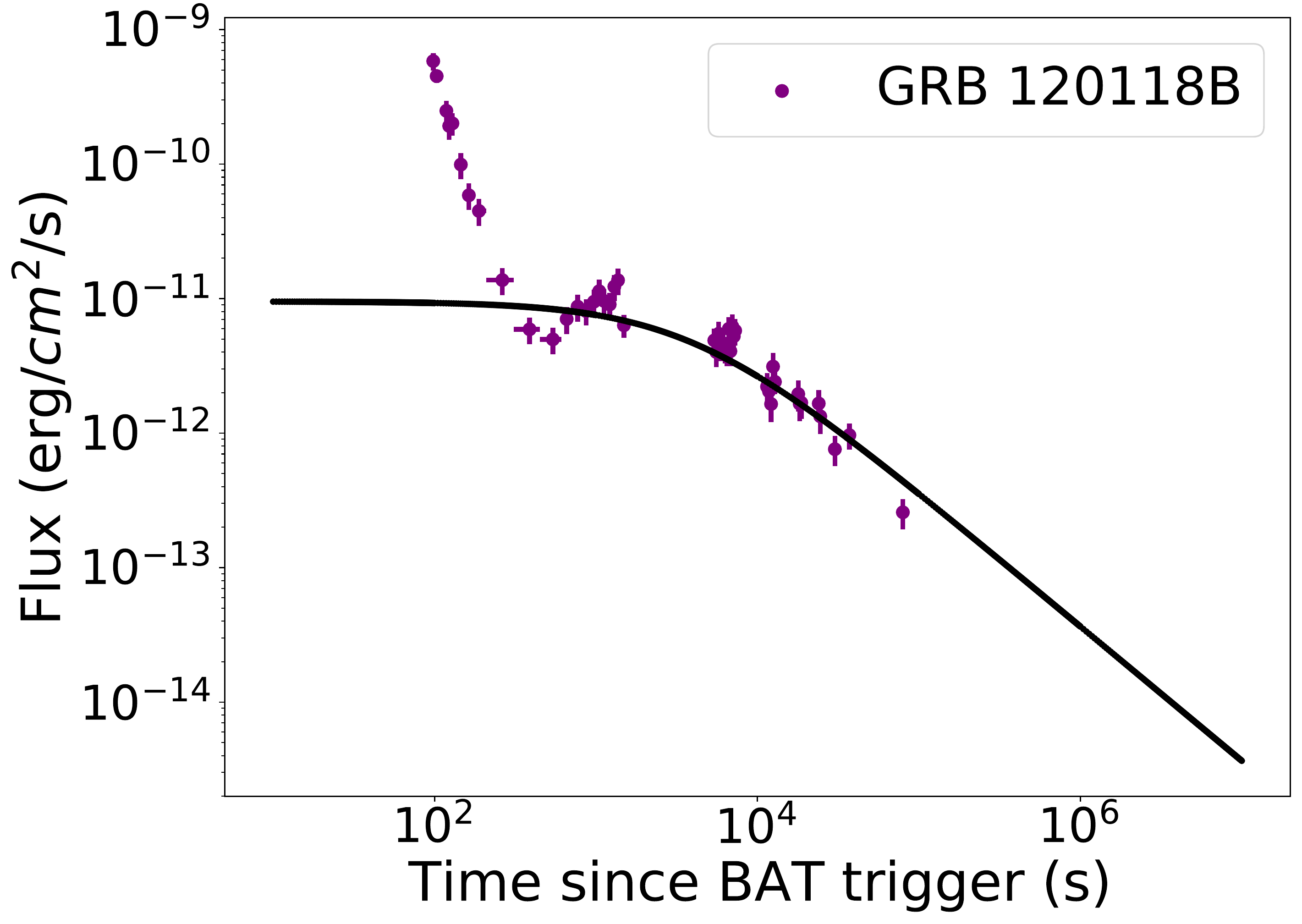}
	\includegraphics[width=0.31\hsize]{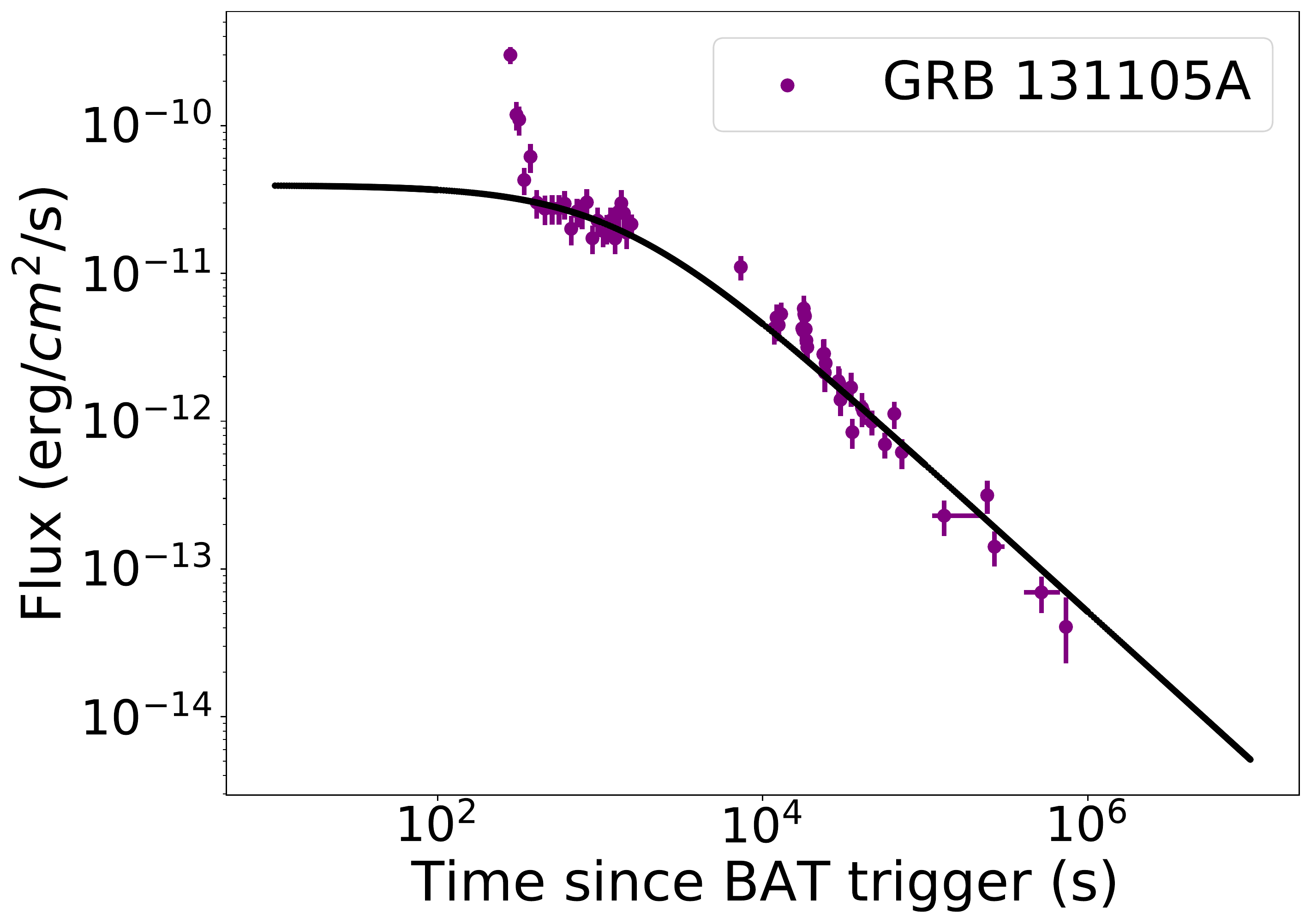}
	\includegraphics[width=0.31\hsize]{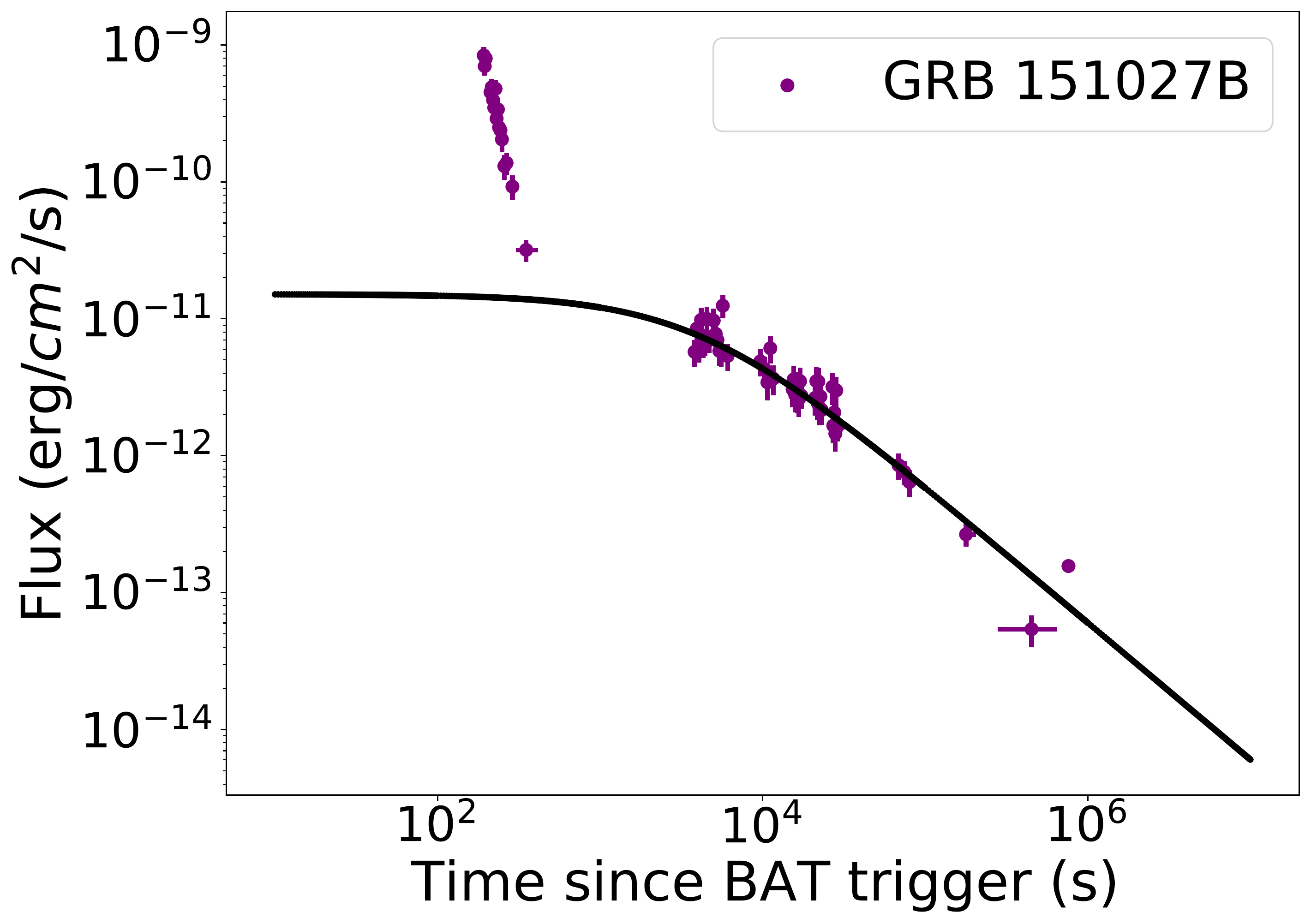}
	\includegraphics[width=0.31\hsize]{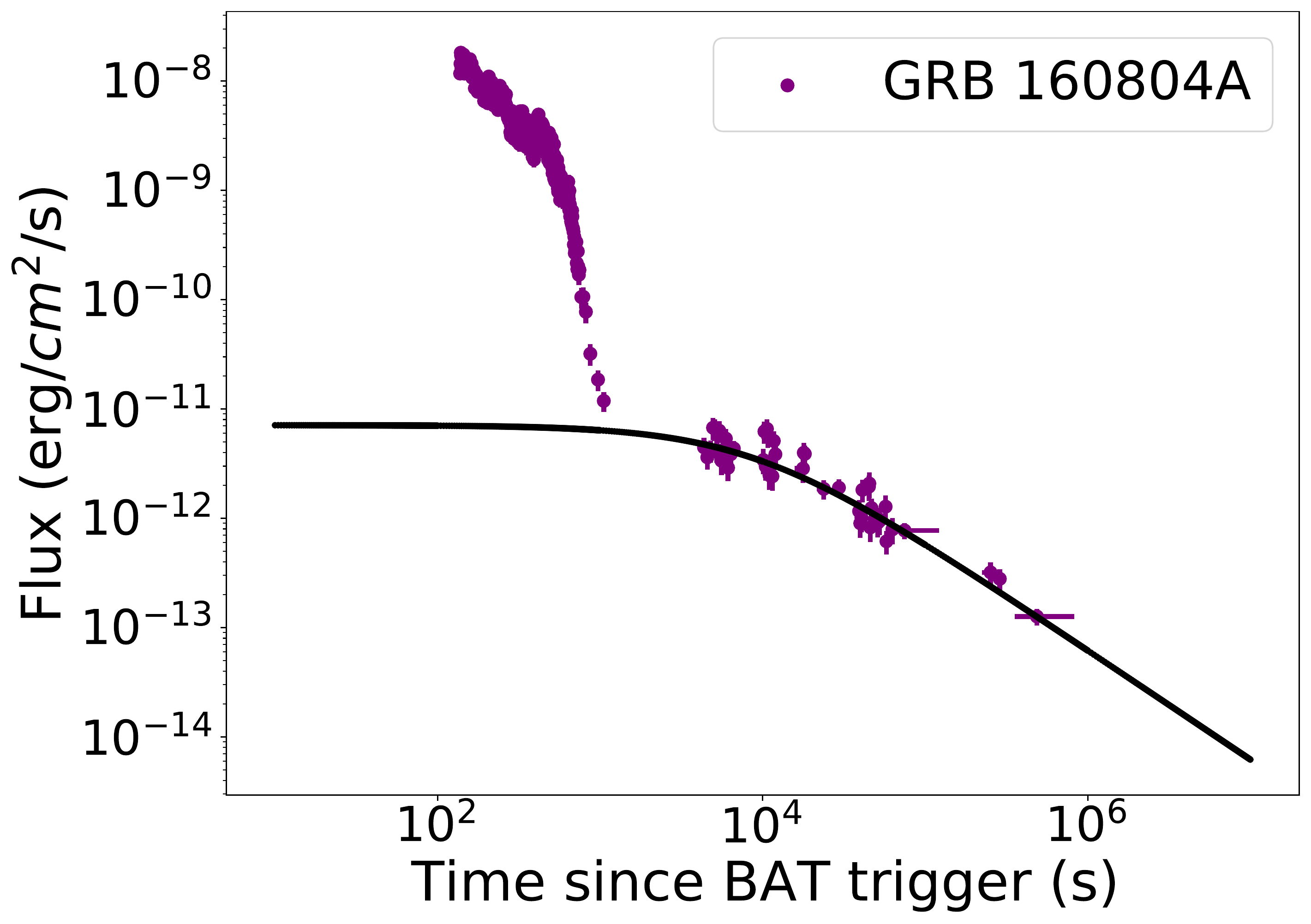}
	\includegraphics[width=0.31\hsize]{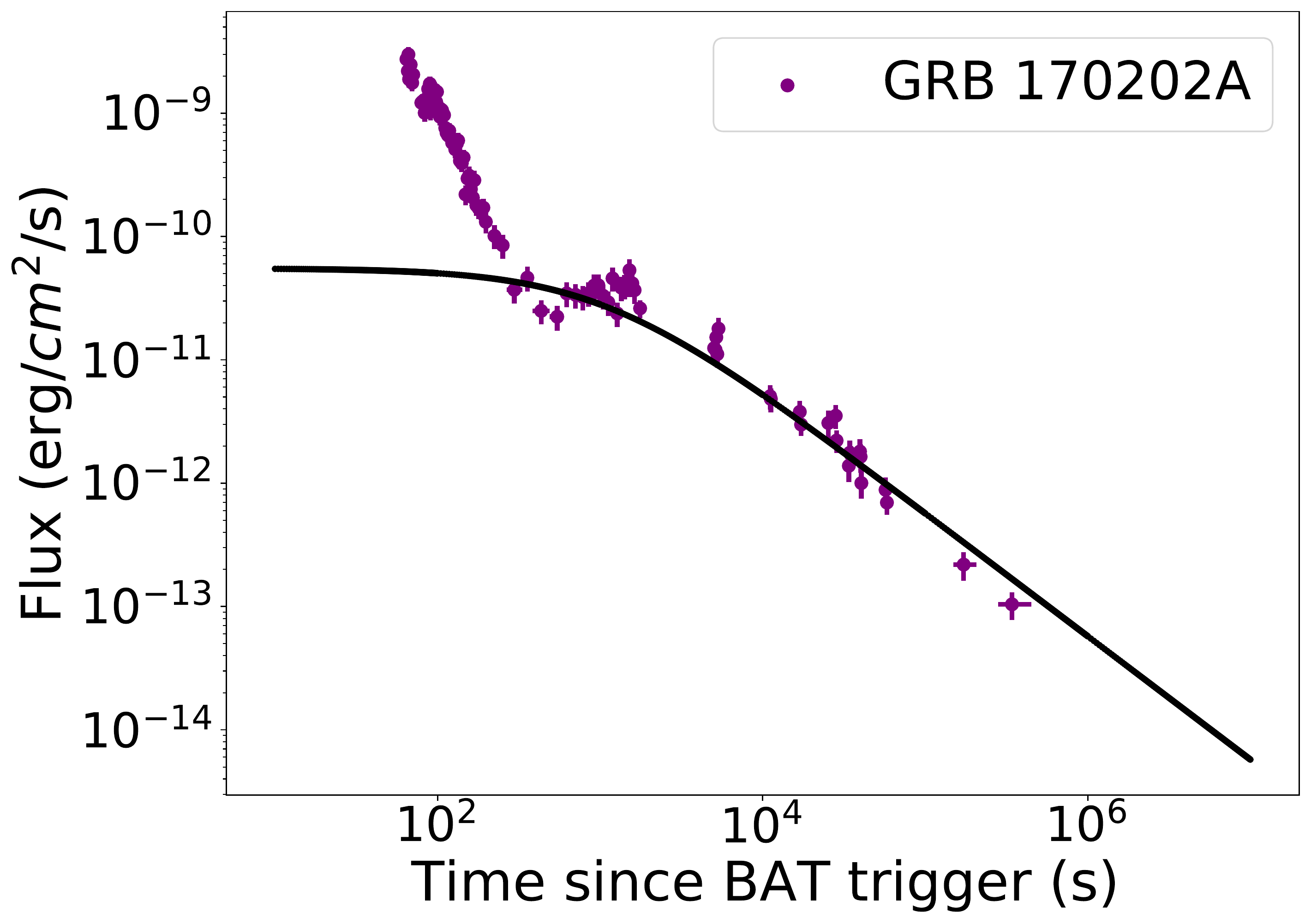}
	\includegraphics[width=0.31\hsize]{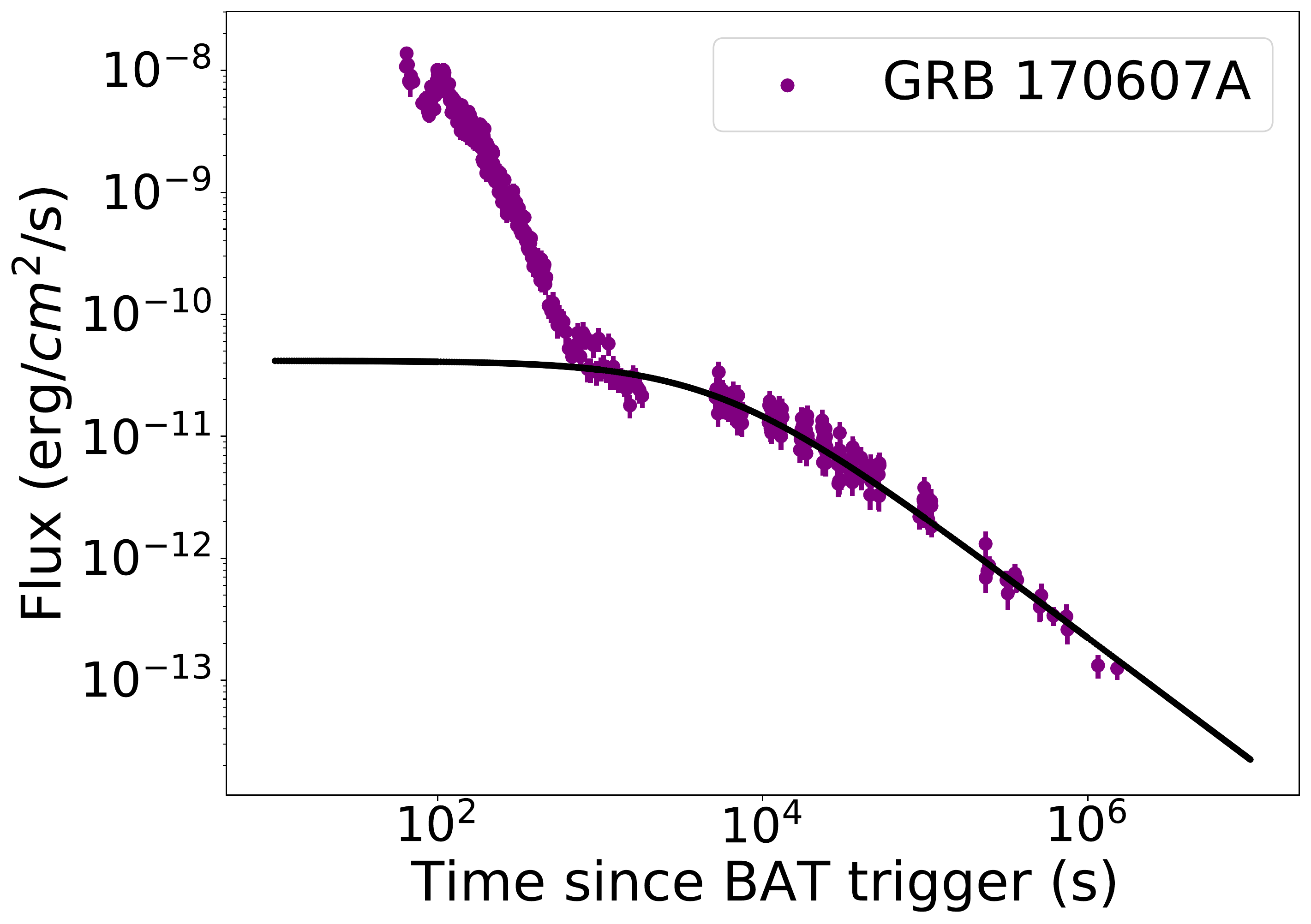}
	\includegraphics[width=0.31\hsize]{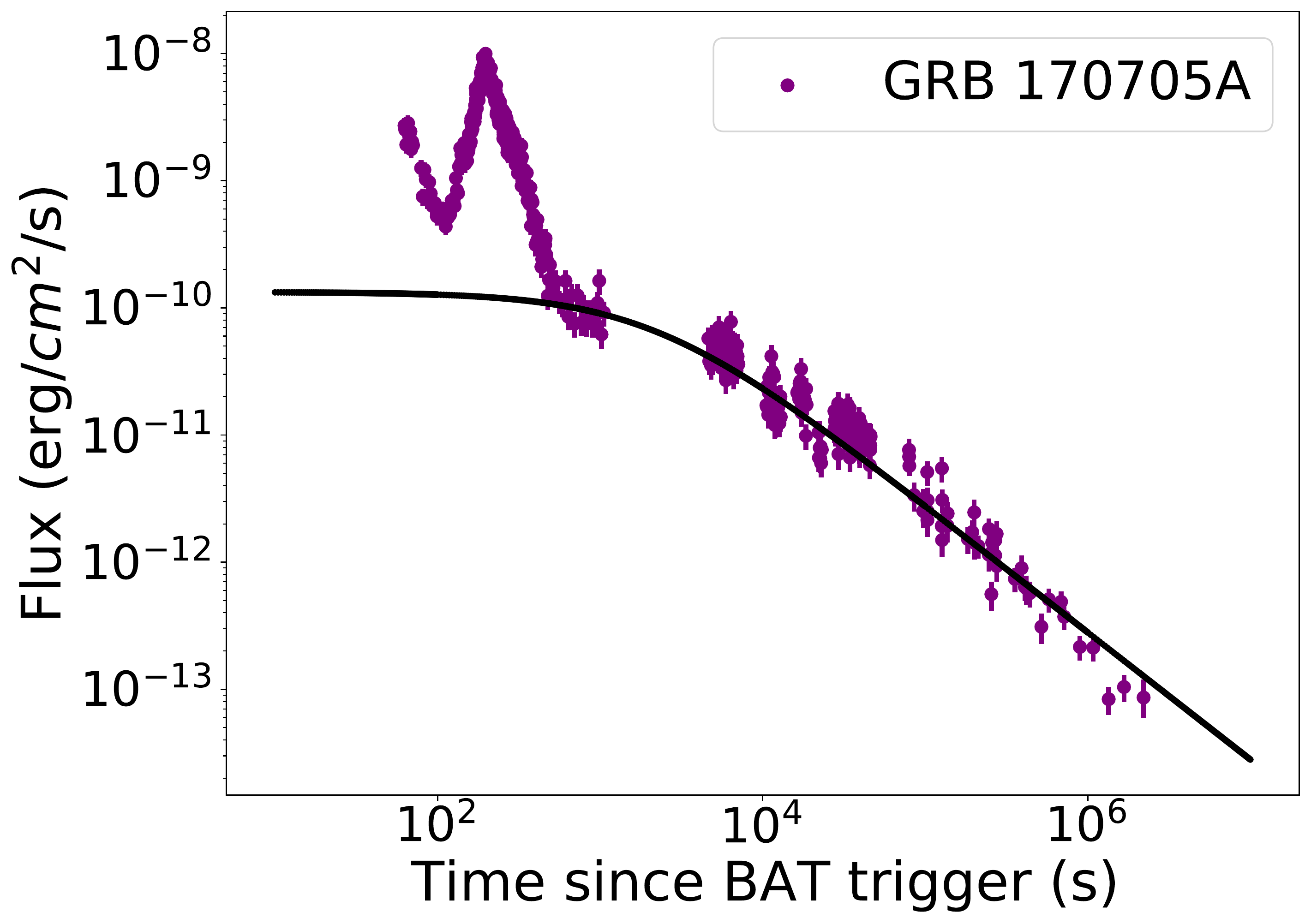}
	\caption{(Continued)}
	\label{F:LGRB2}       
\end{figure}

\label{lastpage}
\end{document}